\documentclass[twocolumn,letterpaper,aps,prd,superscriptaddress,showpacs,nofootinbib,floatfix]{revtex4}
\usepackage{graphicx}	% Include figure files

\usepackage{xspace}	% Include xspace

\newcommand{\pt}{\mbox{$p_T$}\xspace}

\newcommand{\meanpt}{\mbox{$\langle p_T \rangle$}\xspace}

\newcommand \sqs{\mbox{$\sqrt{s}$}\xspace}
\newcommand \sqsn{\mbox{$\sqrt{s_{_{NN}}}$}\xspace}
\newcommand{\pbar}{\mbox{$\overline{p}$}\xspace}
\newcommand{\mt}{\mbox{$m_T$}\xspace}

\newcommand{\xt}{\mbox{$x_T$}\xspace}
\newcommand{\piz}{\mbox{$\pi^{\rm 0}$}\xspace}
\newcommand{\ut} {\mbox{$\langle u_t \rangle$}}
\newcommand{\neff}{\mbox{$n_{\rm eff}$}\xspace}

\begin{document}


\title{Identified charged hadron production in $p+p$ collisions at $\sqs$~=~200 and 62.4~GeV}

\newcommand{\abilene}{Abilene Christian University, Abilene, Texas 79699, USA}
\newcommand{\acadsin}{Institute of Physics, Academia Sinica, Taipei 11529, Taiwan}
\newcommand{\banaras}{Department of Physics, Banaras Hindu University, Varanasi 221005, India}
\newcommand{\barc}{Bhabha Atomic Research Centre, Bombay 400 085, India}
\newcommand{\bnlcoll}{Collider-Accelerator Department, Brookhaven National Laboratory, Upton, New York 11973-5000, USA}
\newcommand{\bnlphys}{Physics Department, Brookhaven National Laboratory, Upton, New York 11973-5000, USA}
\newcommand{\caucr}{University of California - Riverside, Riverside, California 92521, USA}
\newcommand{\charlesczech}{Charles University, Ovocn\'{y} trh 5, Praha 1, 116 36, Prague, Czech Republic}
\newcommand{\ciae}{China Institute of Atomic Energy (CIAE), Beijing, People's Republic of China}
\newcommand{\cns}{Center for Nuclear Study, Graduate School of Science, University of Tokyo, 7-3-1 Hongo, Bunkyo, Tokyo 113-0033, Japan}
\newcommand{\colorado}{University of Colorado, Boulder, Colorado 80309, USA}
\newcommand{\columbia}{Columbia University, New York, New York 10027 and Nevis Laboratories, Irvington, New York 10533, USA}
\newcommand{\czechtech}{Czech Technical University, Zikova 4, 166 36 Prague 6, Czech Republic}
\newcommand{\dapnia}{Dapnia, CEA Saclay, F-91191, Gif-sur-Yvette, France}
\newcommand{\debrecen}{Debrecen University, H-4010 Debrecen, Egyetem t{\'e}r 1, Hungary}
\newcommand{\elte}{ELTE, E{\"o}tv{\"o}s Lor{\'a}nd University, H - 1117 Budapest, P{\'a}zm{\'a}ny P. s. 1/A, Hungary}
\newcommand{\fit}{Florida Institute of Technology, Melbourne, Florida 32901, USA}
\newcommand{\fsu}{Florida State University, Tallahassee, Florida 32306, USA}
\newcommand{\gsu}{Georgia State University, Atlanta, Georgia 30303, USA}
\newcommand{\hiroshima}{Hiroshima University, Kagamiyama, Higashi-Hiroshima 739-8526, Japan}
\newcommand{\ihepprot}{IHEP Protvino, State Research Center of Russian Federation, Institute for High Energy Physics, Protvino, 142281, Russia}
\newcommand{\illuiuc}{University of Illinois at Urbana-Champaign, Urbana, Illinois 61801, USA}
\newcommand{\instpasczech}{Institute of Physics, Academy of Sciences of the Czech Republic, Na Slovance 2, 182 21 Prague 8, Czech Republic}
\newcommand{\isu}{Iowa State University, Ames, Iowa 50011, USA}
\newcommand{\jinrdubna}{Joint Institute for Nuclear Research, 141980 Dubna, Moscow Region, Russia}
\newcommand{\kek}{KEK, High Energy Accelerator Research Organization, Tsukuba, Ibaraki 305-0801, Japan}
\newcommand{\kfki}{KFKI Research Institute for Particle and Nuclear Physics of the Hungarian Academy of Sciences (MTA KFKI RMKI), H-1525 Budapest 114, POBox 49, Budapest, Hungary}
\newcommand{\korea}{Korea University, Seoul, 136-701, Korea}
\newcommand{\kurchatov}{Russian Research Center ``Kurchatov Institute", Moscow, 123098 Russia}
\newcommand{\kyoto}{Kyoto University, Kyoto 606-8502, Japan}
\newcommand{\labllr}{Laboratoire Leprince-Ringuet, Ecole Polytechnique, CNRS-IN2P3, Route de Saclay, F-91128, Palaiseau, France}
\newcommand{\lawllnl}{Lawrence Livermore National Laboratory, Livermore, California 94550, USA}
\newcommand{\losalamos}{Los Alamos National Laboratory, Los Alamos, New Mexico 87545, USA}
\newcommand{\lpc}{LPC, Universit{\'e} Blaise Pascal, CNRS-IN2P3, Clermont-Fd, 63177 Aubiere Cedex, France}
\newcommand{\lund}{Department of Physics, Lund University, Box 118, SE-221 00 Lund, Sweden}
\newcommand{\mass}{Department of Physics, University of Massachusetts, Amherst, Massachusetts 01003-9337, USA }
\newcommand{\muenster}{Institut f\"ur Kernphysik, University of Muenster, D-48149 Muenster, Germany}
\newcommand{\muhlenberg}{Muhlenberg College, Allentown, Pennsylvania 18104-5586, USA}
\newcommand{\myongji}{Myongji University, Yongin, Kyonggido 449-728, Korea}
\newcommand{\nagasaki}{Nagasaki Institute of Applied Science, Nagasaki-shi, Nagasaki 851-0193, Japan}
\newcommand{\newmex}{University of New Mexico, Albuquerque, New Mexico 87131, USA }
\newcommand{\nmsu}{New Mexico State University, Las Cruces, New Mexico 88003, USA}
\newcommand{\ornl}{Oak Ridge National Laboratory, Oak Ridge, Tennessee 37831, USA}
\newcommand{\orsay}{IPN-Orsay, Universite Paris Sud, CNRS-IN2P3, BP1, F-91406, Orsay, France}
\newcommand{\peking}{Peking University, Beijing, People's Republic of China}
\newcommand{\pnpi}{PNPI, Petersburg Nuclear Physics Institute, Gatchina, Leningrad region, 188300, Russia}
\newcommand{\riken}{RIKEN Nishina Center for Accelerator-Based Science, Wako, Saitama 351-0198, Japan}
\newcommand{\rikjrbrc}{RIKEN BNL Research Center, Brookhaven National Laboratory, Upton, New York 11973-5000, USA}
\newcommand{\rikkyo}{Physics Department, Rikkyo University, 3-34-1 Nishi-Ikebukuro, Toshima, Tokyo 171-8501, Japan}
\newcommand{\saispbstu}{Saint Petersburg State Polytechnic University, St. Petersburg, 195251 Russia}
\newcommand{\saopaulo}{Universidade de S{\~a}o Paulo, Instituto de F\'{\i}sica, Caixa Postal 66318, S{\~a}o Paulo CEP05315-970, Brazil}
\newcommand{\seoulnat}{Seoul National University, Seoul, Korea}
\newcommand{\stonybrkc}{Chemistry Department, Stony Brook University, SUNY, Stony Brook, New York 11794-3400, USA}
\newcommand{\stonycrkp}{Department of Physics and Astronomy, Stony Brook University, SUNY, Stony Brook, New York 11794-3400, USA}
\newcommand{\subatech}{SUBATECH (Ecole des Mines de Nantes, CNRS-IN2P3, Universit{\'e} de Nantes) BP 20722 - 44307, Nantes, France}
\newcommand{\tenn}{University of Tennessee, Knoxville, Tennessee 37996, USA}
\newcommand{\titech}{Department of Physics, Tokyo Institute of Technology, Oh-okayama, Meguro, Tokyo 152-8551, Japan}
\newcommand{\tsukuba}{Institute of Physics, University of Tsukuba, Tsukuba, Ibaraki 305, Japan}
\newcommand{\vandy}{Vanderbilt University, Nashville, Tennessee 37235, USA}
\newcommand{\waseda}{Waseda University, Advanced Research Institute for Science and Engineering, 17 Kikui-cho, Shinjuku-ku, Tokyo 162-0044, Japan}
\newcommand{\weizmann}{Weizmann Institute, Rehovot 76100, Israel}
\newcommand{\yonsei}{Yonsei University, IPAP, Seoul 120-749, Korea}
\affiliation{\abilene}
\affiliation{\acadsin}
\affiliation{\banaras}
\affiliation{\barc}
\affiliation{\bnlcoll}
\affiliation{\bnlphys}
\affiliation{\caucr}
\affiliation{\charlesczech}
\affiliation{\ciae}
\affiliation{\cns}
\affiliation{\colorado}
\affiliation{\columbia}
\affiliation{\czechtech}
\affiliation{\dapnia}
\affiliation{\debrecen}
\affiliation{\elte}
\affiliation{\fit}
\affiliation{\fsu}
\affiliation{\gsu}
\affiliation{\hiroshima}
\affiliation{\ihepprot}
\affiliation{\illuiuc}
\affiliation{\instpasczech}
\affiliation{\isu}
\affiliation{\jinrdubna}
\affiliation{\kek}
\affiliation{\kfki}
\affiliation{\korea}
\affiliation{\kurchatov}
\affiliation{\kyoto}
\affiliation{\labllr}
\affiliation{\lawllnl}
\affiliation{\losalamos}
\affiliation{\lpc}
\affiliation{\lund}
\affiliation{\mass}
\affiliation{\muenster}
\affiliation{\muhlenberg}
\affiliation{\myongji}
\affiliation{\nagasaki}
\affiliation{\newmex}
\affiliation{\nmsu}
\affiliation{\ornl}
\affiliation{\orsay}
\affiliation{\peking}
\affiliation{\pnpi}
\affiliation{\riken}
\affiliation{\rikjrbrc}
\affiliation{\rikkyo}
\affiliation{\saispbstu}
\affiliation{\saopaulo}
\affiliation{\seoulnat}
\affiliation{\stonybrkc}
\affiliation{\stonycrkp}
\affiliation{\subatech}
\affiliation{\tenn}
\affiliation{\titech}
\affiliation{\tsukuba}
\affiliation{\vandy}
\affiliation{\waseda}
\affiliation{\weizmann}
\affiliation{\yonsei}
\author{A.~Adare} \affiliation{\colorado}
\author{S.~Afanasiev} \affiliation{\jinrdubna}
\author{C.~Aidala} \affiliation{\columbia} \affiliation{\mass}
\author{N.N.~Ajitanand} \affiliation{\stonybrkc}
\author{Y.~Akiba} \affiliation{\riken} \affiliation{\rikjrbrc}
\author{H.~Al-Bataineh} \affiliation{\nmsu}
\author{J.~Alexander} \affiliation{\stonybrkc}
\author{K.~Aoki} \affiliation{\kyoto} \affiliation{\riken}
\author{L.~Aphecetche} \affiliation{\subatech}
\author{R.~Armendariz} \affiliation{\nmsu}
% ...
\collaboration{PHENIX Collaboration} \noaffiliation

\date{\today}

%%%%%%%%%%%%%%%%%%%%%%%%%%%%%%%%%%%%%%%%%%%%%%%%%%%%%%%%%%%%%%%%%%%%%%%%%%%% |
\begin{abstract}

Transverse momentum distributions and yields for $\pi^{\pm}$, $K^{\pm}$, 
$p$ and $\overline{p}$ in $p+p$ collisions at $\sqrt{s}$~=~200 and 
62.4~GeV at midrapidity are measured by the PHENIX experiment at the 
Relativistic Heavy Ion Collider (RHIC). These data provide important 
baseline spectra for comparisons with identified particle spectra in 
heavy ion collisions at RHIC. We present the inverse slope parameter 
$T_{\rm inv}$, mean transverse momentum $\langle p_T \rangle$ and yield 
per unit rapidity $dN/dy$ at each energy, and compare them to other 
measurements at different $\sqrt{s}$ in $p+p$ and $p+\overline{p}$ 
collisions.  We also present the scaling properties such as $m_T$ 
scaling, $x_T$ scaling on the $p_T$ spectra between different energies.  
To discuss the mechanism of the particle production in $p+p$ collisions, 
the measured spectra are compared to next-to-leading-order or 
next-to-leading-logarithmic perturbative quantum chromodynamics 
calculations.

\end{abstract}

	
% For heavy ion papers we usually use just the one above (max is 4)
%%%%%%%%% Examples for p+p and spin papers include:
% PPG031:  \pacs{14.20.Dh, 13.60.Hb, 21.10.Hw, 25.40.Fq}
% PPG050:  \pacs{14.20.Dh, 25.40.Ep, 13.85.Ni, 13.88.+e}
% PPG037:  \pacs{13.85.Qk, 13.20.Fc, 13.20.He, 25.75.Dw}

% It is optional to also add (uncomment):
% \keywords{}

\maketitle

%%%%%%%%%%%%%%%%%%%
% 1. INTRODUCTION %
%%%%%%%%%%%%%%%%%%%
\section{INTRODUCTION}
\label{sec:intro}

Single particle spectra of identified hadrons in high energy elementary 
collisions have attracted physicists for many decades due to their 
fundamental nature and simplicity. Particle production, in general, can 
be categorized into two different regimes depending on the transverse 
momentum of the hadrons.  One is soft multiparticle production, dominant 
at low transverse momentum ($\pt \leq 2$ GeV/$c$), which corresponds to 
the $\sim1$ fm scale of the nucleon radius described by constituent 
quarks.  Another regime is hard-scattering particle production, evident 
at high transverse momentum ($\pt\geq 2$ GeV/$c$) due to the hard 
scattering of point-like current quarks, which corresponds to a very 
short distance scale $\sim0.1$ fm~\cite{BBK1971} and contributes less 
than a few percent of the cross section for \sqs $\le$~200~GeV.  These 
two different regimes of particle production in $p+p$ collisions 
indicate that ``elementary'' $p+p$ collisions are actually rather 
complicated processes.  It is interesting to know where the ``soft-hard 
transition'' happens, and its beam energy and particle species 
dependences, since they have not yet been fully understood.

%% --- old days, 1950's~ ---
In soft particle production, cosmic ray physicists observed in the 1950s 
that the average transverse momentum of secondary particles is limited 
to $\sim$0.5 GeV/$c$, independent of the primary 
energy~\cite{Nishimura,Cocconi1958}.  Cocconi, Koester and 
Perkins~\cite{Cocconi1961} then proposed the prescient empirical formula 
for the transverse momentum spectrum of meson production:
 \begin{equation}
{d\sigma \over {\pt d\pt}}=A e^{-6\pt} , \qquad
\label{eq:CKP}
 \end{equation}
where $\pt$ is the transverse momentum in GeV/$c$ and $\meanpt = 2/6 
=$0.333 GeV/$c$.  The observation by Orear~\cite{Orear1964} that large 
angle $p+p$ elastic scattering measurements at AGS energies (10 to 30 
GeV in incident energy) ``can be fit by a single exponential in 
transverse momentum, and that this exponential is the very same 
exponential that describes the transverse momentum distribution of pions 
produced in nucleon-nucleon collisions'', led to the 
interpretation~\cite{Hagedorn:1994sc} that particle production is 
``statistical'' with Eq.~\ref{eq:CKP} as a thermal Boltzmann spectrum, 
with 1/6=0.167 GeV/$c$ representing the ``temperature'' $T$ at which the 
mesons or protons are emitted~\cite{ErwinLanderKo}.

%% --- description with mT ---

It was natural in a thermal scenario~\cite{Hagedorn1970, Barshay1972} to 
represent the invariant cross section as a function of the rapidity 
($y$) and the transverse mass ($\mt = \sqrt{\pt^2+m^2}$) with a 
universal temperature parameter $T$. This description explained well the 
observed successively increasing $\meanpt$ of $\pi$, $K$, $p$, $\Lambda$ 
with increasing rest 
mass~\cite{Boeggild1973,Deutschmann1974,Bartke1977}, and had the added 
advantage of explaining, by the simple factor $e^{-6(m_K-m_{\pi})}\sim$ 
12\% , the low value of $\sim$10\% observed for the $K/\pi$ ratio at low 
$\pt$ at ISR energies ($\sqrt{s}\sim$20--60 GeV)~\cite{Alper:1975jm}.

%% --- quark model ---

In 1964, the constituent quark model with SU(3) symmetry was introduced 
to explain the hadron flavor spectrum and the static properties of 
hadrons~\cite{GellMann,Zweig}. Later on, a dynamical model was developed 
to calculate the flavor dependence of identified hadrons in soft 
multi-particle production~\cite{Anisovich1974}, together with the 
inclusive reaction formalism~\cite{Feynman1969,Benecke1969,Mueller}. 
These theoretical studies on the particle production mechanism showed 
that there was much to be learned by simply measuring a single particle 
spectrum, and it brought the study of identified inclusive single 
particle production into the mainstream of $p+p$ physics.

% (removed, TC, May 1, 2010)
%
% However, in the constituent quark model, the 
% ``suppression'' of strange quarks, evident from the small $K/\pi$ ratio, 
% was not explained, but simply quantified~\cite{Anisovich1973} by a parameter 
% $\lambda$, which represents the ratio of the produced numbers of $s\bar{s}$ pairs
% to $u\bar{u}+d\bar{d}$ pairs.
%

%% --- soft multiparticle production (soft (1)) ---
One of the controversial issues in understanding soft multi-particle 
production in the 1950's was whether more than one meson could be 
produced in a single nucleon-nucleon collision (``multiple 
production''), or whether the multiple meson production observed in 
nucleon-nucleus ($p+A$) interactions was the result of several 
successive nucleon-nucleon collisions with each collision producing only 
a single meson (``plural production'')~\cite{Camerini1952}. The issue 
was decided when multiple meson production was first observed in 1954 at 
the Brookhaven Cosmotron in collisions between neutrons with energies up 
to 2.2 GeV and protons in a hydrogen filled cloud 
chamber~\cite{Fowler1954,Hagedorn:1994sc}.

%% --- statistical and hydrodynamical approach (soft (2)) ---

Then the observation of multi-particle production occurring not only in 
nucleon-nucleus ($p+A$) but also in nucleon-nucleon ($p+p$) collisions 
motivated Fermi and Landau to develop the statistical~\cite{Fermi} and 
hydrodynamical~\cite{Landau} approach to multi-particle production. 
Belenkij and Landau observed that although the statistical model of 
Fermi is sufficient to describe the particle numbers in terms of only a 
temperature and a chemical potential, this model has to be extended to 
hydrodynamics, when particle spectra are considered. They also noted 
that the domain of the applicability of ideal relativistic hydrodynamics 
coincides with the domain of the applicability of thermodynamical models 
in high energy $p+p$ collisions~\cite{Landau}.

%% --- p+p physics at ISR and RHIC (hard) ---

On the other hand, understanding of the particle production by hard 
scattering partons has been advanced by the appearance 
of a rich body of data in $p+p$ collisions at the 
CERN-ISR~\cite{Alper:1975jm,Alper_npb87,isr_lambda_pp63} in the 1970s, 
followed by measurements at the Relativistic Heavy Ion Collider 
(RHIC) at \sqs = 200~\cite{PPG030,Adler:2003pb,phenix:ppg099,Abelev:2008ez,Adare:2007dg,Abelev:2006cs,Adams:2006xb,Adams:2006nd} 
and 62.4 GeV~\cite{PPG087} over the last decade.
The hard scattering in $p+p$ collisions has been discovered by the 
observation of an unexpectedly large yield of particles with large 
transverse momentum and the phenomena of dijets at 
ISR~\cite{Darriulat:1980nk}. These observations indicate that hard 
scattering process occurs between the quarks and gluons constituents (or 
partons) inside the nucleons. This scattering process can be described 
by the perturbative quantum chromodynamics (pQCD) because the strong 
coupling constant $\alpha_s$ of QCD becomes small (asymptotically free) 
for large momentum transfer ($Q^2$) parton-parton scatterings. After the 
initial high $Q^2$ parton-parton scatterings, these partons fragment 
into the high $\pt$ hadrons or jets. In fact, at RHIC energies, single 
particle spectra of high $\pt$ hadrons are well described by 
pQCD~\cite{Adare:2007dg,PPG087,Adams:2006nd}. Furthermore, $\xt$ ($= 
2\pt/\sqrt{s}$), which is also inspired by pQCD, is known to be a good 
scaling variable of the particle production at high $\pt$ at both 
ISR~\cite{xt_scaling_1} and RHIC~\cite{PPG087} energies, so that $\xt$ 
scaling can be used to distinguish between the soft and hard particle 
productions.

%% --- comparison with A+A ---

Another important point of measurements in $p+p$ collisions is as a 
baseline for the heavy ion (A$+$A) data.  The nuclear modification factor 
$R_{\rm AA}$, for example, uses $\pt$ spectra in $p+p$ collisions as a 
denominator, and those in A$+$A collisions (with the appropriate scaling 
of number of binary nucleon-nucleon collisions) as a numerator. In 
addition, $\pt$ spectra in $p+p$ provide a reference of bulk properties 
of A$+$A collisions, such as the inverse slope parameter $T_{\rm inv}$, 
mean transverse momentum $\meanpt$, and yield per unit rapidity $dN/dy$. 
These data in $p+p$ collisions can be treated as baseline values for the 
smallest A$+$A collisions.

%% --- what will be shown in this paper, the motivations ---

In this paper, we present measurements of identified charged hadron 
$\pt$ spectra for $\pi^{\pm}$, $K^{\pm}$, $p$ and $\pbar$ at midrapidity 
in $p+p$ collisions at \sqs~=~200 and 62.4 GeV from the PHENIX 
experiment.
First, we compare the results of particle spectra at 200 GeV with those 
at 62.4 GeV as a function of $\pt$, $\mt$, and $\mt-m$ (where $m$ is the 
rest mass). Second, the extracted values from $\pt$ spectra, i.e. 
$T_{\rm inv}$, $\meanpt$, and $dN/dy$, are compared between the two beam 
energies.  For the systematic study of particle production as a function 
of \sqs the data are further compared to measurements in $p+p$ and 
$p+\overline{p}$ collisions at the CERN-ISR and FNAL-Tevatron 
colliders.

From these measurements, we discuss the following key issues:

\begin{description}

\item{\bf hard scattering particle production} -- 
The data are compared with the results of perturbative quantum 
chromodynamics (pQCD) calculations. 

\item{\bf transition from soft to hard physics} -- 
Since the $\pt$ regions presented in this paper can cover the region 
where the soft-hard transition occurs, the scaling properties in $\mt$ 
and $\xt$ with their beam energy and particle species dependences are 
shown.  

\item{\bf comparisons with heavy-ion data as a baseline measurement} 
-- Some of the data in $p+p$ are compared with the existing data in 
Au$+$Au~\cite{PPG026}.

\end{description}

%% --- structure of this paper ---

The paper is organized as follows:  Section~\ref{sec:setup} describes 
the PHENIX detector as it was used in this measurement.  
Section~\ref{sec:analysis} discusses the analysis details, including 
data sets, event selection, track selection, particle identification, 
corrections applied to the data, and systematic uncertainties.  
Section~\ref{sec:results} gives the experimental results for $\pt$ 
spectra for identified charged particles, particle ratios, $\mt$ 
scaling, the excitation function of observables (such as $T_{\rm inv}$, 
$\meanpt$, and $dN/dy$), and $R_{\rm AA}$.  
Section~\ref{sec:discussion} compares 
the results with next-to-leading-order (NLO) and next-to-leading-log 
(NLL) pQCD calculations, and discuss soft and hard particle production, 
and the transition between them.  Section~\ref{sec:summary} gives the 
summary and conclusions.

%%%%%%%%%%%%
% 2. SETUP %
%%%%%%%%%%%%

\section{EXPERIMENTAL SETUP}
\label{sec:setup}

\begin{figure}[tb]
\includegraphics[width=1.01\linewidth]{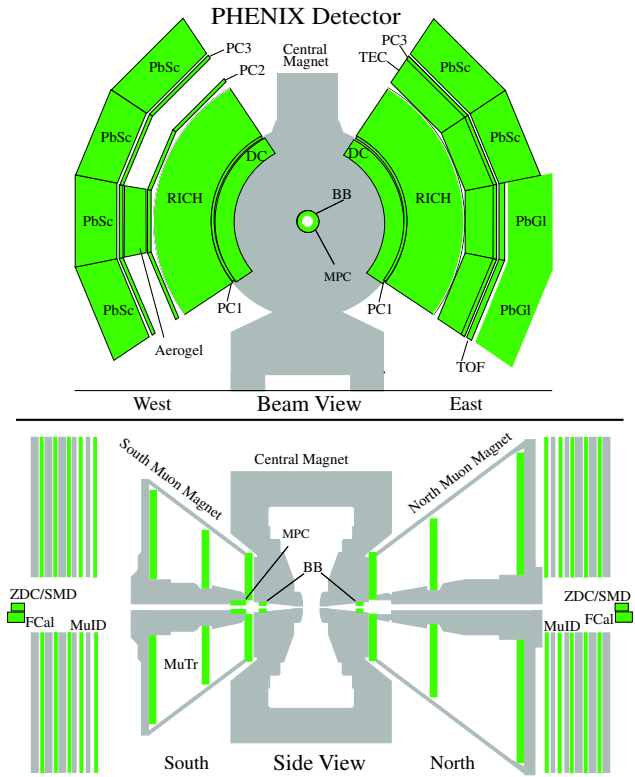}
\caption{\label{fig:phenix_setup} (color online) 
The PHENIX detector configuration for RHIC Run-6 data taking period.
The beam-beam counter (BBC) is labeled as BB on the figure.}
\end{figure}

The PHENIX experiment is designed to perform a broad study of A$+$A, 
$d+A$, and $p+p$ collisions to investigate nuclear matter under extreme 
conditions, as well as to measure the spin structure of the nucleon. It 
is composed of two central arms (called east and west arm, 
respectively), two forward muon arms, and global detectors, as shown in 
Fig.~\ref{fig:phenix_setup}. The central arms are designed to detect 
electrons, photons and charged hadrons in the pseudorapidity range 
$|\eta|$~$<$~0.35. The global detectors measure the start time, 
collision vertex, and charged hadron multiplicity of the interactions in 
the forward pseudorapidity region.  The following sections describe 
those parts of the detector that are used in the present analysis. A 
detailed description of the complete set of detectors can be found 
elsewhere~\cite{PHENIX_overview, 
PHENIX_magnet,PHENIX_PID,PHENIX_inner,PHENIX_tracking}.

%%%%%%%%%%%%%%%%%%%%%%%%%%%%%%%%%%%%%%%%%%%%%%%%%%%%%%%%%%%%% Fig_1

The beam-beam counters (BBC)~\cite{PHENIX_inner} determine the start time 
information for time-of-flight measurements and the collision vertex 
point, as well as providing the main collision trigger.  The two BBCs 
are located at 1.44 m from the nominal interaction point along the beam 
line on each side.  Each BBC comprises 64 $\check{\rm C}$erenkov 
telescopes, arranged radially around the beam line.  The BBCs measure the 
number of charged particles in the pseudorapidity region 
3.0~$<$~$|\eta|$~$<$~3.9.

Charged particle tracks are reconstructed using the central arm 
spectrometers~\cite{PHENIX_tracking}. The east arm spectrometer of the 
PHENIX detector contains the following subsystems used in this analysis: 
drift chamber (DC), pad chamber (PC) and time of flight (TOF). The 
magnetic field for the central arm spectrometers is supplied by the 
central magnet~\cite{PHENIX_magnet} that provides an axial field 
parallel to the beam around the collision vertex.

The drift chambers are the closest tracking detectors to the beam line, 
located at a radial distance of 2.2~m (geometric center, same for other 
detectors). They measure charged particle trajectories in the azimuthal 
direction to determine the transverse momentum of each particle. By 
combining the polar angle information from the first layer of PC, as 
described below, with the transverse momentum, the total momentum $p$ is 
determined. The momentum resolution in $p+p$ collisions is $\delta p/p 
\simeq 0.7\% \oplus 1.0\%\times p$ (GeV/$c$), where the first term is 
due to the multiple scattering before DC and the second term is the 
angular resolution of the DC. The absolute momentum scale is known as 
$\pm$ 0.7\% rms from the reconstructed proton mass using TOF.

The pad chambers are multi-wire proportional chambers that form three 
separate layers of the central tracking system. The first layer (PC1) is 
located at the radial outer edge of each drift chamber at a distance of 
2.49~m, while the third layer is at 4.98~m from the interaction point. 
The second layer is located at a radial distance of 4.19~m in the west 
arm only. PC1 and DC, along with the vertex position measured by BBC, 
are used in the global track reconstruction to determine the polar angle 
of each charged track.

The time-of-flight detector serves as the primary particle 
identification device for charged hadrons by measuring the stop time. 
The start time is given by BBC. TOF is located at a radial distance of 
5.06~m from the interaction point in the east central arm. This contains 
960 scintillator slats oriented along the azimuthal direction. It is 
designed to cover $|\eta|< 0.35$ and $\Delta\phi=45^{o}$ in azimuthal 
angle. The intrinsic timing resolution is $\sigma \simeq 115$ ps, which 
in combination with the BBC timing resolution of 60 ps, allows for a 
2.6$\sigma$ $\pi/K$ separation at $\pt \simeq 2.5$~GeV/$c$, and 
$K/p$ separation out to $\pt=4.5$~GeV/$c$, using an asymmetric 
particle-identification (PID) cut, as described below.

%%%%%%%%%%%%%%%%%%%%
% 3. DATA ANALYSIS %
%%%%%%%%%%%%%%%%%%%%
\section{DATA ANALYSIS}
\label{sec:analysis}

The two RHIC data sets analyzed are 2005 data for $p+p$ collisions at 
$\sqs=~$ 200~GeV and 2006 data for $p+p$ collisions at $\sqs=$ 62.4 GeV.  
Each data set was analyzed separately by taking into account the 
different run conditions and accelerator performance.  In this section, 
we explain the event selection, track reconstruction, particle 
identification, and corrections to obtain the $\pt$ spectra.  The event 
normalization and systematic uncertainties are also presented.

%------------------
% Event selection 
%------------------

\subsection{Event selection}

We use the PHENIX minimum bias trigger events, which are determined by a 
coincidence between north and south BBC signals, requiring at least one 
hit on both sides of BBCs. Due to the limited acceptance, approximately 
only half of the $p+p$ inelastic cross section can be measured by BBC. 
The PHENIX minimum bias data, triggered by BBC in $p+p$ collisions 
within a vertex cut of $\pm 30$ cm, include $\sigma_{\rm BBC} = $~23.0 
$\pm 2.2$~mb at $\sqs=$ 200~GeV and $\sigma_{\rm BBC} = $~13.7 $\pm 
1.5$~mb at $\sqs=$ 62.4~GeV (see Section~\ref{sec:analysis:norm}). We 
analyze 9.2$\times 10^{8}$ minimum bias events for the 2005 $p+p$ data 
at $\sqs=~$ 200~GeV, which is more than 30 times larger than the 2003 
data set~\cite{PPG030} and 2.14$\times 10^{8}$ times larger than the 
minimum bias events for the 2006 data at $\sqs=~$62.4 GeV.

%------------------
% Tracking and PID
%------------------
\subsection{Track reconstruction and particle identification}
% Tracking _________

As in previous publications~\cite{PHENIX_recoNIM, PPG026}, charged 
particle tracks are reconstructed by DC based on a combinatorial Hough 
transform, which gives the angle of the track in the main bend plane. 
PC1 is used to measure the position of the hit in the longitudinal 
direction along the beam axis. When combined with the location of the 
collision vertex along the beam axis, the PC1 hit gives the polar angle 
of the track. Only tracks with valid information from both DC and PC1 
are used in the analysis. To associate a track with a hit on TOF, the 
track is projected to its expected hit location on TOF. We require 
tracks to have a hit on TOF within $\pm$2$\sigma$ of the expected hit 
location in both the azimuthal and beam directions. The track 
reconstruction efficiency is approximately 98\% in $p+p$ collisions. 
Finally, a cut on the energy loss in the TOF scintillator is applied to 
each track. This $\beta$-dependent energy loss cut is based on a 
parameterization of the Bethe-Bloch formula. The flight path length is 
calculated from a fit to the reconstructed track trajectory in the 
magnetic field. The background due to random association of DC/PC1 
tracks with TOF hits is reduced to a negligible level when the mass cut 
used for particle identification is applied.

% PID __________

\begin{figure}[htb]
\includegraphics[width=1.01\linewidth]{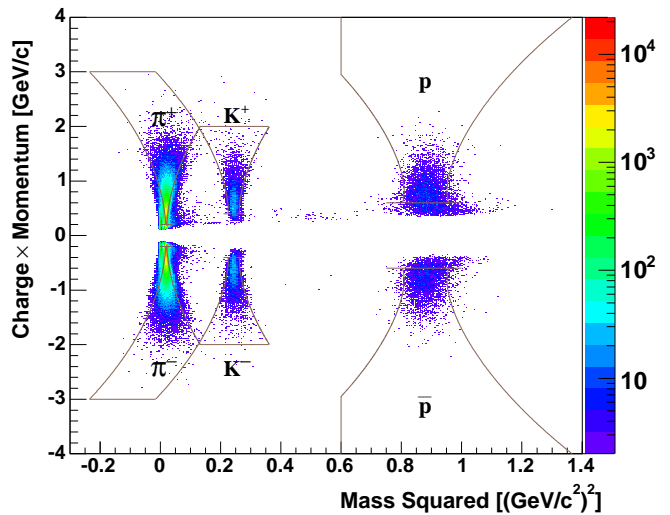}
\caption{(color online) 
Momentum multiplied by charge versus mass squared
distribution in $p+p$ collisions at $\sqrt{s}$~=~62.4~GeV.
The lines indicate the PID cut boundaries (2$\sigma$) for pions, kaons, 
and protons (antiprotons) from left to right, respectively.}
\label{fig:PID}
\end{figure}

Charged particles are identified using the combination of three 
measurements:  time-of-flight from the BBC and TOF, momentum from the DC, 
and flight-path length from the collision vertex point to the TOF hit 
position.  The square of mass is derived from
 \begin{equation}
m^{2} = \frac{p^{2}}{c^{2}} \Bigl[ {\Bigl( \frac{t_{\rm tof}}{L/c} \Bigr)}^{2} -1 \Bigr],
\label{eq:m2}
 \end{equation}
where $p$ is the momentum, $t_{\rm tof}$ is the time of flight, $L$ is 
the flight path length, and $c$ is the speed of light. The charged 
particle identification is performed using cuts in $m^{2}$ and momentum 
space. In Fig.~\ref{fig:PID}, a plot of momentum multiplied by charge 
versus $m^{2}$ is shown together with applied PID cuts as solid curves. 
We use 2$\sigma$ standard deviation PID cuts in $m^{2}$ and momentum 
space for each particle species. The PID cut is based on a 
parameterization of the measured $m^{2}$ width as a function of momentum
 \begin{eqnarray}
{\sigma^2_{m^2}} &=& \frac{\sigma_{\alpha}^2} {K_{1}^{2}} (4m^{4}p^{2}) +
                     \frac{\sigma_{ms}^2} {K_{1}^{2}}
                           \Bigl[ 4m^{4} \Bigl( 1+\frac{m^2}{p^2} \Bigr) \Bigl] \nonumber \\
                 &+& \frac{\sigma_{t}^2 c^2} {L^2}
                           \bigl[ 4p^{2} \bigl( m^2 + p^2 \bigr) \bigr],
\label{eq:pid}
\end{eqnarray}
where $\sigma_\alpha$ is the angular resolution, $\sigma_{ms}$ is the 
multiple scattering term, $\sigma_{t}$ is the overall time-of-flight 
resolution, $m$ is the centroid of $m^{2}$ distribution for each 
particle species, and $K_1$ is the magnetic field integral constant term 
of 101~mrad$\cdot$GeV. The parameters for PID are $\sigma_\alpha = 
0.99$~mrad, $\sigma_{ms} = 1.02$~mrad$\cdot$GeV, and $\sigma_{t} = 
130$~ps. For pion identification above 2~GeV/$c$, we apply an asymmetric 
PID cut to reduce kaon contamination of pions. As shown by the lines in 
Fig.~\ref{fig:PID}, the overlap regions which are within the 2$\sigma$ 
cuts for both pions and kaons are excluded. The lower momentum cut-offs 
are 0.3~GeV/$c$ for pions, 0.4~GeV/$c$ for kaons, and 0.5~GeV/$c$ for 
protons and antiprotons. The lower momentum cut-off value for $p$ and 
$\pbar$ is larger than for pions and kaons due to the larger 
energy loss effect.

For kaons, the upper momentum cut-off is 2~GeV/$c$ since the $\pi +p$ 
contamination level for kaons is $\approx$ 8\% at that momentum. The 
upper momentum cut-off for pions is $\pt = $ 3~GeV/$c$ where the $K+p$ 
contamination reaches $\approx$ 3\%. Electron (positron) and decay muon 
background at very low $\pt$ ($<$ 0.3~GeV/$c$) are well separated from 
the pion mass-squared peak. For protons the upper momentum cut-off is 
set at 4.5~GeV/$c$.  For protons and antiprotons an additional cut,
$m^{2}>0.6({\rm GeV}/c^{2})^{2}$, is introduced to 
reduce the contamination.  The contamination background on each particle 
species is subtracted statistically after applying these PID cuts.

%%%%%%%%%%%%%%%%%%%%%%%%%%%%%%%%%%%%%%%%%%%%%%%%%%%%%%%%%%%%% Fig_2

%------------------------
% Efficiency corrections 
%------------------------

\subsection{Efficiency corrections}

We use the PHENIX-integrated-simulation application, which is 
a {\sc geant}~\cite{geant} based Monte Carlo (MC) simulation program of 
the PHENIX detector, to correct for geometrical acceptance, 
reconstruction efficiency, in-flight decay for $\pi$ and $K$, multiple 
scattering effect, and nuclear interactions with materials in the 
detector (including $\pbar$ absorption). Single particle tracks are 
passed from {\sc geant} to the PHENIX event reconstruction 
software~\cite{PHENIX_recoNIM}. In this simulation, the BBC, DC, and TOF 
detector responses are tuned to match the real data. For example, dead 
areas of DC and TOF are included, and momentum and time-of-flight 
resolutions are tuned. The track association to TOF in both azimuth and 
along the beam axis as a function of momentum and the PID cut boundaries 
are parameterized to match the real data. A fiducial cut is applied to 
choose identical active areas on TOF in both the simulation and data.

We generate 1$\times 10^{7}$ single particle events for each particle 
species ($\pi^{\pm}$, $K^{\pm}$, $p$ and $\pbar$) with flat $\pt$ 
distributions for high $\pt$ (2--4 GeV/$c$ for pions and kaons, 2--8 
GeV/$c$ for $p$ and $\pbar$) with enhancement at low $\pt$ ($<$ 2 
GeV/$c$). Weighting functions to the $\pt$ distributions are also used 
to check the effect of steepness, which is less than $\sim$1\% level on 
the final yields in the measured $\pt$ range. The rapidity range is set 
to be wider than the PHENIX acceptance, i.e. flat in -0.6 $<$ $y$ $<$ 
0.6 ($\Delta y$ = 1.2) to deal with particles coming from outside (the 
denominator of Eq.~\ref{eq:acc} is weighted with a factor 1/$\Delta y$ = 
1/1.2 in order to normalize the yield for unit rapidity). The 
efficiencies are determined in each $\pt$ bin by dividing the 
reconstructed output by the generated input as expressed as follows:

\begin{equation}
\epsilon(\pt) = \frac{\mbox{\rm \# of reconstructed MC tracks}}
                                 {\mbox{\rm \# of generated MC tracks}}.
\label{eq:acc}
\end{equation}

The resulting correction factors $C_{\rm eff}(\pt)$ (= 
1/$\epsilon(\pt)$) is multiplied to the raw $\pt$ spectra for each $\pt$ 
bin and for each individual particle species (see 
Section~\ref{subsec:Invariant_cross_section}).

%-----------------------
% Feed-down corrections
%-----------------------

\subsection{Feed-down corrections}
\label{sec:analysis:feed}

The proton and antiproton $\pt$ spectra are corrected for feed-down from 
weak decays of hyperons. The detailed procedure of the feed-down 
correction can be found in~\cite{PPG030}. We include the following decay 
modes: $\Lambda \rightarrow p \pi^{-}$, $\Sigma^{+} \rightarrow p 
\pi^{0}$, and $\Lambda$ production from $\Sigma^{0}$, $\Xi^{0}$, 
$\Xi^{-}$. The feed-down contributions for antiproton yields are also 
estimated using the above decay modes for antiparticles.

In order to estimate the fractions of protons and antiprotons from weak 
decays of hyperons in the measured proton and antiproton $\pt$ spectra, 
we use three input $\Lambda$ and $\overline{\Lambda}$ $\pt$ spectra: 

\begin{enumerate}
\item measured $\Lambda$ and $\overline{\Lambda}$ $\pt$ spectra in 
PHENIX in $p+p$ collisions at $\sqrt{s} = 200$ and 62.4 GeV,
\item measured $p$ ($\pbar$) distributions scaled with measured 
$\Lambda$ ($\overline{\Lambda}$) distributions~\cite{Abelev:2006cs}, and
\item measured $p$ ($\pbar$) distributions scaled with ISR $\Lambda$ 
($\overline{\Lambda}$) distributions~\cite{isr_lambda_pp63}.
\end{enumerate}

\begin{figure}[htb]
\includegraphics[width=0.9\linewidth]{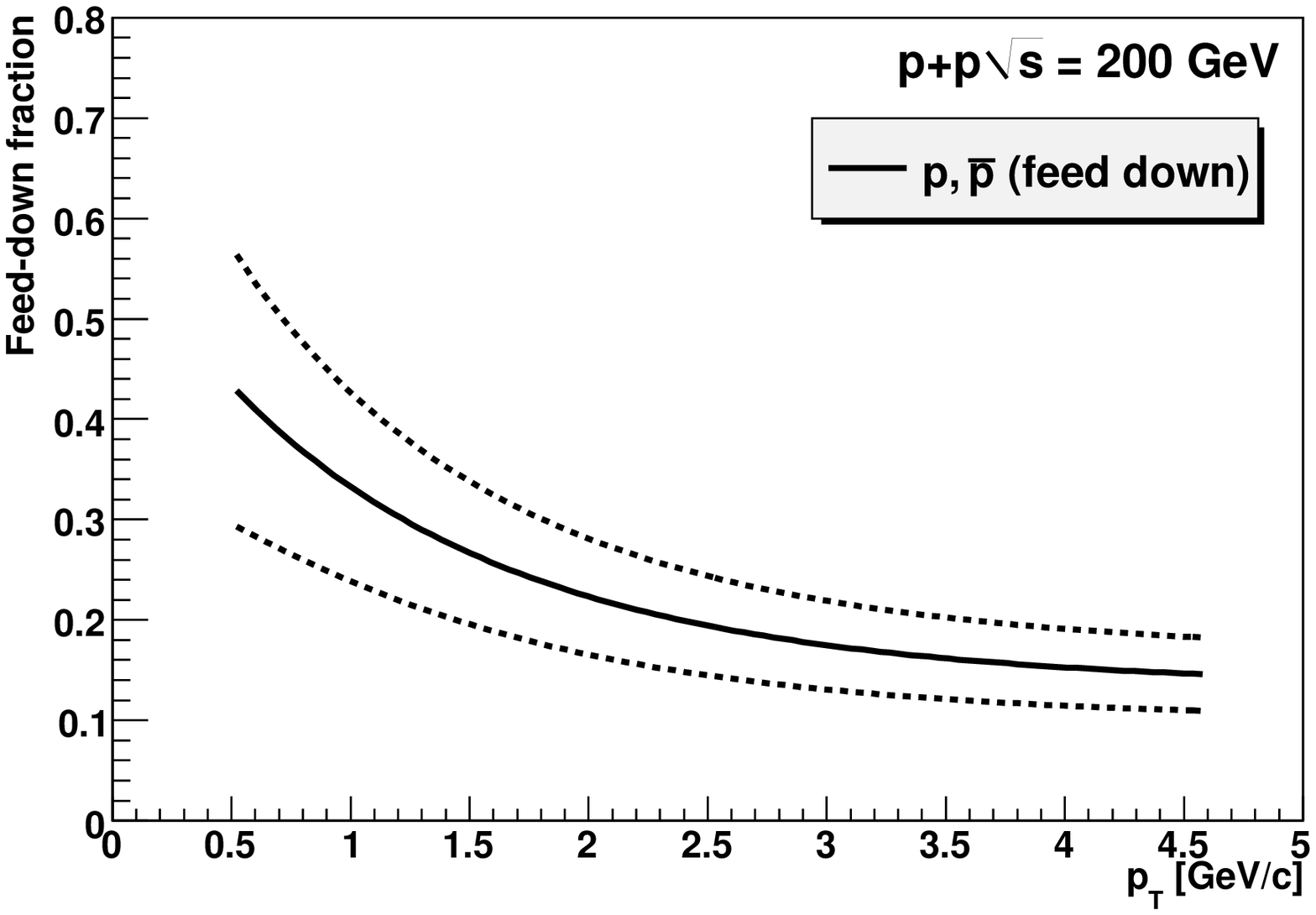}
\includegraphics[width=0.9\linewidth]{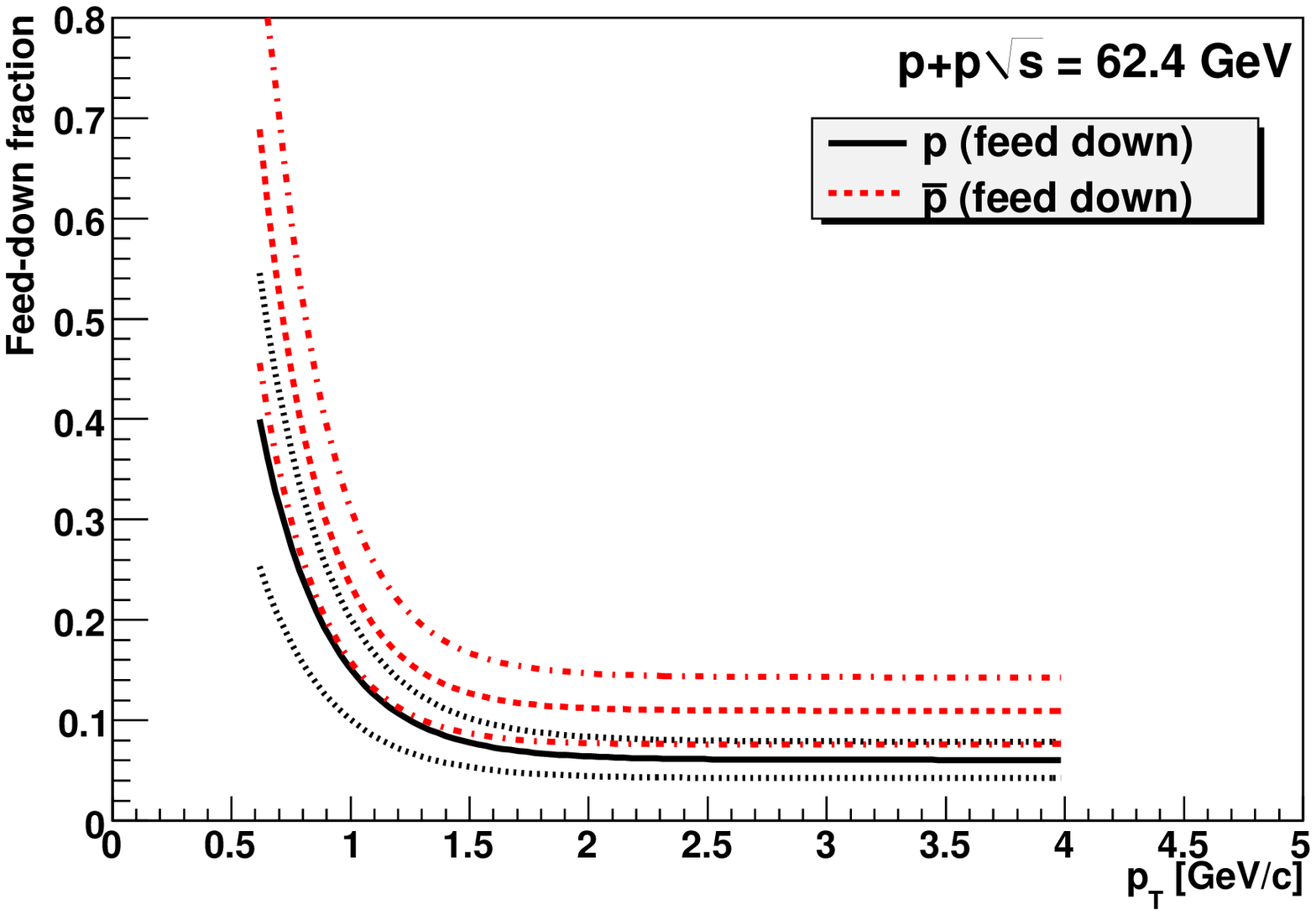}
\caption{(color online) 
Fraction of feed-down protons and antiprotons as a function of $\pt$ with systematic uncertainties.
Top: 200 GeV $p+p$ (positive and negative functions are common). Bottom: 62.4 GeV $p+p$.}
\label{fig:frac_w_syserr_FD}
\end{figure}

Using each input above, proton and antiproton spectra from weak decays 
are calculated by using Monte Carlo simulation to take into account 
decay kinematics, the PHENIX track reconstruction efficiency and 
experimental acceptance.  Then systematic uncertainties are evaluated 
from different $\Lambda$ and $\overline{\Lambda}$ spectra inputs. The 
resulting uncertainties on the final proton and antiproton spectra are 
of the order of 20--30\% at $\pt$ = 0.6 GeV/$c$ and 2--5\% at $\pt$ = 4 
GeV/$c$. The fractional contribution of the feed-down protons 
(antiprotons) to the total measured proton (antiproton) spectra, 
$\delta_{\rm feed}(\pt)$, is approximately 10--20 (5--15)~\% at 
$\pt$ = 4 GeV/$c$ for 200 GeV $p+p$ (62.4 GeV $p+p$) and it shows an 
increase at lower $\pt$ as shown in Fig.~\ref{fig:frac_w_syserr_FD}. The 
correction factor for the feed-down correction can be expressed as: 
$C_{\rm feed} (\pt) = 1 - \delta_{\rm feed}(\pt)$, which is multiplied 
to the raw $\pt$ spectra (see 
Section~\ref{subsec:Invariant_cross_section}).

The feed-down correction for protons is different from that for 
antiprotons in 62.4 GeV, because of the difference in $\Lambda/p$ and 
$\overline{\Lambda}/\pbar$ ratio at this beam energy. At 62.4 GeV, 
$\Lambda/p$ ratio is ~0.2 while $\overline{\Lambda}/\pbar$ ratio is 
~0.4~\cite{isr_lambda_pp63}, so that the feed-down contribution for 
antiprotons is bigger than that for protons. At 200 GeV, these two 
ratios are almost same~\cite{Abelev:2006cs}, therefore the feed-down 
corrections for $p$ and $\pbar$ become identical.

%%%%%%%%%%%%%%%%%%%%%%%%%%%%%%%%%%%%%%%%%%%%%%%%%%%%%%%%%%%%% Fig_3

%-----------------------------
% Cross section normalization 
%-----------------------------
\subsection{Cross section normalization}
\label{sec:analysis:norm}

The BBC counter serves a dual function as both the minimum bias trigger 
and the calibrated luminosity monitor. The luminosity $\cal L$ is 
defined as the interaction rate for a given cross section: $dN/dt = 
{\cal L} \sigma$ and the total number of events for a given cross 
section is
 \begin{equation}
N = \sigma \times \int{\cal L} dt,
\end{equation}
where $\int{\cal L} dt$ is the integrated luminosity. To connect the 
number of minimum bias triggered events and the integrated luminosity, 
$\sigma_{\rm BBC}$ is introduced where $1/\sigma_{\rm BBC}$ corresponds 
to the integrated luminosity per minimum bias triggered event 
(Eq.~\ref{eq:sigma_bbc}).
   \begin{equation}
     N_{\rm BBC}=\sigma_{\rm BBC} \times \int{\cal L}dt ,
     \label{eq:sigma_bbc}
   \end{equation}
where $N_{\rm BBC}$ is the number of minimum bias events and 
$\int{\cal L}dt$ is the corresponding integrated luminosity.
$\sigma_{\rm BBC}$ is measured by a Van der Meer scan method (Vernier scan) in PHENIX~\cite{PAC_vernier,PPG087}.

Vernier scans were performed for $\sqs=$ 200 and 62.4 GeV data sets. The 
obtained $\sigma_{\rm BBC}$ are $23.0 \pm 2.2$ mb and $13.7 \pm 1.5$ mb 
for $\sqs=$ 200 and 62.4 GeV respectively. The quoted uncertainty is a 
systematic uncertainty. These numbers were reported in our measurements 
of $\pi^{0}$ production~\cite{Adare:2007dg, PPG087}.

\begin{figure}[htb]
\includegraphics[width=1.0\linewidth]{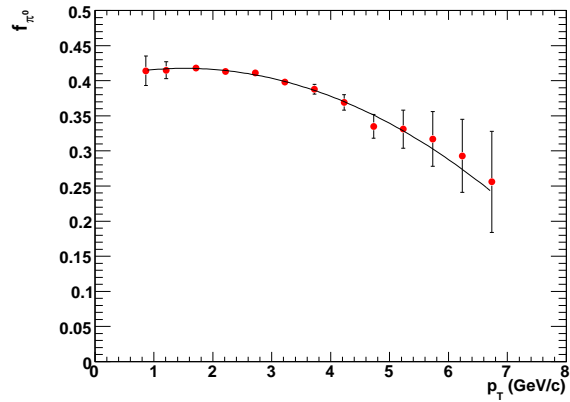}
\caption{(color online) 
Fraction of the inclusive $\pi^{0}$ yield which satisfied the
BBC trigger condition in 62.4 GeV $p+p$. Data points are 
from Fig. 1 of~\cite{PPG087}.}
\label{fig:trig_bias}
\end{figure}

Since the minimum bias trigger registers only half of the $p+p$ 
inelastic cross section, it is expected that there is a trigger bias 
against particles in the central spectrometers. This was checked with 
$\pi^0$'s in the electromagnetic calorimeter with high $\pt$ photon 
triggered events, and with charged tracks in accelerator's beam crossing 
(clock) triggered events. The trigger bias $\epsilon_{\rm bias}$ 
determined from the ratio ($f_{\pi^{0}}$) of the number of $\pi^0$ in 
the high $\pt$ photon triggered sample with and without the BBC trigger 
requirement\cite{PPG087}. We assume $\epsilon_{\rm bias}$ is process 
dependent and so that it is measured as $\epsilon_{\rm bias} = 
f_{\pi^{0}}$. This ratio, $f_{\pi^{0}}$, is $0.79 \pm 0.02$ independent 
of the transverse momentum for $\sqs$ = 200 GeV. At 62.4 GeV, the 
trigger bias was found to be transverse momentum 
dependent~\cite{PPG087}. Fig.~\ref{fig:trig_bias} shows that the trigger 
bias $f_{\pi^{0}}$, and $f_{\pi^{0}}$ is $\approx$40\% up to 
$\pt \approx$3 GeV/$c$, and monotonically decreases to 25\% at $\pt 
\approx$7 GeV/$c$.  As described in the previous PHENIX 
publication~\cite{PPG087}, this decrease can be understood by the fact 
that most of the energy is used for the production of high-energy jets 
which contain the measured high $\pt$ $\piz$ and charged hadrons, and 
there is not enough energy left to produce particles for $\sqs =$ 62.4 GeV 
$p+p$ collisions at the forward rapidity (3.0~$<$~$|\eta|$~$<$~3.9) where 
the BBC is located. This drop can be seen only for 62.4 GeV data. Also we 
assume no particle species dependence for this trigger bias. We use this 
$\pt$ dependent trigger bias correction for charged hadrons, by using 
fitted coefficients of a second order polynomial, as shown in 
Fig.~\ref{fig:trig_bias}.

With those values, the invariant yield per BBC trigger count ($Y/N_{\rm 
BBC}$) is related to the invariant cross section ($\sigma$) using 
   \begin{equation}
     \sigma=(Y/N_{\rm BBC})\times(\sigma_{\rm BBC}/\epsilon_{\rm bias}).
     \label{eq:calcxsec}
   \end{equation}   

%%%%%%%%%%%%%%%%%%%%%%%%%%%%%%%%%%%%%%%%%%%%%%%%%%%%%%%%%%%%% Fig_4

%-------------------------
% Systematic uncertainties
%-------------------------
\subsection{Systematic uncertainties}

In order to estimate the systematic uncertainties, the $\pt$ spectra with 
slightly different analysis cuts from what we use for the final results 
are prepared, and these spectra are compared to those with the standard 
analysis cuts.  We checked the following analysis cuts: (1) fiducial, 
(2) track association windows, and (3) PID.

For each spectrum with modified cuts, the same changes in the cuts are 
made in the Monte Carlo simulation. The fully corrected spectra with 
different cut conditions are divided by the spectra with the baseline 
cut condition, resulting in uncertainties associated with each cut 
condition as a function of $\pt$. The obtained uncertainties are added 
in quadrature. Tables~\ref{tab:sys_error_pp200} and 
\ref{tab:sys_error_pp62} show the systematic uncertainties on $\pt$ 
spectra for each data set. There are three categories of systematic 
uncertainty: Type A is a point-to-point error uncorrelated between $\pt$ 
bins, type B is $\pt$ correlated, where all points move in the same 
direction but not by the same factor, while in type C all points move by 
the same factor independent of $\pt$~\cite{Adare:2008cg}.
%
% (i)  all points move by the same factor ($\pt$ independently), but the direction is case by case (same direction or random variation), 
% (ii) all points move in the same direction, but not by the same factor (i.e. $\pt$ dependently).
%
In this study, the systematic uncertainties on feed-down correction and 
PID contamination correction are Type B, other systematic uncertainties 
on applied analysis cuts are Type C. There are two types of the 
PID-related uncertainties. One is the systematic uncertainty of the 
yield extraction, which is evaluated by changing the PID boundary in 
$m^2$ vs momentum plane. The other is the systematic uncertainty of the 
particle contamination, which is evaluated by using the contamination 
fraction. The fraction is estimated by fitting $m^2$ distributions on 
each $\pt$ slice under the several conditions as follows: (1) fixed 
parameters for $p$ and $\pbar$ mass centroid and width, (2) $p$ and 
$\pbar$ mass centroid free with fixed mass width, (3) $p$ and $\pbar$ 
mass width free with fixed mass centroid.

The systematic uncertainty on the BBC cross section is 9.7\% and 11\% 
for $\sqs=$ 200 and 62.4 GeV respectively. The systematic uncertainty on 
the trigger bias is 3\% and 1--5\% for $\sqs=$ 200 and 62.4 GeV 
respectively (see Section~\ref{sec:analysis:norm}). These uncertainties 
on normalization (Type C) are not included in 
Tables~\ref{tab:sys_error_pp200} and \ref{tab:sys_error_pp62}.  All the 
figures and tables including the tables in Appendix \ref{appendix} do 
not include the normalization uncertainties, unless explicitly noted.

%%%%%%%%%%%%%%%%%%%%%%%%%%%%%%%%%%%%%%%%%%%%%%%%%%%%  Table_I
\begin{table}[t]
\caption{Systematic uncertainties on the $\pt$ spectra for $\sqs$~=~200 
GeV $p+p$ given in percent.
The number in parenthesis includes the $\pt$ dependence of the uncertainties for PID cut, feed-down correction, PID contamination correction.}
\begin{ruledtabular} \begin{tabular}{lllllll}
Source                  & $\pi^{+}$  & $\pi^{-}$  & $K^{+}$    & $K^{-}$   & $p$      & $\overline{p}$ \\ \hline
Fiducial Cut            &  5         &  5        &  4         &  5        &  4       &  5      \\
Track Matching          &  4         &  4        &  5         &  4        &  4       &  4      \\
PID Cut                 &  3         &  3        &  2         &  2        &  2--8     &  2--10   \\
Efficiency Correction   &  2         &  2        &  2         &  2        &  2       &  2      \\
Feed-down Correction    &  -         &  -        &  -         &  -        &  4--25    &  4--25      \\
PID Contamination       &  -         &  -        &  -         &  -        &  0--2     &  0--2      \\ \hline
Total                   &  7         &  7        &  7         &  7        &  6 (8--25) &  7 (9--25)    \\
\end{tabular} \end{ruledtabular}
\label{tab:sys_error_pp200}
\end{table}
%

%-------------------------
% Invariant cross section
%-------------------------
\subsection{Invariant cross section}
\label{subsec:Invariant_cross_section}
The differential invariant cross section is determined as 
\begin{equation}
  E\frac{d^{3}\sigma}{dp^{3}} = \frac{1}{2\pi \pt} \frac{\sigma_{\rm BBC}}{N_{\rm BBC}C_{\rm bias}^{\rm BBC}(\pt)}
                                C_{\rm eff}(\pt) C_{\rm feed}(\pt) \frac{d^{2}N}{d\pt dy},
 \label{eq:yield}
\end{equation}
where $\sigma$ is the cross section, $\pt$ is transverse momentum, and 
$y$ is rapidity, $N_{\rm BBC}$ is the number of minimum bias events, 
$\sigma_{\rm BBC}$ is the minimum bias cross section measured by BBC, 
$C_{\rm eff}(\pt)$ is the acceptance correction factor including 
detector efficiency, $C_{\rm bias}^{\rm BBC}(\pt)$ is the trigger bias, 
$C_{\rm feed}(\pt)$ is the feed-down correction factor only for protons 
and antiprotons, and $N$ is the number of measured tracks.

%%%%%%%%%%%%%%%%%%%%%%%%%%%%%%%%%%%%%%%%%%%%%%%%%%%%  Table_II
\begin{table}[t]
\caption{Systematic uncertainties on the $\pt$ spectra for $\sqs$~=~62.4 
GeV $p+p$ given in percent.
The number in parenthesis includes the $\pt$ dependence of the uncertainties for feed-down correction, PID contamination correction.}
\begin{ruledtabular} \begin{tabular}{lllllll}
Source                & $\pi^{+}$  & $\pi^{-}$  & $K^{+}$  & $K^{-}$   & $p$        & $\overline{p}$ \\ \hline
Fiducial Cut          &  6         &  5        &  6       &  5       &  7         &  5      \\
Track Matching        &  2         &  2        &  3       &  3       &  3         &  3      \\
PID Cut               &  2         &  2        &  3       &  3       &  4         &  4      \\
Efficiency Correction &  2         &  2        &  2       &  2       &  2         &  2      \\
Feed-down Correction  &  -         &  -        &  -       &  -       &  1--16      &  3--50      \\
PID Contamination     &  -         &  -        &  0--5     &  0--5     &  -         &  -      \\ \hline
Total                 &  7         &  6        &  7       &  7       &  9 (9--18)  &  7 (8--50)    \\
\end{tabular} \end{ruledtabular}
\label{tab:sys_error_pp62}
\end{table}

%% %\clearpage

%\clearpage

%%%%%%%%%%%%%%
% 4. RESULTS %
%%%%%%%%%%%%%%
\section{RESULTS}
\label{sec:results}

In this section, we show the transverse momentum distributions and 
yields for $\pi^{\pm}$, $K^{\pm}$, $p$ and $\pbar$ in $p+p$ collisions 
at $\sqs$~=~200 and 62.4~GeV at midrapidity measured by the PHENIX 
experiment. We also present the transverse mass ($m_T$) spectra, the 
inverse slope parameter $T_{\rm inv}$, mean transverse momentum $\meanpt$, 
yield per unit rapidity $dN/dy$, and particle ratios at each energy, and 
compare them to other measurements at different $\sqs$ in $p+p$ and 
$p+\pbar$ collisions. The measured $T_{\rm inv}$, $\meanpt$, and $dN/dy$ in 
$p+p$ 200 GeV are also compared with those in published results in 
Au$+$Au at 200 GeV.  The nuclear modification factor $R_{\rm AA}$ for 200 
GeV Au$+$Au using the present study in $p+p$ 200 GeV are also presented.

%------------
% pT spectra
%------------
\subsection{$\pt$ spectra}
\label{sec:pt}

\begin{figure*}[t]
\vspace{-0.3cm}
\hspace{0.3cm}
\includegraphics[width=0.45\linewidth]{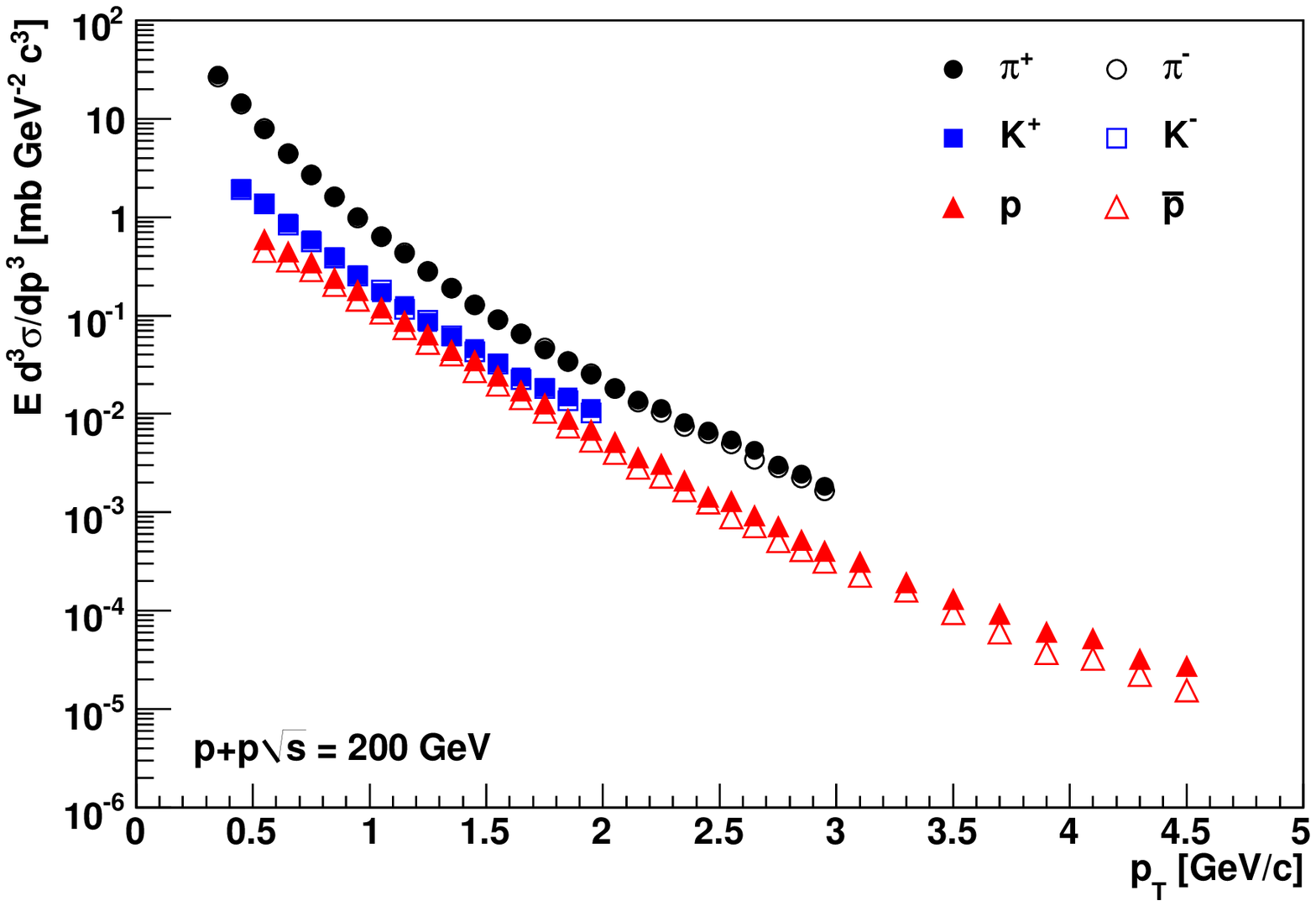}
\hspace{0.3cm}
\includegraphics[width=0.45\linewidth]{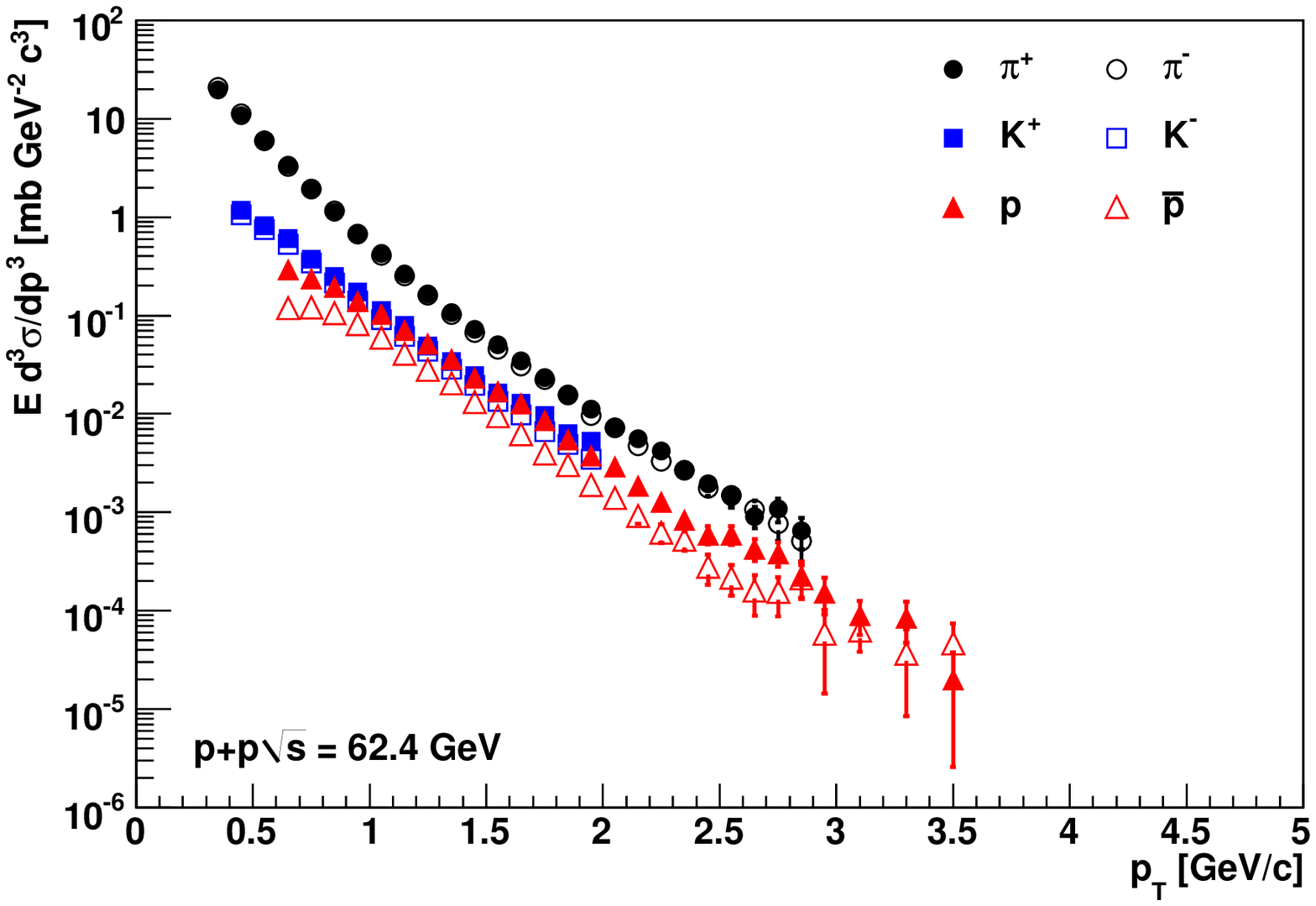}
\includegraphics[width=0.48\linewidth]{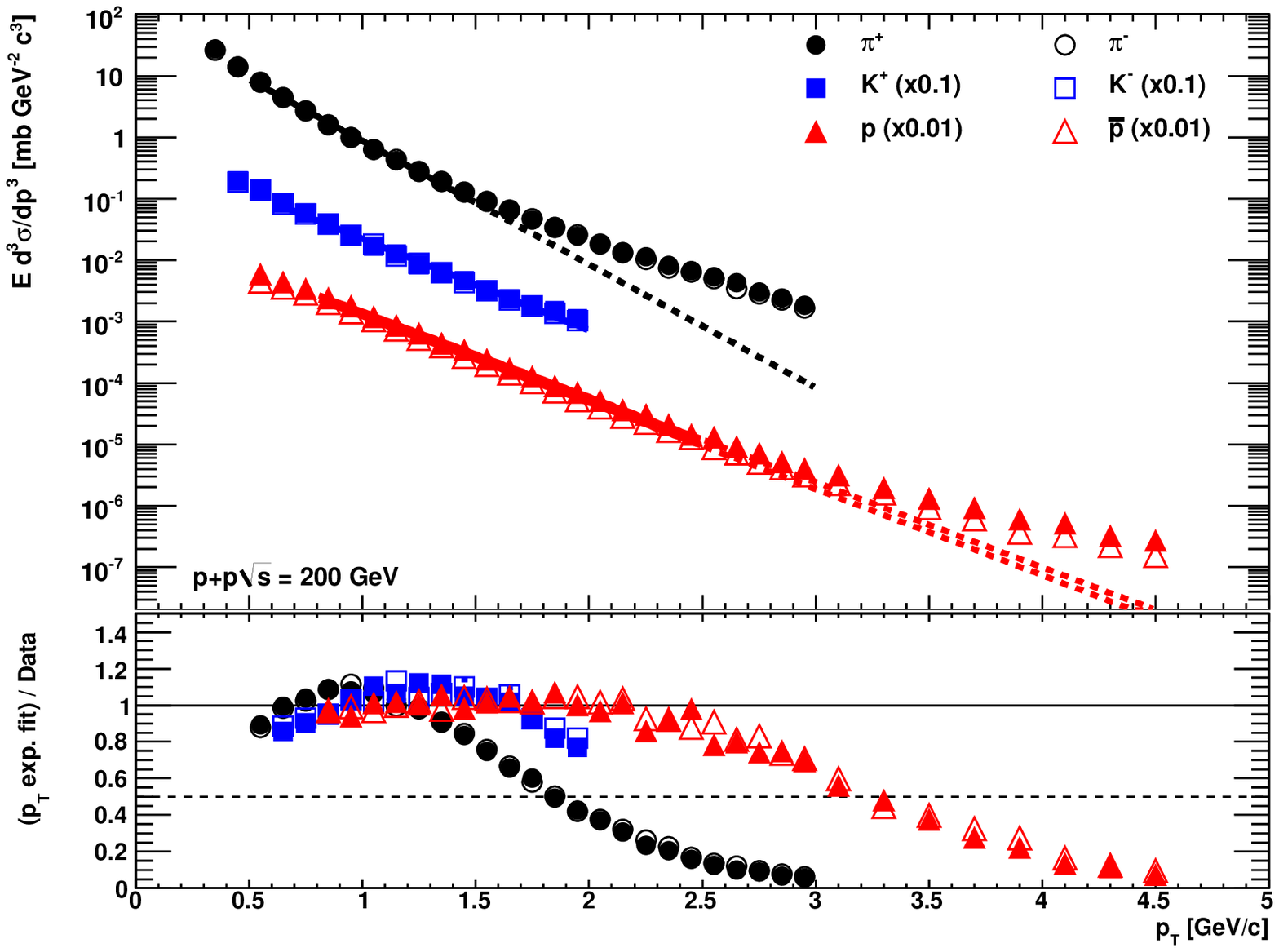}
\includegraphics[width=0.48\linewidth]{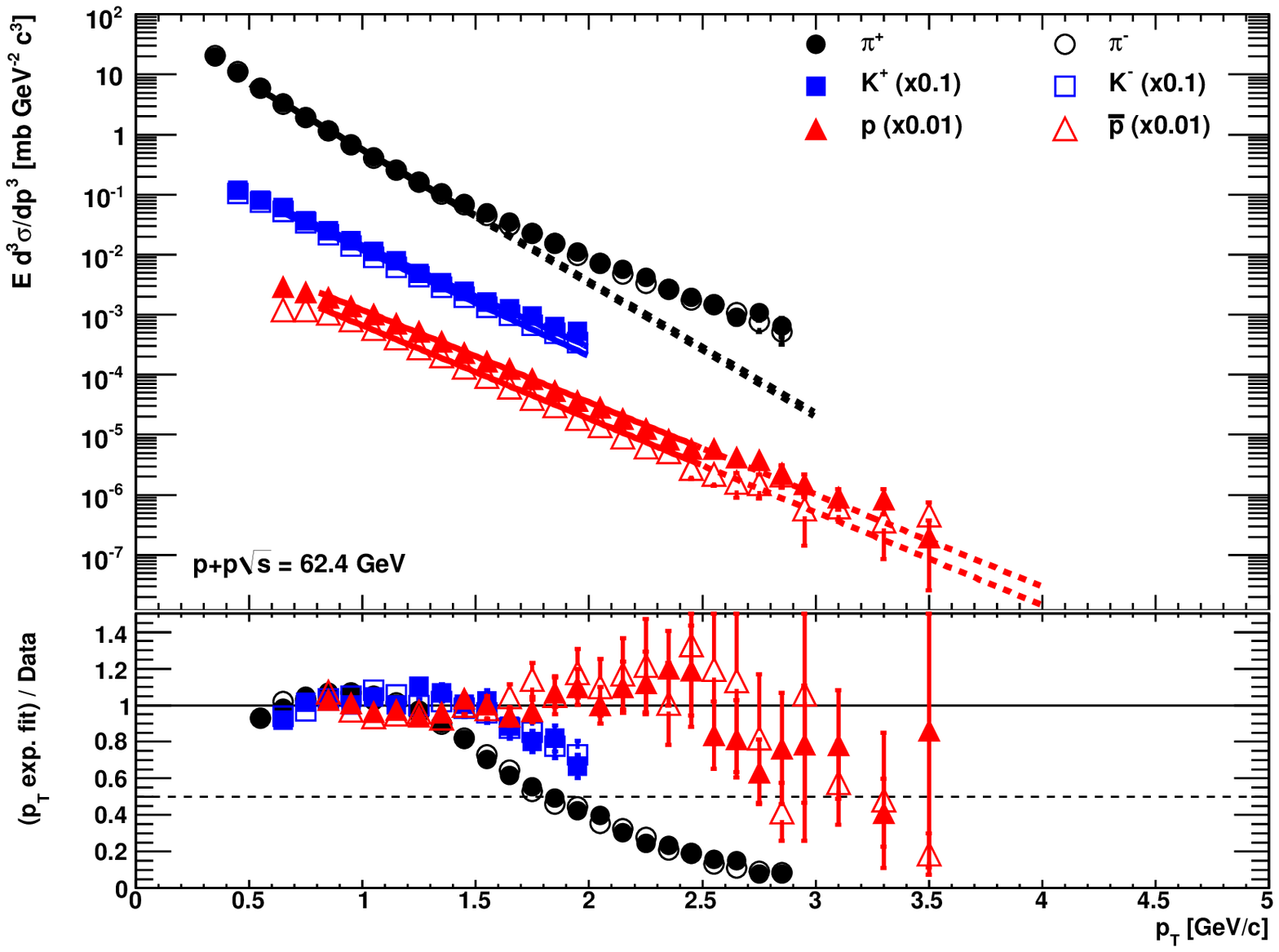}
\includegraphics[width=0.55\linewidth]{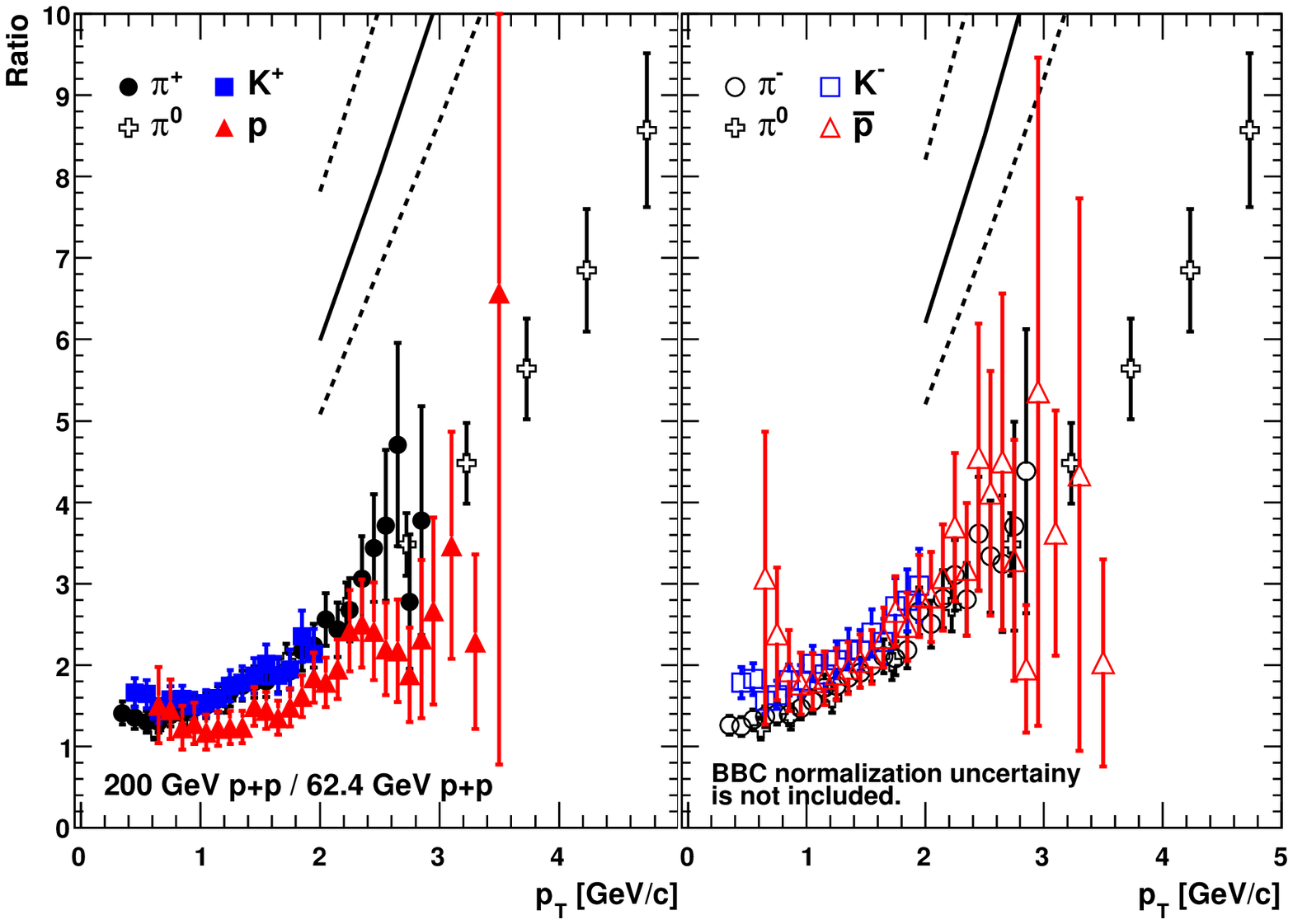}
\caption{(color online) 
(top, middle) Transverse momentum distributions for 
$\pi^{\pm}$, $K^{\pm}$, $p$, and $\pbar$
in $p+p$ collisions at $\sqs=$ (left) 200 and (right) 62.4 GeV at midrapidity.
Only statistical uncertainties are shown.
(middle plots) Each spectrum is fitted with an exponential function.
(lower panels of middle plots) Ratio of the exponential 
fit to data for each particle species.
(bottom) Ratios of $\pt$ spectra for $\pi^{\pm}$, 
$\pi^{0}$~\cite{Adare:2007dg, PPG087}, 
$K^{\pm}$, $p$, and $\pbar$ in 200 GeV $p+p$ collisions to those in 
62.4 GeV $p+p$ collisions. 
% The $\pi^{0}$ spectra are measured at 
% PHENIX~\cite{Adare:2007dg, PPG087}. 
Statistical and systematic uncertainties are combined in quadrature. 
The trigger cross section uncertainty is not included. The lines 
represent the NLO pQCD calculations (DSS fragmentation function) for 
pions with different factorization, fragmentation, and renormalization 
scales (which are equal)~\cite{W_Vogelsang_private}.
}
\label{fig:pt_pikp}
\end{figure*}

Figure~\ref{fig:pt_pikp} shows transverse momentum spectra for 
$\pi^{\pm}$, $K^{\pm}$, $p$, and $\pbar$ 
in 200 and 62.4 GeV $p+p$ collisions.  Feed-down correction for weak 
decays is applied for $p$ and $\pbar$, and the same correction factors 
are consistently used for all figures through Section~\ref{sec:results} 
unless otherwise specified. The systematic uncertainty on the BBC cross 
section is 9.7\% and 11\% for $\sqs=$ 200 and 62.4 GeV respectively.
Each of $\pt$ spectra is fitted with an exponential functional form:
  \begin{equation}
  \frac{1}{2\pi \pt}\frac{d^{2}\sigma}{dyd\pt} = A \exp{ \Bigl(- \frac{\pt}{T} \Bigr) }
  \label{eq:pt_exp_fun}
  \end{equation}
where $A$ is a normalization factor and $T$ is an inverse slope parameter 
for $\pt$. The fitting parameters and $\chi^{2}$/NDF by using 
Eq.~\ref{eq:pt_exp_fun} for $\pi^{\pm}$, $K^{\pm}$, $p$, and $\pbar$ in 
200 and 62.4 GeV $p+p$ collisions, are tabulated in 
Table~\ref{tab:tinv_fit_pt}.  The fitting range is fixed as 
$\pt$ = 0.5--1.5 GeV/$c$ for $\pi^{\pm}$, 0.6--2.0 GeV/$c$ for 
$K^{\pm}$, and 0.8--2.5 GeV/$c$ for $p$, $\pbar$ at both collision 
energies.

Figure~\ref{fig:pt_pikp} shows that pions, protons, and 
antiprotons exhibit an exponential spectral shape at low $\pt$ and a 
power-law shape at high $\pt$, while kaons are exponential in the 
measured $\pt$ range. The transition from exponential to power-law can 
be better seen at $\pt \sim$ 2 GeV/$c$ for pions, and at $\pt \sim$ 3 
GeV/$c$ for protons and antiprotons at both energies. The fractions of 
soft and hard components gradually change in the transition region.

%%%%%%%%%%%%%%%%%%%%%%%%%%%%%%%%%%%%%%%%%%%%%%%%%%%%%%%%%%%%% Fig_5
%\label{fig:pt_w_line_pikp}
%\label{fig:ratio_pt_pp200_pp62_w_FDcor}

%%%%%%%%%%%%%%%%%%%%%%%%%%%%%%%%%%%%%%%%%%%%%%%%%%%%%%%%%%%%% Fig_6
\begin{figure*}[t]
\includegraphics[width=0.49\linewidth]{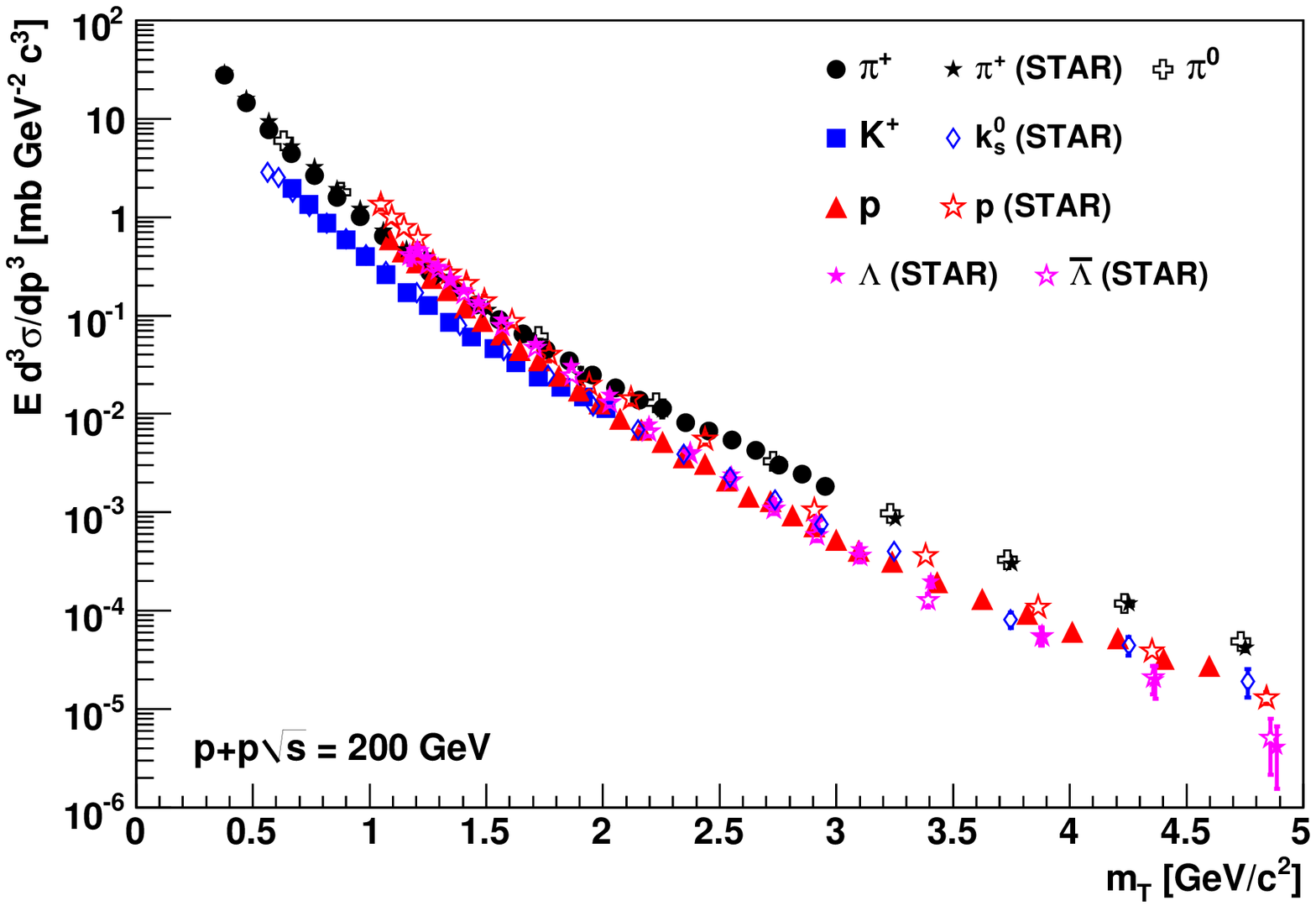}
\includegraphics[width=0.49\linewidth]{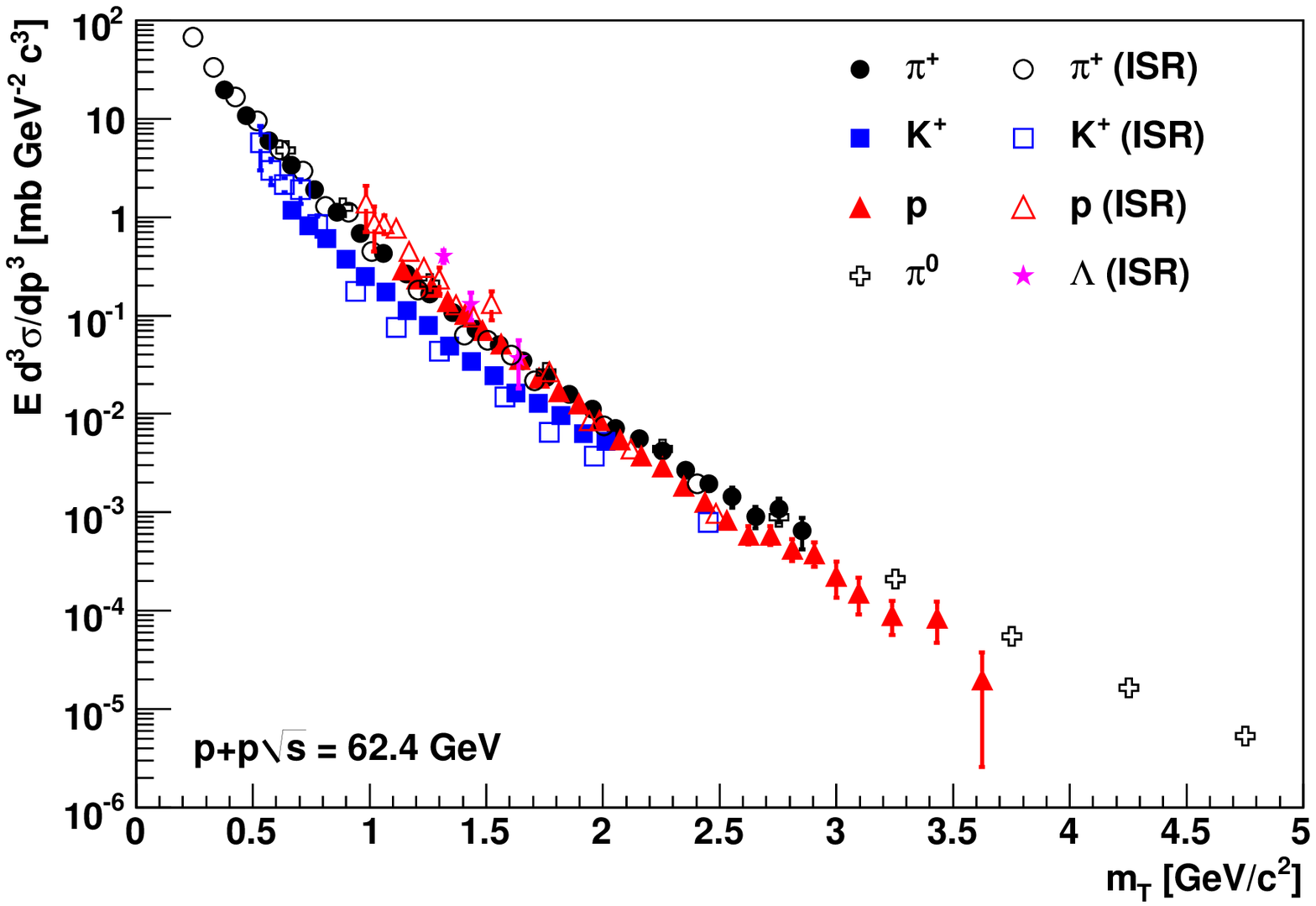}
\includegraphics[width=0.49\linewidth]{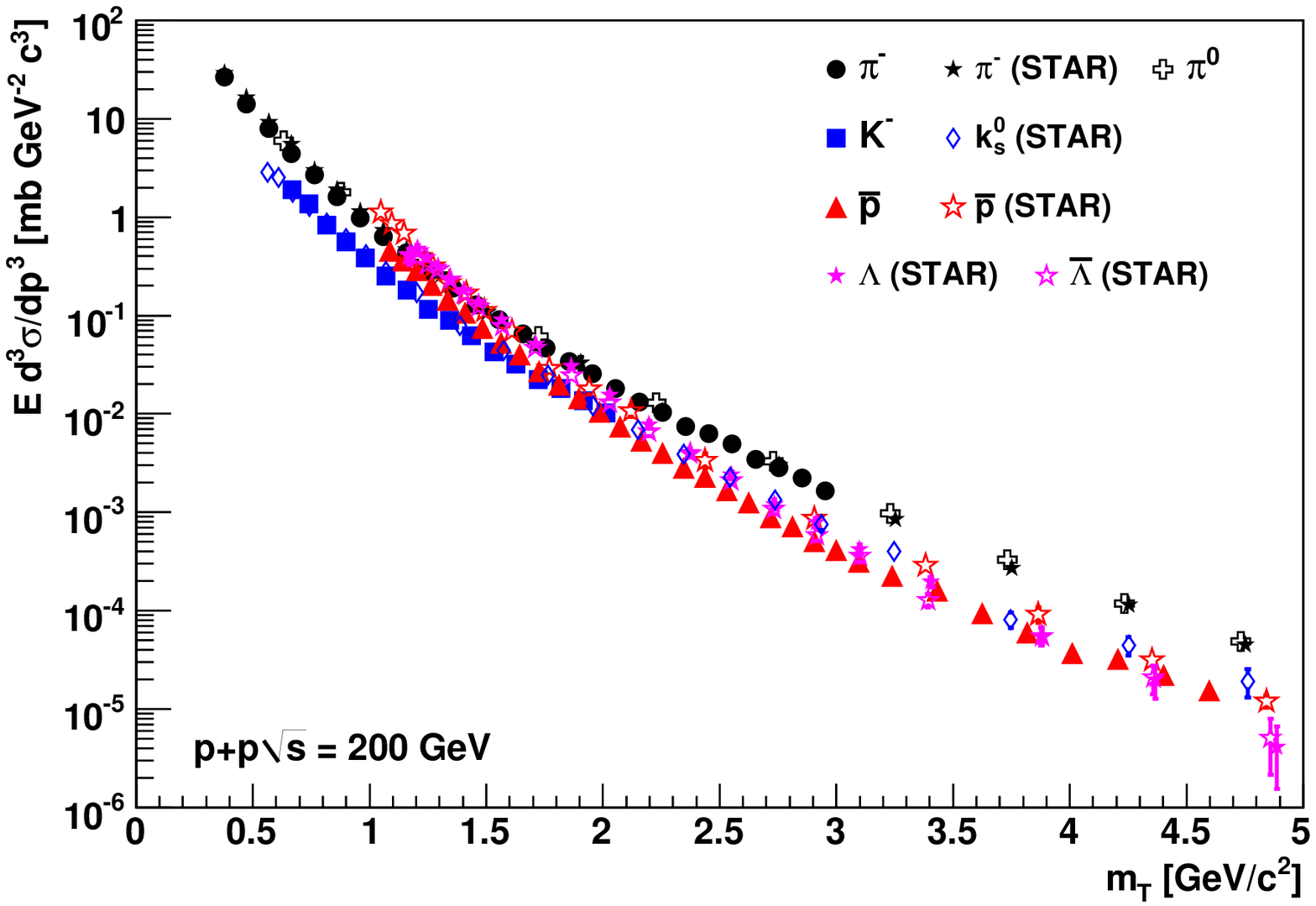}
\includegraphics[width=0.49\linewidth]{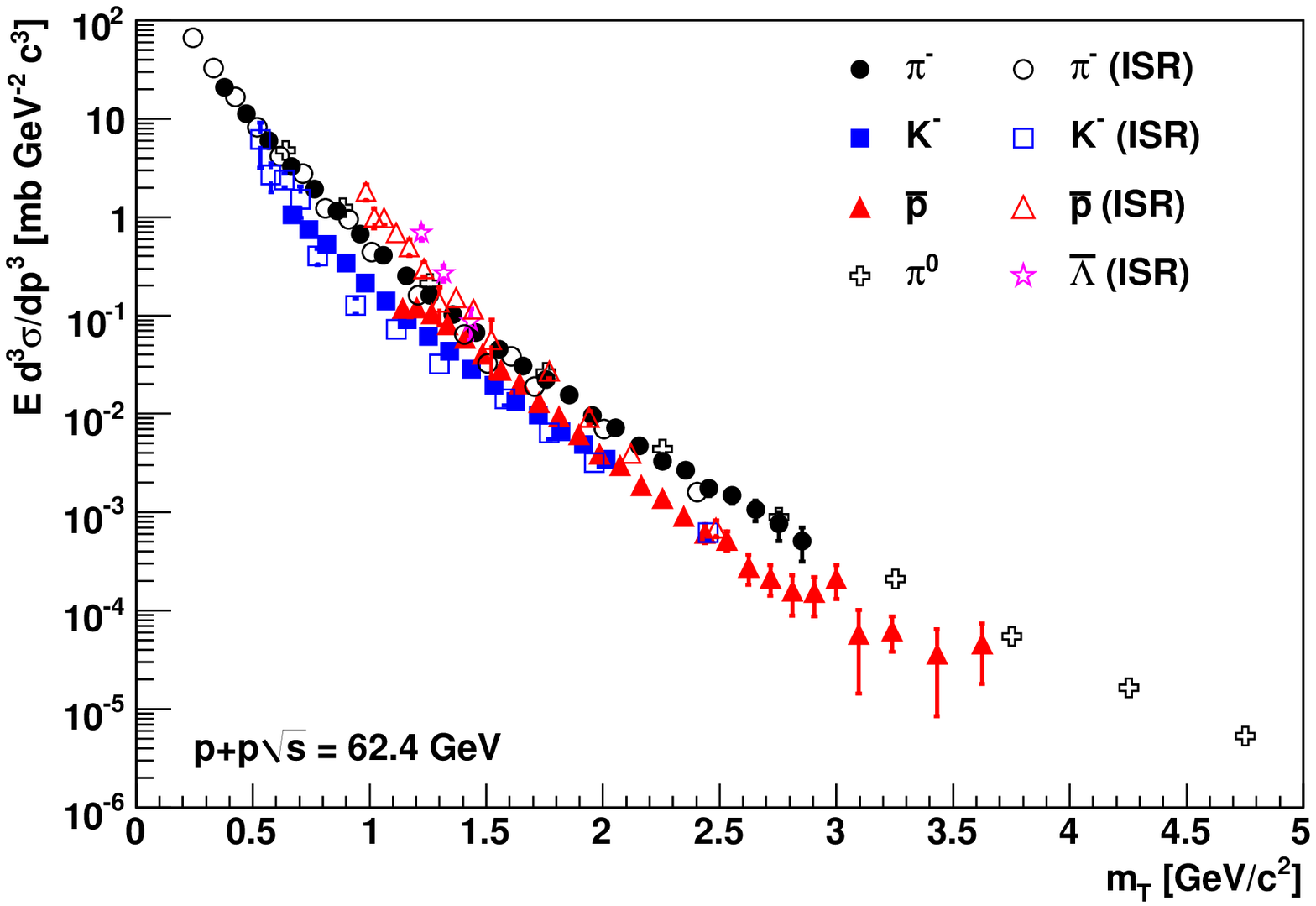}
\caption{(color online) 
Transverse mass distributions 
for $\pi^{\pm}$, $\pi^{0}$, $K^{\pm}$, $p$, and $\pbar$ in $p+p$ 
collisions at $\sqs = $ (left) 200 and (right) 62.4 GeV at midrapidity 
for (upper) positive and (lower) negative hadrons.
Only statistical uncertainties are shown.
The references for STAR data are
$\pi^{\pm}$, $p$, $\pbar$~\cite{Adams:2006nd} and
$K^{0}_s$, $\Lambda$, $\overline{\Lambda}$~\cite{Abelev:2006cs}.
The references for ISR data are
$\pi^{\pm}$, $K^{\pm}$, $p$, $\pbar$~\cite{Alper:1975jm}
and $\Lambda$, $\overline{\Lambda}$~\cite{isr_lambda_pp63}.
}
\label{fig:mt_pikp}
%\label{fig:mt_pikp_run5pp200}
%\label{fig:mt_pikp_run6pp62}
\end{figure*}

%%%%%%%%%%%%%%%%%%%%%%%%%%%%%%%%%%%%%%%%%%%%%%%%%%%%%%%%%%%%% Fig_7
\begin{figure*}[t]
\includegraphics[width=0.49\linewidth]{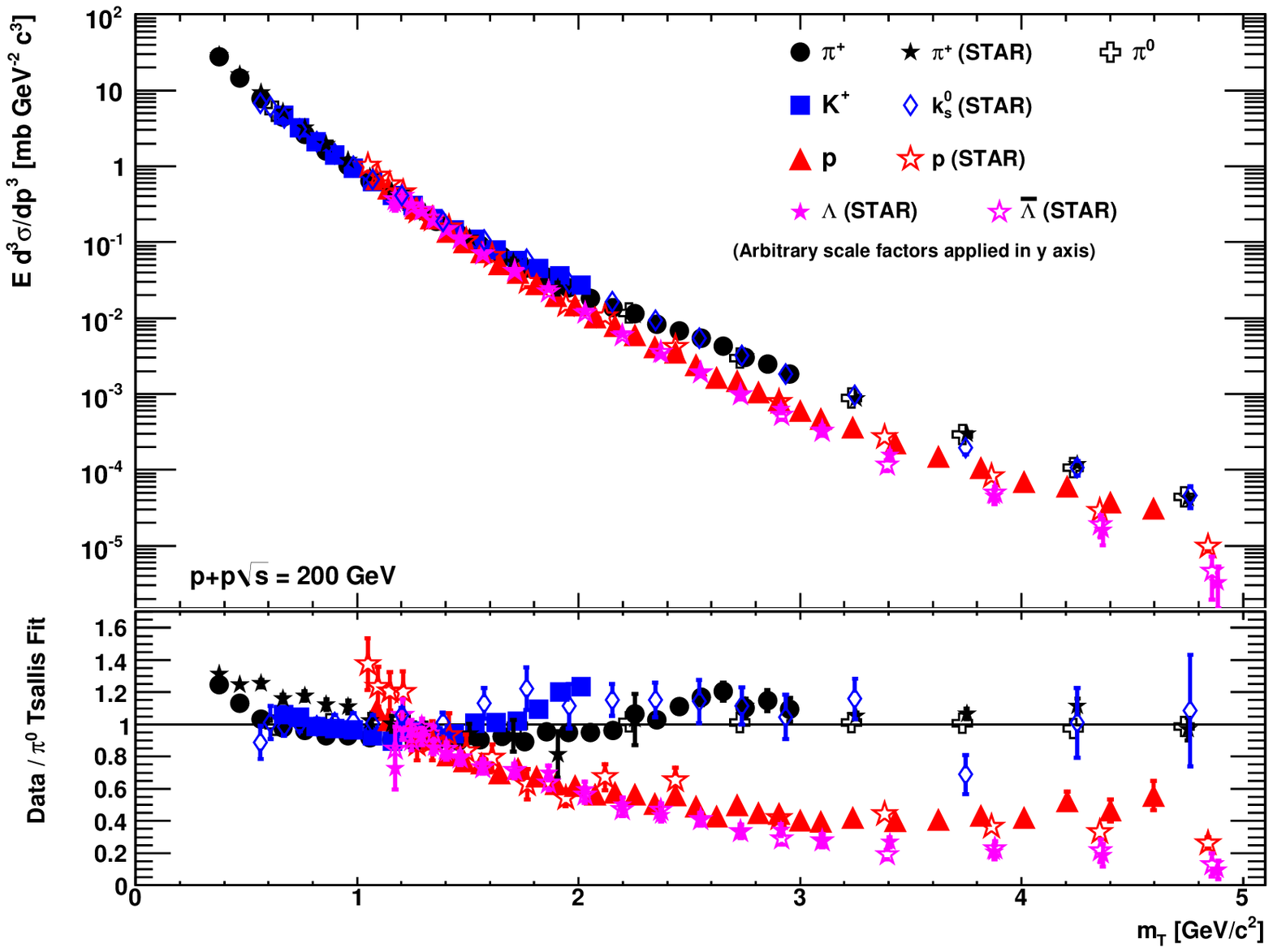}
\includegraphics[width=0.49\linewidth]{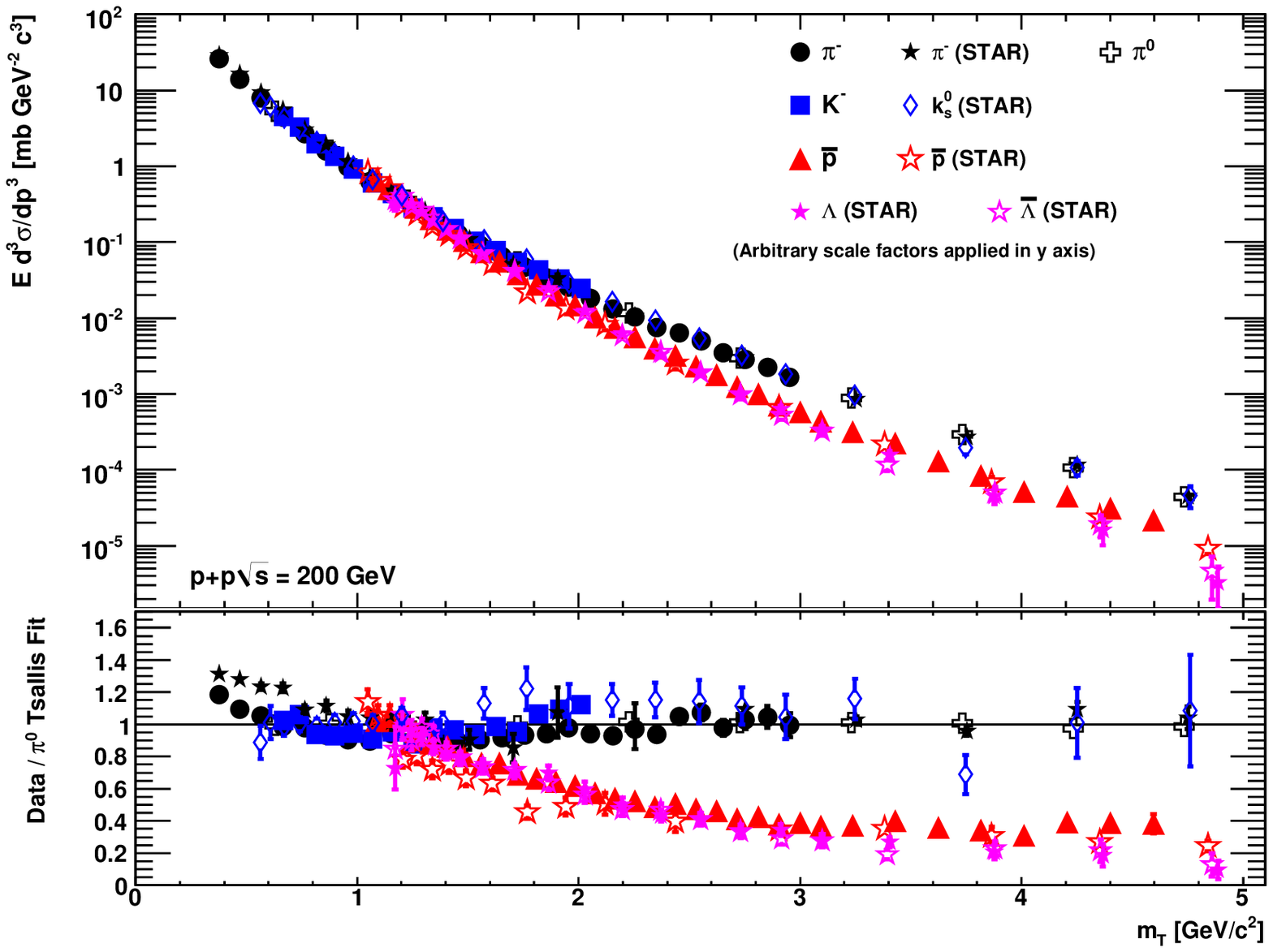}
\includegraphics[width=0.49\linewidth]{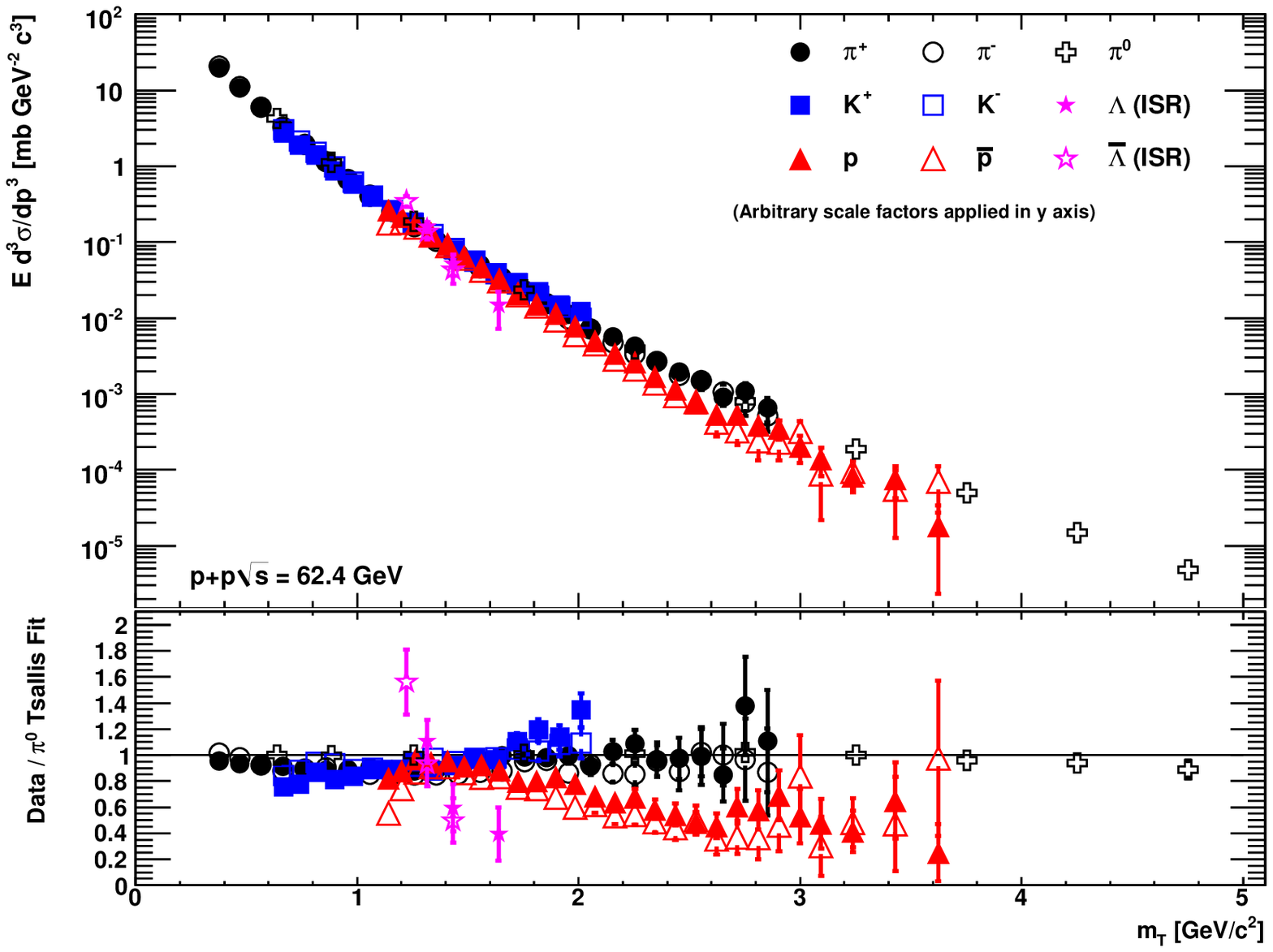}
\caption{(color online) 
Scaled transverse mass distributions
for $\pi^{\pm}$, $\pi^{0}$, $K^{\pm}$, $p$, and $\pbar$ in $p+p$ 
collisions at $\sqs = $ (upper) 200 and (lower) 62.4 GeV at midrapidity 
for (upper left) positive, (upper right) negative, and (lower) $\pm$ hadrons.
Only statistical uncertainties are shown.
(upper) The STAR spectra 
for $K^{0}_s$, $\Lambda$, $\overline{\Lambda}$ are from~\cite{Abelev:2006cs}.
(lower) The ISR spectra for $\Lambda$, $\overline{\Lambda}$ are 
from~\cite{isr_lambda_pp63}.
Arbitrary scaling factors are applied to match the yield of other particles
to that of charged pions in the range of $\mt$ = 1.0--1.5 GeV/$c^2$.
The lower panels of each plot show the ratio to $\pi^0$ Tsallis fit.
}
\label{fig:mt_scaled_pikp}
%\label{fig:mt_scaled_pikp_run5pp200}
%\label{fig:mt_scaled_pikp_run6pp62}
\end{figure*}

Ratios of the $\pt$ spectra at 200 GeV to those at 62.4 GeV are shown in 
the bottom plot of Fig.~\ref{fig:pt_pikp}.  The left panel shows the 
ratios for positively charged particles, and the right panel is those 
for the negatively charged particles. The data for neutral 
pions~\cite{Adare:2007dg, PPG087} are also shown on both panels. The 
ratios show a clear increase as a function of $\pt$ for all the ratios. 
Since hard scattering is expected to be the dominant particle production 
process at high $\pt$, this strong $\pt$ dependence indicates two 
features: (1) spectral shape is harder for 200 GeV compared to those for 
62.4 GeV, and (2) a universal shape for all particle species up to 
$\pt =$ 2--3 GeV/$c$. In the same figure, the results from NLO pQCD 
calculations with DSS fragmentation function~\cite{W_Vogelsang_private} 
for pions with different factorization, fragmentation, and 
renormalization scales (which are equal) are also shown.
The agreement is relatively poor, due to the disagreement between the 
NLO pQCD calculation with DSS fragmentation function and measurement for 
pions at $\sqrt{s}=62.4$ GeV. As we will discuss in detail in 
Section~\ref{sec:pQCD}, it is found that NLL pQCD gives a better 
description of the data for $p+p$ at 62.4 GeV.

%\clearpage

Please note that each line in pQCD is calculated for each $\mu$ (= 
$\pt$/2, $\pt$, 2$\pt$) value. The hard scale resides in the hard 
scattering, which is expected to be the same regardless of hadron 
species. The theoretical uncertainty in the ratio of NLO (200 GeV / 62.4 
GeV) significantly cancels. The same comparison of ratio for NLL cannot 
be made due to the unreliability of resummation in NLL pQCD at 200 GeV 
in the low $\pt$ region~\cite{W_Vogelsang_private}.

%\newpage

%------------
% mT spectra
%------------
\subsection{$\mt$ spectra}
\label{sec:mt}

In $p+p$($\pbar$) collisions at high energies, the transverse mass 
($\mt$) spectra of identified hadrons show a universal scaling behavior, 
and this fact is known as $\mt$ scaling. In order to check the $\mt$ 
scaling and to gain a further insight on the particle production 
mechanism especially at high $\pt$ at RHIC energies, transverse mass 
spectra in 200 and 62.4 GeV $p+p$ collisions are shown in 
Fig.~\ref{fig:mt_pikp}. The 
data for $\pi^{\pm}$, $K^{\pm}$, $p$, and $\pbar$ in 200 and 62.4 GeV 
are from this study. The $\pi^{0}$ spectra are taken from the PHENIX 
measurements~\cite{Adare:2007dg,PPG087}. From the STAR experiment, 
$\pi^{\pm}$, $p$, and $\pbar$ spectra in 200 GeV $p+p$ are taken from 
~\cite{Adams:2006nd}, and $K^{0}_s$, $\Lambda$, and $\overline{\Lambda}$ 
spectra in 200 GeV $p+p$ are taken from ~\cite{Abelev:2006cs}. The 
$\pi^{\pm}$, $K^{\pm}$, $p$, $\pbar$ spectra in 63 GeV $p+p$ are 
from~\cite{Alper:1975jm}, and $\Lambda$, $\overline{\Lambda}$ 
spectra in 63 GeV $p+p$ are from the ISR 
experiment~\cite{isr_lambda_pp63}. For both energies one can see 
``almost'' the same spectral shape for all particle species. 
However, this scaling is not perfect, because it is violated by the 
yields that are different for each particle species. In order to see 
the approximate scaling behavior and its violation more clearly, one 
needs to introduce certain factors, that normalize these yields to 
make them coincide at a certain value of $\mt$.

%\clearpage

%%%%%%%%%%%%%%%%%%%%%%%%%%%%%%%%
% T_inv (pt spectra) fit table %
%%%%%%%%%%%%%%%%%%%%%%%%%%%%%%%% 
%%%%%%%%%%%%%%%%%%%%%%%%%%%%%%%%%%%%%%%%%%%%%%%%%%%%  Table_III
\begin{table}[ht]
\caption{Fitting results for $A$, $T$ of Eq.~\ref{eq:pt_exp_fun} 
for $\pt$ spectra for $\pi^{\pm}$, $K^{\pm}$, $p$, and $\pbar$ in 
200 and 62.4 GeV $p+p$ collisions. The fitting range is fixed as $\pt$ = 
0.5--1.5 GeV/$c$ for $\pi^{\pm}$,
0.6--2.0 GeV/$c$ for $K^{\pm}$, and 
0.8--2.5 GeV/$c$ for $p$, $\pbar$
at both collision energies.}
\begin{ruledtabular} \begin{tabular}{ccccc}
$\sqrt{s}$ & hadron & $A$ & $T$   &  $\chi^2$/NDF \\ 
   (GeV)   & & & (GeV/$c$) & \\ \hline
200 & $\pi^{+}$ & 80.1   $\pm$ 7.2 & 0.220      $\pm$ 0.004 & 11.5/8 \\
    & $\pi^{-}$ & 80.7   $\pm$ 7.5 & 0.220      $\pm$ 0.004 & 13.5/8  \\
    & $K^{+}$ & 6.45     $\pm$ 0.50 & 0.296     $\pm$ 0.005 & 29.4/12 \\
    & $K^{-}$ & 6.62     $\pm$ 0.51 & 0.293     $\pm$ 0.004 & 18.8/12 \\
    & $p$     & 3.24     $\pm$ 0.38 & 0.318     $\pm$ 0.006 & 3.3/15  \\
    & $\pbar$ & 2.83     $\pm$ 0.35 & 0.318     $\pm$ 0.006 & 2.8/15  \\
62.4 & $\pi^{+}$ & 78.0  $\pm$ 7.0     & 0.203  $\pm$ 0.003 & 9.0/8   \\
     & $\pi^{-}$ & 81.0  $\pm$ 6.2     & 0.200  $\pm$ 0.003 & 11.1/8  \\
     & $K^{+}$   & 6.17  $\pm$ 0.52    & 0.264  $\pm$ 0.004 & 15.6/12 \\
     & $K^{-}$   & 6.01  $\pm$ 0.49    & 0.254  $\pm$ 0.004 & 10.0/12 \\
     & $p$       & 4.61  $\pm$ 0.48    & 0.275  $\pm$ 0.005 & 2.8/15  \\
     & $\pbar$   & 2.95  $\pm$ 0.36    & 0.267  $\pm$ 0.005 & 2.9/15  \\ 
\end{tabular} \end{ruledtabular}
\label{tab:tinv_fit_pt}
\end{table}
%====

Figure~\ref{fig:mt_scaled_pikp} shows the $\mt$ spectra with such 
scaling factors implemented. These normalization scaling factors are 
determined to match the yield of each particle species to that of 
charged pions in the range of $\mt$ = 1.0--1.5 GeV/$c^{2}$, i.e. factors 
which scale the normalization (vertical axis) as in the legends of 
Fig.~\ref{fig:mt_scaled_pikp}.  These scaling factors are 
tabulated in Table~\ref{tab:mt_scale_factor}.  The bottom panels on 
the plots in Fig.~\ref{fig:mt_scaled_pikp} are the ratio of data to the 
fitting result using Tsallis function for $\pi^{0}$ data in 200 
GeV~\cite{Adare:2007dg} and 62.4 GeV~\cite{PPG087}.  Above $\mt >$ 1.5 
GeV/$c^2$, these figures indicate a clear separation between meson and 
baryon spectra.  The meson spectra are apparently harder than the baryon 
spectra in this representation.  This effect can be seen more clearly on 
the $\sqs$ = 200 GeV dataset, than on data measured at 62.4 GeV. Such a 
baryon-meson splitting in $\mt$ spectra has been reported by the STAR 
experiment in $p+p$ collisions at $\sqs = $ 200 
GeV~\cite{Abelev:2006cs}.  That paper argued for a given jet energy, 
mesons might be produced with higher transverse momentum than baryons, 
because meson production in jet fragmentation requires only a (quark, 
antiquark) pair, while baryon production requires a (diquark, 
antidiquark) pair.

However, such a rescaling along the vertical axis is arbitrary. Since it 
is known that the shapes of the power-law tail for different particle 
species at high $\pt$ in 200 GeV data are very 
similar~\cite{phenix:ppg099}, one can show by using different scaling 
factors $\mt$ spectra match at higher $\mt$. In this case, $\mt$ spectra 
for baryons overshoot those for mesons at low $\mt$. The interpretation 
of $\mt$ scaling is not settled yet, but from the experimental results 
presented here, it is clear that the spectral shape for mesons and 
baryons are very similar for both low and high $\mt$ regions, but 
different for the intermediate $\mt$ (1.5--2.5 GeV/$c^2$) region. In 
Section~\ref{sec:mtm}, we will discuss the spectral shape at low $\mt$ 
in detail, by taking into account the hadron mass effect.
 %
% ========================
% mt scaling factor table
% ========================
%%%%%%%%%%%%%%%%%%%%%%%%%%%%%%%%%%%%%%%%%%%%%%%%%%%%  Table_IV
\begin{table}[htb]
\caption{Normalization scaling factors for $\mt$ spectra for 
Fig.~\ref{fig:mt_scaled_pikp}. The scaling factors for the 
STAR experiment are determined from~\cite{Adams:2006nd,Abelev:2006cs} 
and those for the ISR results are determined from~\cite{isr_lambda_pp63}.}
\begin{ruledtabular} \begin{tabular}{cccccccccccc}
$\sqrt{s}$ & Exp. & $\pi^{+}$  & $\pi^{-}$  & $\piz$  & $K^{+}$  & $K^{-}$  
& $K^{0}_{s}$  & $p$    & $\pbar$  & $\Lambda$  & $\overline{\Lambda}$ \\ 
(GeV) & & & & & & & & & & &  \\ \hline 
200  & PHENIX & 1.0  & 1.0  & 0.9  & 2.4  & 2.4  & - & 1.15 & 1.4  & -   & - \\
200  & STAR  & 1.0  & 1.0  & -  & -  & -  & 2.4   & 0.75 & 0.75  & 0.8  & 0.9  \\
62.4 & PHENIX & 1.0  & 1.0  & 0.9  & 2.32  & 2.88  & -  & 0.9 & 1.5  & -   & - \\
63   & ISR  & -   & -   & -  & -  & -  & -   & -  & -  & 0.4  & 0.5  \\
\end{tabular} \end{ruledtabular}
\label{tab:mt_scale_factor}
\end{table}
% ======

%%%%%%%%%%%%%%%%%%%%%%%%%%%%%%%%%%%%%%%%%%%%%%%%%%%%%%%%%%%%% Fig_8
\begin{figure}[htb]
\includegraphics[width=1.01\linewidth]{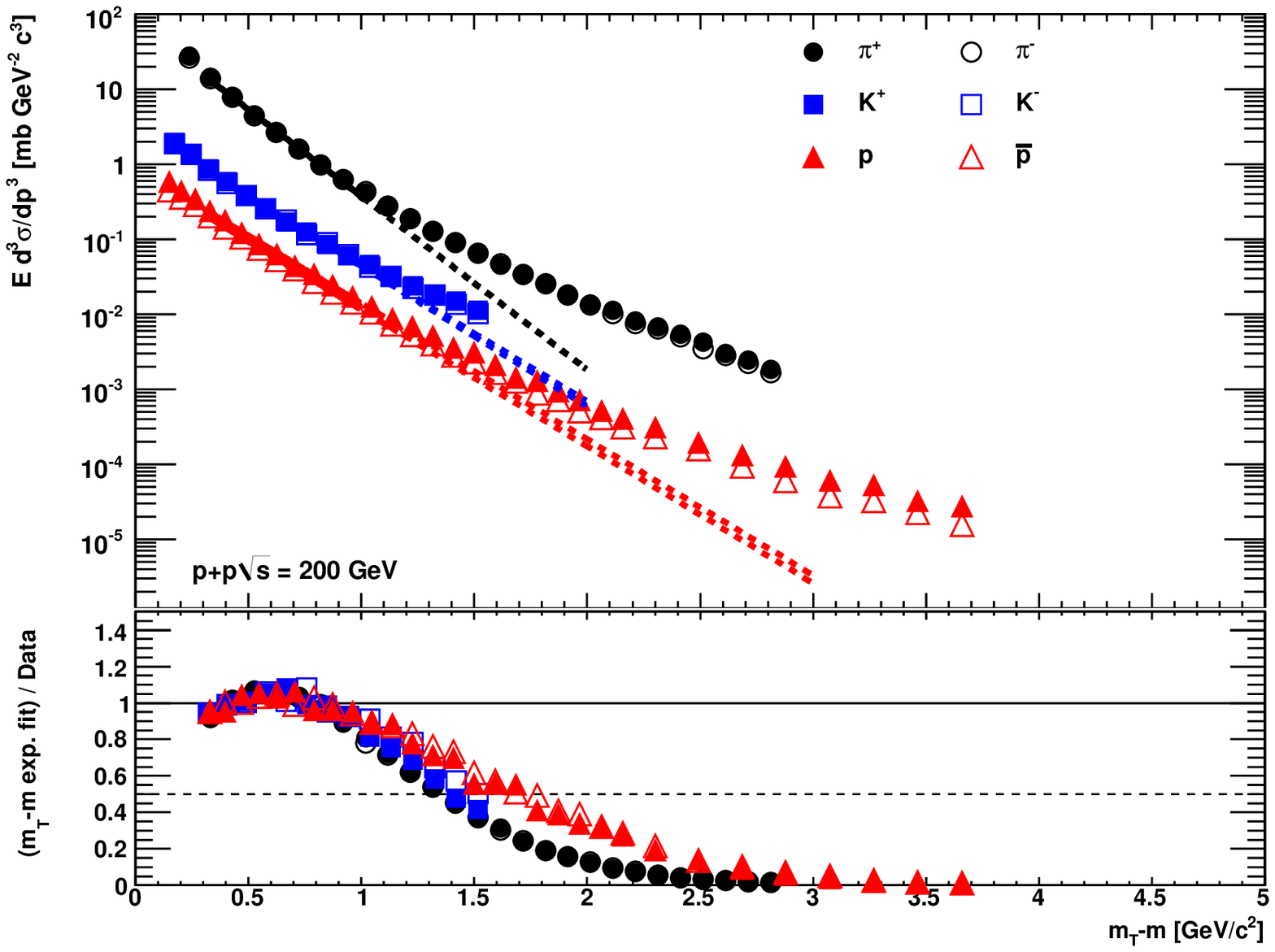}
\includegraphics[width=1.01\linewidth]{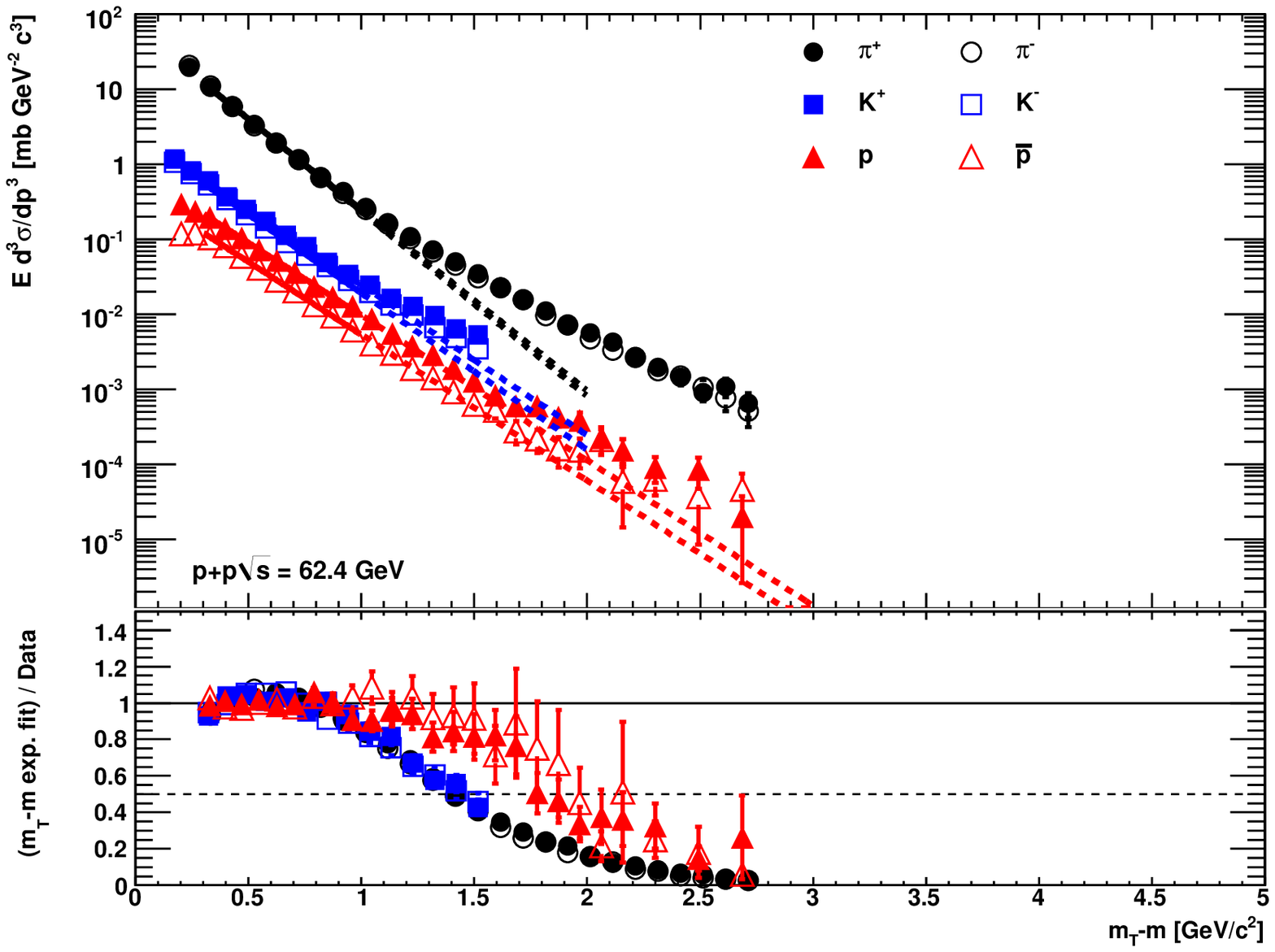}
\caption{(color online) 
$\mt-m$ spectra for $\pi^{\pm}$, $K^{\pm}$, $p$, and $\pbar$ 
in $p+p$ collisions at $\sqs = $ (upper) 200 and (lower) 62.4 GeV at midrapidity.
Only statistical uncertainties are shown.
Each spectrum is fitted with an exponential form of 
Eq.~(\ref{eq:mt-m_exp_fun})
in the range of $\mt-m$ = 0.3--1.0 GeV/$c^{2}$. 
Solid lines represent the functions in the fitted range, dashed lines
show the extrapolation of these functions beyond this range.
(lower panels) Ratio of the exponential fit to data for each particle species.}
\label{fig:mtm_w_line_pikp}
\end{figure}

%%%%%%%%%%%%%%%%%%%%%%%%%%%%%%%%%%%%%%%%%%%%%%%%%%%%%%%%%%%%% Fig_9
\begin{figure}
\includegraphics[width=1.05\linewidth]{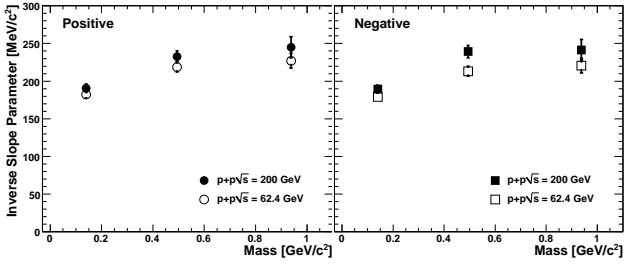}
\caption{Inverse slope parameter $T_{\rm inv}$ for $\pi^{\pm}$, $K^{\pm}$, $p$ and $\pbar$
in $p+p$ collisions at $\sqs =$ 200 GeV and 62.4 GeV. The fitting range is $\mt-m$ = 
0.3--1.0 GeV/$c^{2}$ for all particle species at both collision 
energies.
The errors are statistical and systematic combined. The statistical errors are negligible.
}
\label{fig:tinv_vs_mass}
\end{figure}

%%%%%%%%%%%%%%%%%%%%%%%%%%%%%%%%%%%%%%%%%%%%%%%%%%%%%%%%%%%%% Fig_10
\begin{figure*}[t]
\includegraphics[width=1.05\linewidth]{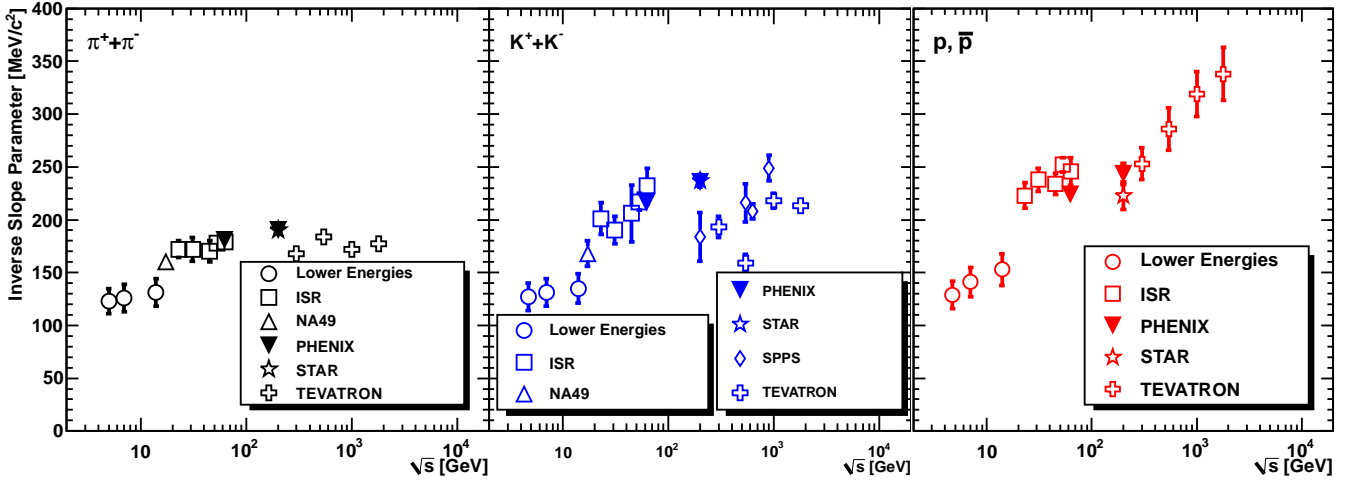}
\caption{(color online) 
Inverse slope parameter $T_{\rm inv}$ for $\pi^{+}+\pi^{-}$, 
$K^{+}+K^{-}$, $p$ and $\pbar$ in $p+p$ and $p+\pbar$ collisions versus 
collision energy $\sqs$. The fitting range is $\mt-m$ = 0.3--1.0 
GeV/$c^{2}$ for all particle species for all collision systems. The 
errors are statistical and systematic combined. The statistical errors 
are negligible. The other experimental data are taken 
from~\cite{Rossi:1974if, Alper_npb87, e735_1993, compilation_kaon, 
Adams:2006nd}.}
 \label{fig:tinv_vs_roots}
\end{figure*}

%%%%%%%%%%%%%%%%%%%%%%%%%%%%%%%%%%%%%%%%%%%%%%%%%%%%%%%%%%%%% Fig_11
\begin{figure}[t]
\includegraphics[width=1.02\linewidth]{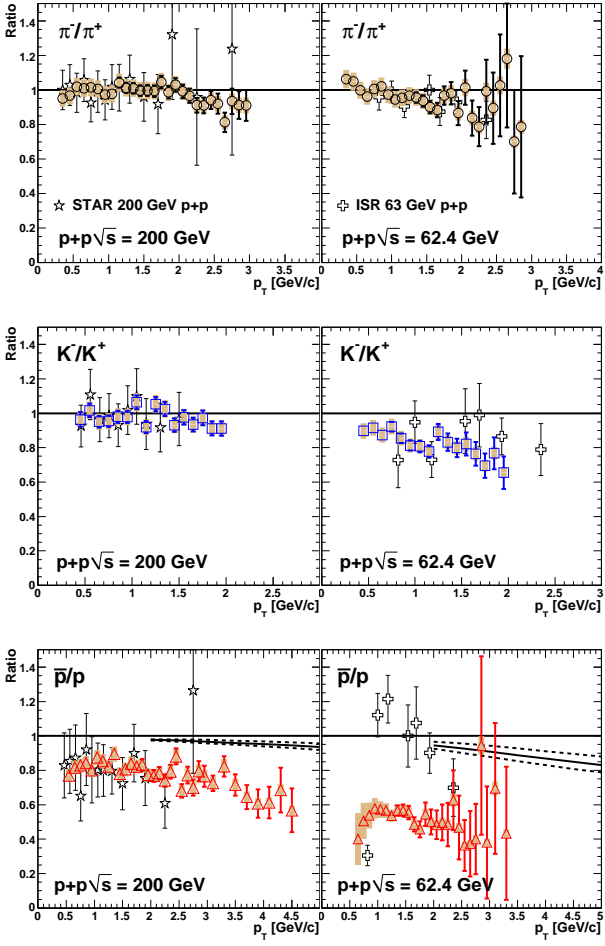}
\caption{(color online) 
$\pi^{-}/\pi^{+}$, $K^{-}/K^{+}$, $\pbar/p$ ratios as a function 
of $\pt$ in $p+p$ collisions at $\sqs =$ (left) 200 and (right) 62.4 GeV.
Systematic uncertainties are shown as vertical shaded band. 
STAR data are from~\cite{STAR-pp-spectra-2} and ISR data are from~\cite{Alper_npb87}.
For $\pbar/p$ ratios, the NLO pQCD calculations using the DSS fragmentation 
functions~\cite{W_Vogelsang_private} are also shown as (solid lines) $\mu = \pt$ and 
(dashed lines) $\mu = 2\pt$, $\pt$/2.}
\label{fig:ratio_negpos_run5pp200_run6pp62}
\end{figure}

%%%%%%%%%%%%%%%%%%%%%%%%%%%%%%%%%%%%%%%%%%%%%%%%%%%%%%%%%%%%% Fig_12
\begin{figure}[t]
\includegraphics[width=1.01\linewidth]{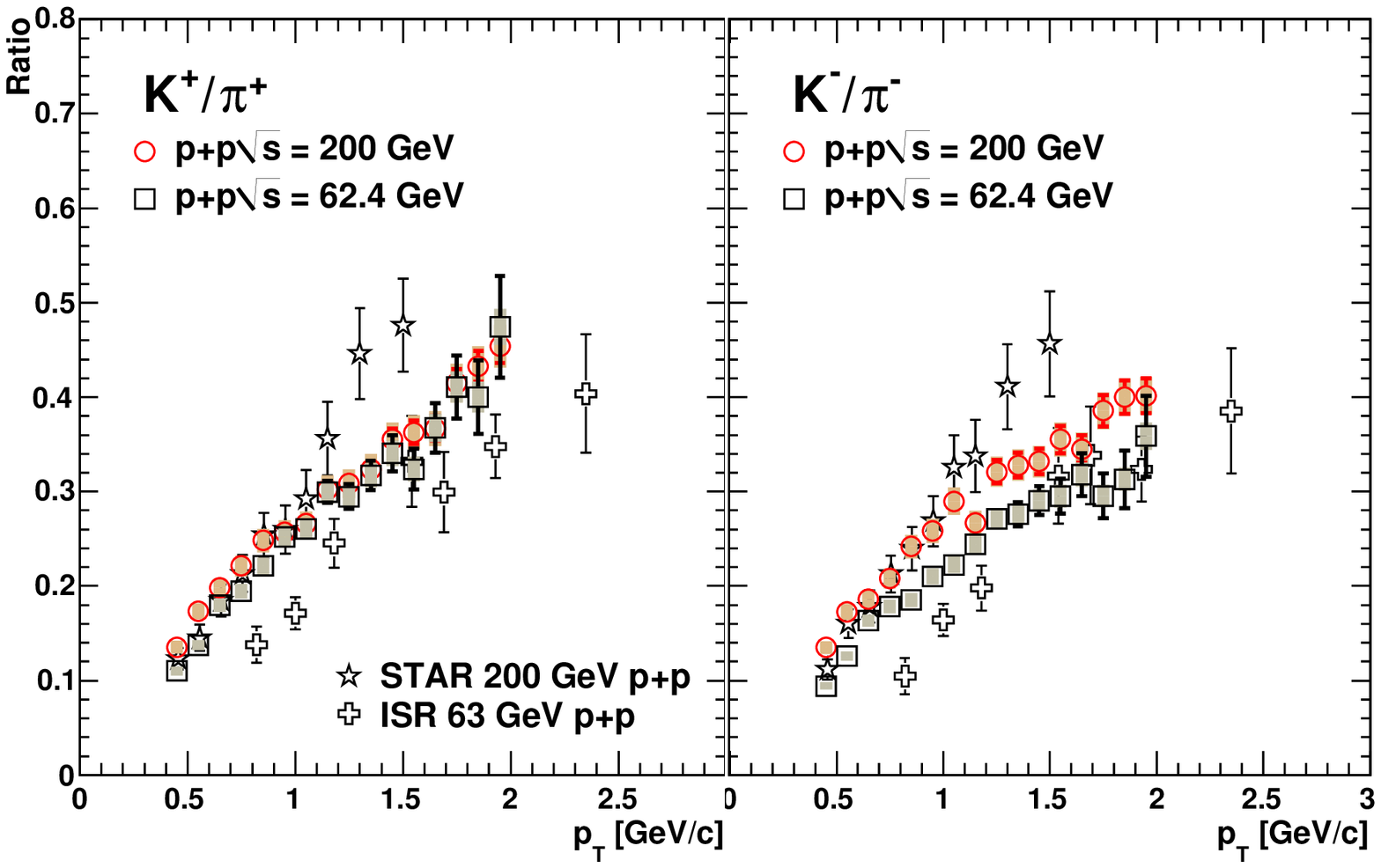}
\includegraphics[width=1.01\linewidth]{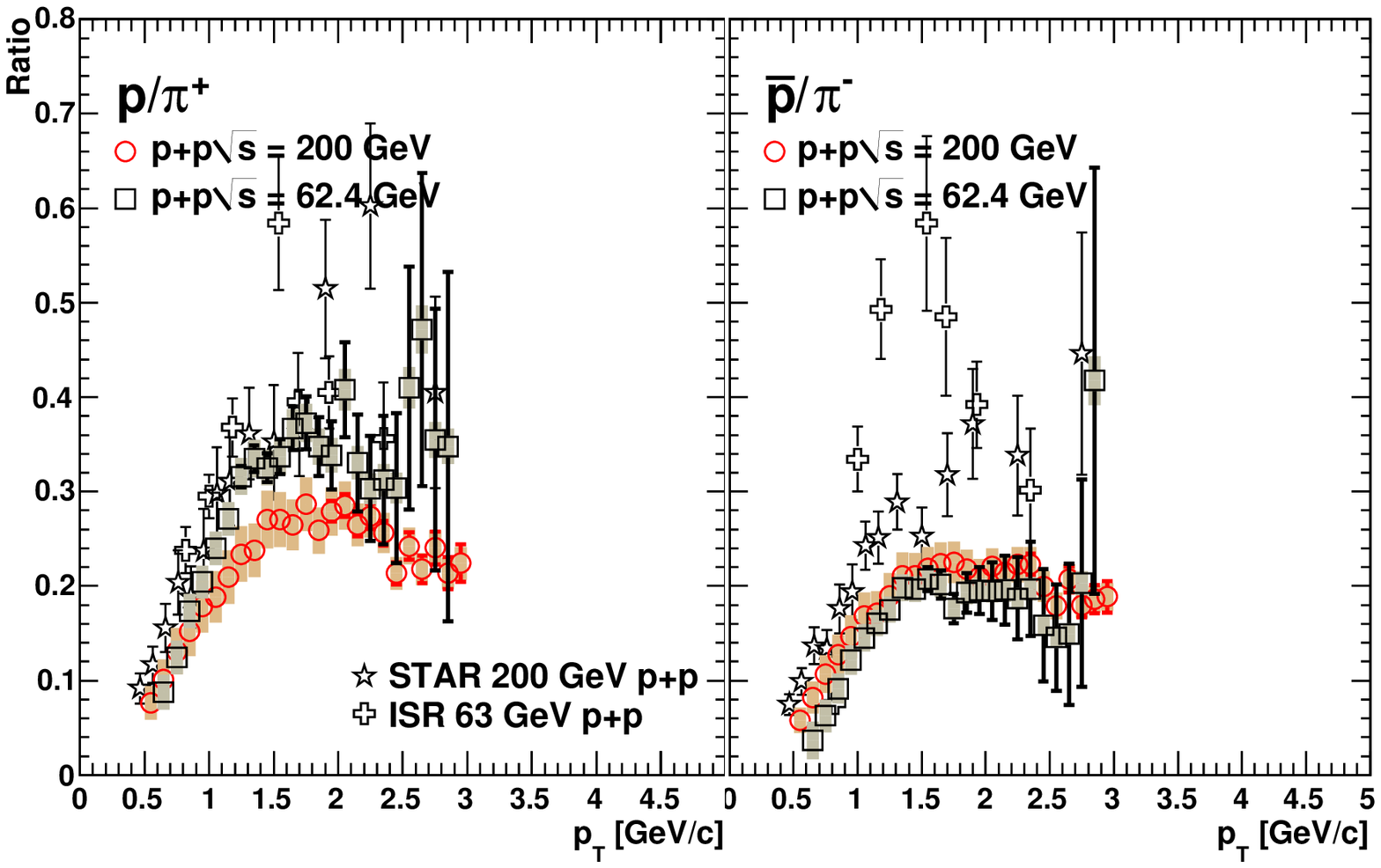}
\includegraphics[width=1.01\linewidth]{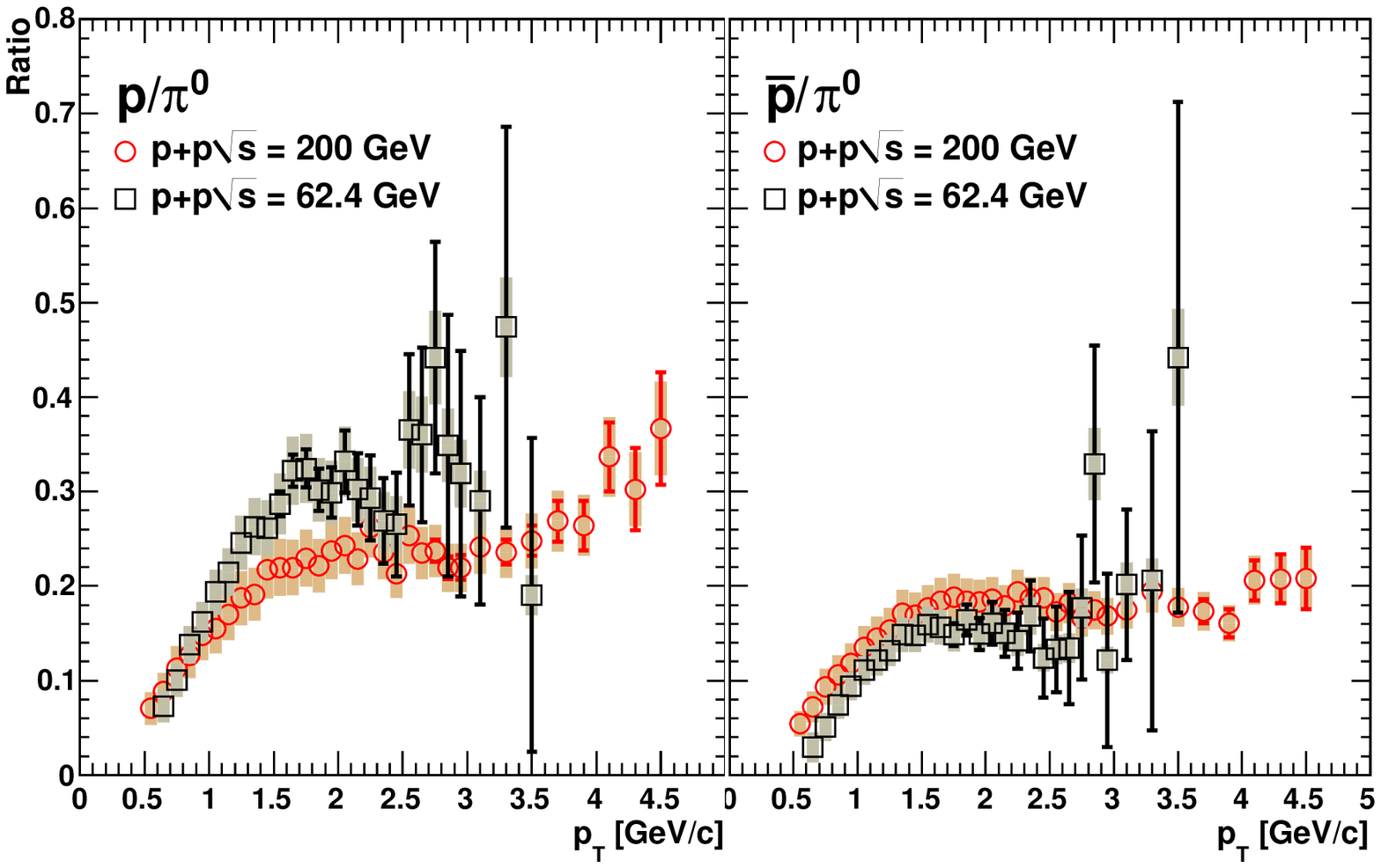}
\caption{(color online) 
(top) $K^{+}/\pi^{+}$ and $K^{-}/\pi^{-}$ ratios,   
(middle) $p/\pi^{+}$ and (bottom) $\pbar/\pi^{-}$ ratios, 
and (bottom) $p/\piz$ and $\pbar/\piz$ ratios 
as a function of $\pt$ in $p+p$ collisions 
at $\sqs =$ (left) 200 and (right) 62.4 GeV.
Systematic uncertainties are shown as vertical shaded bands.}
\label{fig:ratios}
\end{figure}   

%%\clearpage

%---------------
% mt-m spectra
%---------------
\subsection{$\mt-m$ spectra}
\label{sec:mtm}

Figure~\ref{fig:mtm_w_line_pikp} shows the $\mt-m$ spectra for 
$\pi^{\pm}$, $K^{\pm}$, $p$, and $\pbar$ in 200 and 62.4 GeV $p+p$ 
collisions, respectively. When analyzing these $\mt-m$ spectra of 
various identified hadrons, one discusses the spectral shape mainly in 
the low $\mt-m$ region. Each of these spectra is fitted with an 
exponential functional form:
  \begin{equation}
  \frac{1}{2\pi \mt}\frac{d^{2}\sigma}{dyd\mt} 
   = A \exp{ \Bigl( - \frac{\mt-m}{T_{\rm inv}} \Bigr) }
  \label{eq:mt-m_exp_fun}
  \end{equation}
where $A$ is a normalization factor and $T_{\rm inv}$ is called the inverse 
slope parameter. The fitting parameters and $\chi^{2}$/NDF by using 
Eq.~\ref{eq:mt-m_exp_fun} for $\pi^{\pm}$, $K^{\pm}$, $p$, and $\pbar$ 
in 200 and 62.4 GeV $p+p$ collisions, are tabulated in 
Table~\ref{tab:tinv_fit_pt}. The fitting range is fixed as $\mt-m$ = 
0.3--1.0 GeV/$c^{2}$ for all particle species at both collision 
energies. We obtain smaller $\chi^{2}$/NDF for protons and antiprotons 
than those for pions and kaons, because of the larger systematic 
uncertainties for $p$ and $\pbar$ at low $\pt$ due to the uncertainties 
of weak decay feed-down corrections. As seen in 
Fig.~\ref{fig:mtm_w_line_pikp} the spectra are exponential in the low 
$\mt-m$ range. At higher transverse mass, the spectra become less 
steep, corresponding to an emerging power law behavior. The transition 
from exponential to power law can be seen at $\mt-m$ = 1--2 GeV/$c^2$ 
for all particle species.

%%%%%%%%%%%%%%%%%%%
% T_inv fit table %
%%%%%%%%%%%%%%%%%%% 
%%%%%%%%%%%%%%%%%%%%%%%%%%%%%%%%%%%%%%%%%%%%%%%%%%%%  Table_V
\begin{table}[htb]
\caption{Fitting results for $A$, $T_{\rm inv}$ of Eq.~\ref{eq:mt-m_exp_fun} 
for $\pi^{\pm}$, $K^{\pm}$, $p$, and $\pbar$ in 200 and 62.4 GeV $p+p$ 
collisions. The fitting range is fixed as $\mt-m$ = 0.3--1.0 GeV/$c^{2}$ 
for all particle species at both collision energies.}
 \begin{ruledtabular} \begin{tabular}{ccccc}
$\sqrt{s}$ & hadron & $A$  & $T_{\rm inv}$  &  $\chi^{2}$/NDF \\ 
(GeV) & & & (GeV/$c^2$) & \\ \hline 
200 & $\pi^{+}$ & 73.4     $\pm$ 7.1    &  0.190  $\pm$ 0.005                    & 5.6/5  \\
    & $\pi^{-}$ & 74.8     $\pm$ 7.2    &  0.189  $\pm$ 0.005                    & 3.1/5  \\
    & $K^{+}$   & 3.25     $\pm$ 0.29   &  0.232  $\pm$ 0.007                    & 3.6/6  \\
    & $K^{-}$   & 2.99     $\pm$ 0.27   &  0.239  $\pm$ 0.008                    & 3.6/6  \\
    & $p$       & 0.85    $\pm$ 0.14  &  0.245  $\pm$ 0.014                    & 1.0/7 \\
    & $\pbar$   & 0.74    $\pm$ 0.13  &  0.241  $\pm$ 0.014                    & 0.5/7 \\
62.4 & $\pi^{+}$ & 61.7     $\pm$  5.9  &  0.182  $\pm$  0.005                   &  3.1/5 \\
    & $\pi^{-}$ & 65.2     $\pm$  5.3  &  0.179  $\pm$  0.004                   &  4.7/5 \\
    & $K^{+}$   & 2.44     $\pm$  0.22  &  0.219  $\pm$  0.007                   &  2.6/6 \\
    & $K^{-}$   & 2.21     $\pm$  0.20  &  0.213  $\pm$  0.006                   &  4.6/6 \\
    & $p$       & 0.81    $\pm$  0.10 &  0.227  $\pm$  0.010                   &  1.1/7 \\
    & $\pbar$   & 0.49    $\pm$  0.07 &  0.221  $\pm$  0.010                   &  0.3/7 \\ 
\end{tabular} \end{ruledtabular}
\label{tab:tinv_fit}
\end{table}
%====

%%%%%%%%%%%%%%%%%%%%%%%%%%%%%%%%%%%%%%%%%%%%%%%%%%%%  Table_VI
\begin{table*}[tb]
\caption{Fitting results from using the Tsallis distribution (Eq.~\ref{eq:levy_fun}) for 
$\pi^{\pm}$, $K^{\pm}$, $p$ and $\pbar$ in $p+p$ collisions at $\sqs$~=~200 
and 62.4~GeV.}
\begin{ruledtabular} \begin{tabular}{cccccc}
$\sqrt{s}$ (GeV) & hadron & $dN/dy$  &  $q$  & $C$  &  $\chi^{2}$/NDF \\ \hline 
200 & $\pi^{+}$ & 0.963    $\pm$  0.071     &  8.24  $\pm$ 0.33     &  0.115  $\pm$  0.006  &  4.3/23  \\
    & $\pi^{-}$ & 0.900    $\pm$  0.063     &  8.95  $\pm$ 0.39     &  0.123  $\pm$  0.006  &  3.2/23  \\
    & $K^{+}$   & 0.108    $\pm$  0.006     &  6.25  $\pm$ 0.64     &  0.137  $\pm$  0.011  &  1.6/13  \\
    & $K^{-}$   & 0.103    $\pm$  0.005     &  7.00  $\pm$ 0.78     &  0.147  $\pm$  0.011  &  2.9/13  \\
    & $p$       & 0.044   $\pm$  0.004    &  11.1  $\pm$ 1.6      &  0.184  $\pm$  0.014  &  4.1/22  \\
    & $\pbar$   & 0.037   $\pm$  0.003    &  12.0  $\pm$ 1.8      &  0.186  $\pm$  0.014  &  1.3/22  \\
62.4 & $\pi^{+}$ & 0.782    $\pm$ 0.056      & 12.1  $\pm$ 0.9       & 0.133   $\pm$ 0.007   &  4.6/22   \\
    & $\pi^{-}$ & 0.824    $\pm$ 0.053      & 11.9  $\pm$ 0.7       & 0.128   $\pm$ 0.006   &  4.8/22   \\
    & $K^{+}$   & 0.076   $\pm$ 0.003     & 10.2  $\pm$ 1.8      & 0.165   $\pm$ 0.012   &  4.9/13   \\
    & $K^{-}$   & 0.067   $\pm$ 0.003     & 11.6  $\pm$ 2.1      & 0.164   $\pm$ 0.011   &  2.2/13   \\
    & $p$       & 0.040   $\pm$ 0.003     & 24.5  $\pm$ 9.9       & 0.201   $\pm$ 0.015   &  7.1/21   \\
    & $\pbar$   & 0.022   $\pm$ 0.002     & 32.5  $\pm$ 21.0      & 0.202   $\pm$ 0.018   &  7.9/21   \\
\end{tabular} \end{ruledtabular}
\label{tab:tsallis_fit}
\end{table*}
%===

%%\clearpage

%-----------------
% Inverse slope T
%-----------------

The dependence of $T_{\rm inv}$ on hadron mass is shown in 
Fig.~\ref{fig:tinv_vs_mass}. These slope parameters are almost 
independent of the energy of $\sqs = 62.4$ and $200$ GeV. The inverse 
slope parameters of kaons is similar to that of protons while the slope 
parameter of pions has slightly smaller values. It may be possible that 
lower $T_{\rm inv}$ values for pions are due to pions from resonance decays 
(e.g. $\rho$, $\Lambda$), although such an effect is reduced by the 
lower transverse momentum cut. An alternative explanation is that 
hydrodynamical collective behavior may develop even in the small $p+p$ 
system, which we will explore at Section~\ref{sec:flow}.

%\newpage

In Fig.~\ref{fig:tinv_vs_roots}, the collision energy dependence of 
$T_{\rm inv}$ is shown by compiling results from past 
experiments~\cite{Rossi:1974if, Alper_npb87, e735_1993, compilation_kaon, 
Adams:2006nd}. The values of $T_{\rm inv}$ reported here are obtained by 
fitting all the $\pt$ spectra in the same way. The fitting range is 
$\mt-m$ = 0.3--1.0 GeV/$c^{2}$ for all particle species in all collision 
systems. The $T_{\rm inv}$ values for RHIC energies are consistent with 
earlier experimental results at other energies~\cite{Rossi:1974if, 
Alper_npb87, e735_1993, compilation_kaon}. For both pions and kaons, the 
inverse slope parameters increase with collision energy from $T_{\rm inv}$ 
= 120 MeV/$c^2$ to 170 MeV/$c^2$ (240 MeV/$c^2$) for pions (kaons) up to 
\sqs~=~200 GeV.  According to Tevatron data, $T_{\rm inv}$ seems to be 
saturated at $\sqs$ above 200 GeV.  The inverse slope parameters of 
protons and antiprotons indicate an increase at lower $\sqs$ which keeps 
on increasing even at Tevatron energies.  We look forward to data from the 
LHC to further clarify these issues.

%\clearpage

%-------
% Particle Ratio
%-------
\subsection{Particle ratios}
\label{sec:ratio}

Figures~\ref{fig:ratio_negpos_run5pp200_run6pp62} and \ref{fig:ratios} 
show particle ratios such as 
antiparticle-to-particle, K/$\pi$, $p/\pi$ as a function of $\pt$.  
The $\pi^-/\pi^+$ and $K^-/K^+$ ratios show a flat $\pt$ dependence 
at both 200 and 62.4 GeV energies. The $\pi^-/\pi^+$ ratio is almost 
unity at both energies. The $K^-/K^+$ ratio is consistent with unity 
at $\sqs = 200$ GeV, while it decreases to 0.8--0.9 in the measured 
$\pt$ range at 62.4 GeV. On the other hand, $\pbar/p$ ratio seems to 
be a decreasing function of $\pt$ at 200 GeV, from the value of 
$\approx$ 0.8 at $\pt = 1.0$ GeV/$c$ to 0.6 at $\pt = 4.5$ GeV/$c$. 
Note that we fit the $\pbar/p$ ratio for 200 GeV $p+p$ from 
$\pt$ = 1--4.5 GeV/$c$ to a linear function, $a+b \pt$, which gives 
$a$ = 0.93 $\pm$ 0.02, $b$ = -0.07 $\pm$ 0.01. This decrease, also seen 
at lower $\sqs$~\cite{Alper_npb87}, might be the result of a difference 
of fragmentation between quark jet and gluon jet at high $\pt$ region as 
suggested by the DSS fragmentation functions~\cite{W_Vogelsang_private}. 
However, the NLO pQCD calculation using the DSS fragmentation functions 
(lines on the panels for $\pbar/p$ ratios) shows this effect is in 
disagreement with the measured $\pbar/p$ ratios. At 62.4 GeV, we can not 
conclude on the significance of the decrease of the $\pbar/p$ ratios as 
a function of $\pt$ due to large statistical fluctuations. It is 
important to note agreement of the ISR measurements of the anti-particle 
to particle ratios as a function of $\pt$ at $\sqs = 62.4$ GeV 
(Fig.~\ref{fig:ratio_negpos_run5pp200_run6pp62}) with the present 
measurements except for the $\pbar/p$ ratio where there is a large 
discrepancy. The $\pbar/p$ ratio integrated over all $\pt$ decreases from 
0.8 at 200 GeV to 0.5 at 62.4 GeV (see further discussion in 
Section~\ref{sec:meanpt_dndy}).  At low $\pt$, the large systematic 
uncertainties of the $\pbar/p$ ratio are due to the uncertainties of the 
weak decay feed down corrections.

Figure~\ref{fig:ratios} presents the ratios of $K^{+}/\pi^{+}$, 
$K^{-}/\pi^{-}$, $p/\pi^{+}$, $p/\pi^{0}$, $\pbar/\pi^{-}$, and 
$\pbar/\pi^{0}$ as a function of $\pt$. Both the $K^{+}/\pi^{+}$ and the 
$K^{-}/\pi^{-}$ ratios increase with increasing $\pt$ up to the $\pt = 
2$ GeV/$c$ limit of the measurement. Both the $p/\pi^{0}$ and the 
$\pbar/\pi^{0}$ ratios seem to increase with $\pt$ for $\pt >$ 2 
GeV/$c$, although the $\pbar/\pi^{0}$ ratio is relatively flat at $\sqs$ 
= 200 GeV in the same transverse momentum region. Clearly, better 
statistics are required to reach a firm conclusion.  As a function of 
$\sqs$ the $K^{+}/\pi^{+}$, $\pbar/\pi^{-}$ and $\pbar/\pi^0$ ratios do 
not change significantly, while the $K^{-}/\pi^{-}$ ratio increases and 
the $p/\pi^{+}$ and $p/\pi^{0}$ ratios decrease significantly for $\pt > 
1$ GeV/$c$ as the collision energy is increased from $\sqs = 62.4$ GeV 
to 200 GeV.
 
%\clearpage

%\clearpage

%-------------
% <pT> and dN/dy
%-------------
\subsection{$\meanpt$ and $dN/dy$}
\label{sec:meanpt_dndy}
Mean transverse momentum $\meanpt$ and particle yield per unit rapidity 
$dN/dy$ are determined by integrating the measured $\pt$ spectrum for 
each particle species. For the unmeasured $\pt$ region, we fit the 
measured $p_T$ spectrum with a Tsallis function~\cite{tsallis_fun} given 
below, as in a related publication~\cite{phenix:ppg099}, and also with 
an $\mt$ exponential function, then extrapolate the obtained function to 
the unmeasured $\pt$ region. The $\pt$ ranges for fitting are: 0.4--3.0 
GeV/$c$ for pions, 0.4--2.0 GeV/$c$ for kaons, 0.5--4.0 GeV/$c$ for 
protons and antiprotons.

The final yield $dN/dy$ is calculated by taking the sum of the yield 
from the data, and the yield from the functional form in the unmeasured 
$\pt$ region. The total inelastic cross sections are assumed to be 42.0 
mb and 35.6 mb for 200 GeV and 62.4 GeV respectively. For $\meanpt$, we 
integrate the measured $\pt$ spectrum with $\pt$ weighting, and then 
divide it by the obtained $dN/dy$. The final values are obtained by 
averaging the results of the two fits. The systematic uncertainties are 
evaluated as half of the difference between these fitting values.

%%% Functions %%%
\begin{itemize}
\item {Tsallis distribution is given by Eq.~\ref{eq:levy_fun} below. In this fitting
form, the free parameters are $dN/dy$, $q$ and $C$, while the mass $m$
is fixed to the hadron mass. The fitting results are given in Table~\ref{tab:tsallis_fit}.
  \begin{eqnarray}
  && \frac{1}{2\pi \pt}\frac{d^{2}N}{dyd\pt} = \nonumber \\
  && \frac{dN}{dy}\frac{(q-1)(q-2)}{2\pi q C \bigl[qC + m(q-2) \bigr]} \Bigl[1 + \frac{\mt-m}{qC} \Bigr]^{-q} ~~
  \label{eq:levy_fun}
  \end{eqnarray}
}
\item {Exponential distribution in $\mt$ is given by Eq.~\ref{eq:mt_exp_fun} below.
The free fit parameters are the normalization constant $A$ and the
inverse slope parameter $T_{\rm inv}$.
  \begin{equation}
  \frac{1}{2\pi \pt}\frac{d^{2}N}{dyd\pt} = 
A \exp{ \Bigl( -\frac{\mt}{T_{\rm inv}} \Bigr) }
  \label{eq:mt_exp_fun}
  \end{equation}

  \begin{equation}
  \frac{dN}{dy} = 2\pi A (m T_{\rm inv} + T_{\rm inv}^{2})
  \label{eq:dndy_from_mt_exp_fun}
  \end{equation}

}
\end{itemize}

% =================
% Tsallis fit table
% ==================

\begin{figure}[htb]
\includegraphics[width=1.05\linewidth]{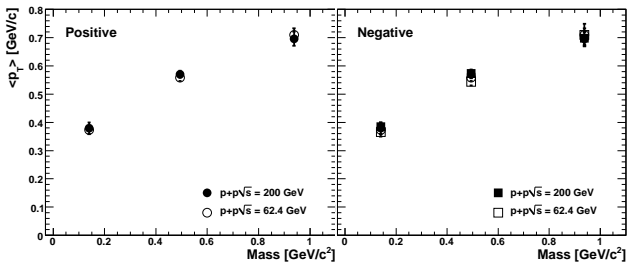}
\caption[]{Mean transverse momentum $\meanpt$ as a function of mass in $p+p$ collisions at $\sqs =$ 200 and 62.4 GeV.
The errors are statistical and systematic combined. The statistical errors are negligible.}
\label{fig:meanpt_vs_mass}
\end{figure}

The obtained $\meanpt$ values are summarized in 
Table~\ref{tab:meanpt_dndy}. They are plotted in 
Fig.~\ref{fig:meanpt_vs_mass}, which indicates a clear increase of 
$\meanpt$ with hadron mass. The values at 200 GeV are almost same as 
those for 62.4 GeV data. If the spectral shape is a pure exponential, 
$\meanpt$ should be equal to $2T_{\rm inv}$ analytically. By comparing 
Tables~\ref{tab:tinv_fit} and ~\ref{tab:meanpt_dndy}, the measured 
$\meanpt$ is almost $2T_{\rm inv}$ for pions. But for kaons and 
(anti)protons, the measured $\meanpt$ is systematically larger than $2 
~T_{\rm inv}$. This demonstrates that the spectral shape at low $\pt$ is not 
a pure exponential especially for kaons and (anti)protons.

\begin{figure*}[htb]
\includegraphics[width=1.0\linewidth]{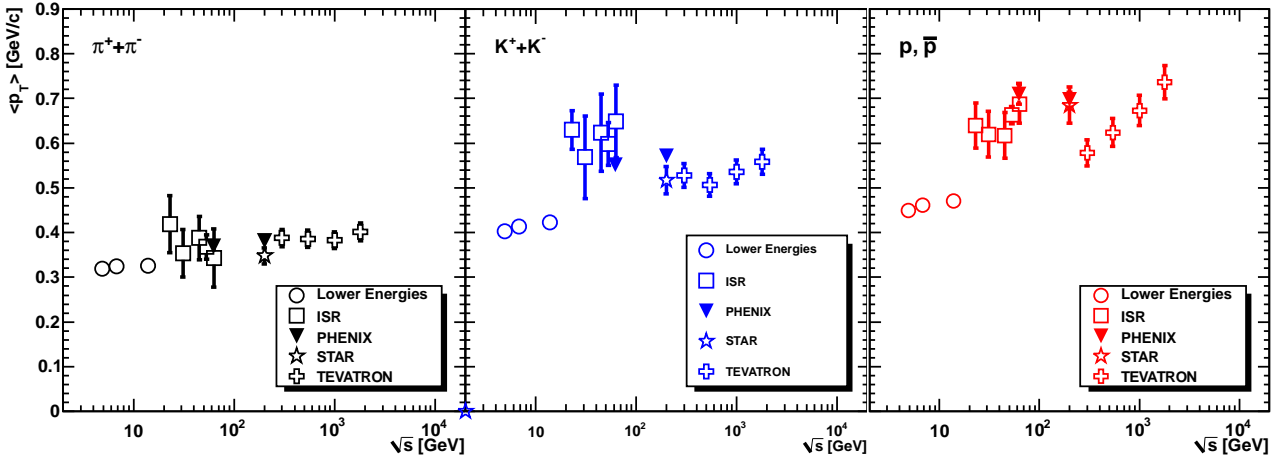}
\caption[]{(color online) 
Mean transverse momentum $\meanpt$ for $\pi^{+}+\pi^{-}$, $K^{+}+K^{-}$, $p$ and $\pbar$
as a function of $\sqs$ in $p+p$ and $p+\pbar$ 
collisions~\cite{Rossi:1974if, Alper_npb87, e735_1993, Alexopoulos:1990hn, Adams:2006nd}.
The errors are statistical and systematic combined. The statistical errors are negligible.}
\label{fig:meanpt_vs_roots}
\end{figure*}

The collision energy dependence of $\meanpt$ for each particle type is 
shown in Fig.~\ref{fig:meanpt_vs_roots}. Data shown here are: lower 
energy data~\cite{Rossi:1974if}, ISR data~\cite{Alper_npb87}, Tevatron 
data~\cite{e735_1993, Alexopoulos:1990hn}, and RHIC data from 
STAR~\cite{Adams:2006nd} and PHENIX (present study).  The $\meanpt$ 
values for all the other experiments have been determined by fitting 
the $\pt$ spectra.  For pions, the 
$\meanpt$ shows a linear increase in $\ln(\sqs)$.  For kaons and 
(anti)protons the increase is much faster than those for pions. 
However, systematic issues at both lower and higher center-of-mass 
energies remain to be resolved.

\begin{figure}[htb]
\includegraphics[width=1.0\linewidth]{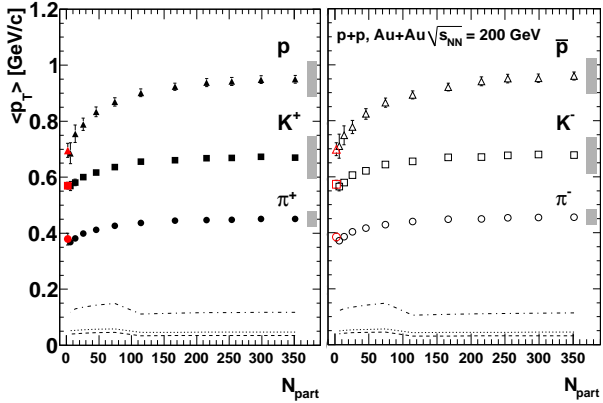}
\caption[]{(color online) 
Mean transverse momentum as a function of centrality ($N_{\rm part}$) for
pions, kaons, protons and anti-protons in in $p+p$ (present analysis, red color) 
and Au$+$Au~\cite{PPG026} (black color) 
collisions at $\sqrt{s_{NN}}$ = 200~GeV. The left (right) panel shows the $\meanpt$ 
for positive (negative)
particles. The error bars are statistical errors. The systematic errors from
cuts conditions are shown as shaded boxes on the right for each particle species.
The systematic errors from extrapolations, which are scaled by a factor of two for
clarity, are shown in the bottom for protons and anti-protons (dashed-dot lines),
kaons (dotted lines), and pions (dashed lines).}
\label{fig:meanpt_Npart}
\end{figure}

Figure~\ref{fig:meanpt_Npart} shows the dependence of $\meanpt$ on the 
centrality of the collisions (given by the number of participating 
nucleons, $N_{\rm part}$) for $\pi^{\pm}$, $K^{\pm}$, $p$ and $\pbar$ in 
Au$+$Au collisions at $\sqrt{s_{NN}}$ = 200 GeV~\cite{PPG026} as 
compared to minimum bias $p+p$ collisions at $\sqs$ = 200 GeV~(present 
analysis). The error bars in the figure represent the statistical 
errors. The systematic errors from cut conditions are shown as shaded 
boxes on the right for each particle species. The systematic errors from 
extrapolations, which are scaled by a factor of two for clarity, are shown 
in the bottom for each particle species. It is found that $\meanpt$ for 
all particle species increases from the most peripheral to midcentral 
collisions, and appears to saturate from the midcentral to central 
collisions. The $\meanpt$ in $p+p$ are consistent with the expectation 
from the $N_{\rm part}$ dependence in Au$+$Au, and are 
similar to the values in peripheral Au$+$Au.

%%%%%%%%%%%%%%%%%%%%%%%%%%%%%%%%%%%%%%%%%%%%%%%%%%%%%%%%%%%%% Fig_13

%%%%%%%%%%%%%%%%%%%%%%%%%%%%%%%%%%%%%%%%%%%%%%%%%%%%%%%%%%%%% Fig_14

%%%%%%%%%%%%%%%%%%%%%%%%%%%%%%%%%%%%%%%%%%%%%%%%%%%%%%%%%%%%% Fig_15

%%%%
%%%%%%%%%%%%%%%%%%%%%%%%%%%%%%%%%%%%%%%%%%%%%%%%%%%%  Table_VII
\begin{table}[htb]
\caption{Mean transverse momentum ($\meanpt$) and particle yield ($dN/dy$) for $\pi^{\pm}$, $K^{\pm}$, 
$p$ and $\pbar$ in $p+p$ collisions at $\sqs$~=~200 and 62.4~GeV.
The errors are statistical and systematic combined,but the statistical errors are negligible.}
\begin{ruledtabular} \begin{tabular}{cccc}
$\sqrt{s}$ &  hadron & $\meanpt$  & $dN/dy$  \\
(GeV) & & (GeV/$c$) &  \\ \hline 
200 & $\pi^{+}$ &  0.379  $\pm$  0.021   & 0.842  $\pm$  0.127 \\     
    & $\pi^{-}$ &  0.385  $\pm$  0.014   & 0.810  $\pm$  0.096 \\
    & $K^{+}$   &  0.570  $\pm$  0.012   & 0.099  $\pm$  0.010 \\
    & $K^{-}$   &  0.573  $\pm$  0.014   & 0.096  $\pm$  0.009 \\ 
    & $p$       &  0.696  $\pm$  0.025   & 0.043  $\pm$  0.003 \\
    & $\pbar$   &  0.698  $\pm$  0.023   & 0.035  $\pm$  0.002 \\
62.4 & $\pi^{+}$ & 0.373    $\pm$   0.013  & 0.722    $\pm$  0.066  \\
     & $\pi^{-}$ & 0.366    $\pm$   0.016  & 0.750    $\pm$  0.079  \\
     & $K^{+}$   & 0.558    $\pm$   0.012  & 0.072    $\pm$  0.004  \\
     & $K^{-}$   & 0.544    $\pm$   0.013  & 0.064    $\pm$  0.004  \\
     & $p$       & 0.710    $\pm$   0.023  & 0.034    $\pm$  0.002  \\
     & $\pbar$   & 0.709    $\pm$   0.040  & 0.018    $\pm$  0.001   \\
\end{tabular} \end{ruledtabular}
\label{tab:meanpt_dndy}
\end{table}
%===

%\clearpage

\begin{figure}[htb]
\includegraphics[width=1.05\linewidth]{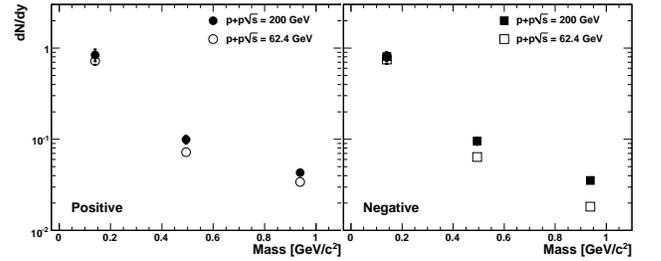}
\caption[]{Particle yield $dN/dy$ as a function of mass in $p+p$ collisions at $\sqs =$ 200 and 62.4 GeV.
The errors are statistical and systematic combined. The statistical errors are negligible.}
\label{fig:dndy_vs_mass}
\end{figure}

The $dN/dy$ values at midrapidity are summarized in 
Table~\ref{tab:meanpt_dndy}. They are plotted in 
Fig.~\ref{fig:dndy_vs_mass} as a function of hadron mass for both 200 
and 62.4 GeV collision energies. There are differences in the yield 
between 200 and 62.4 GeV especially for kaons and antiprotons continuing 
the trend observed at lower $\sqs$~\cite{Alper_npb87}. It is interesting 
to note that even in the situation that $dN/dy$ is different between 
$\sqs=$ 200 and 62.4 GeV, $\meanpt$ is quite similar for both energies.

\begin{figure*}[htb]
\includegraphics[width=0.9\linewidth]{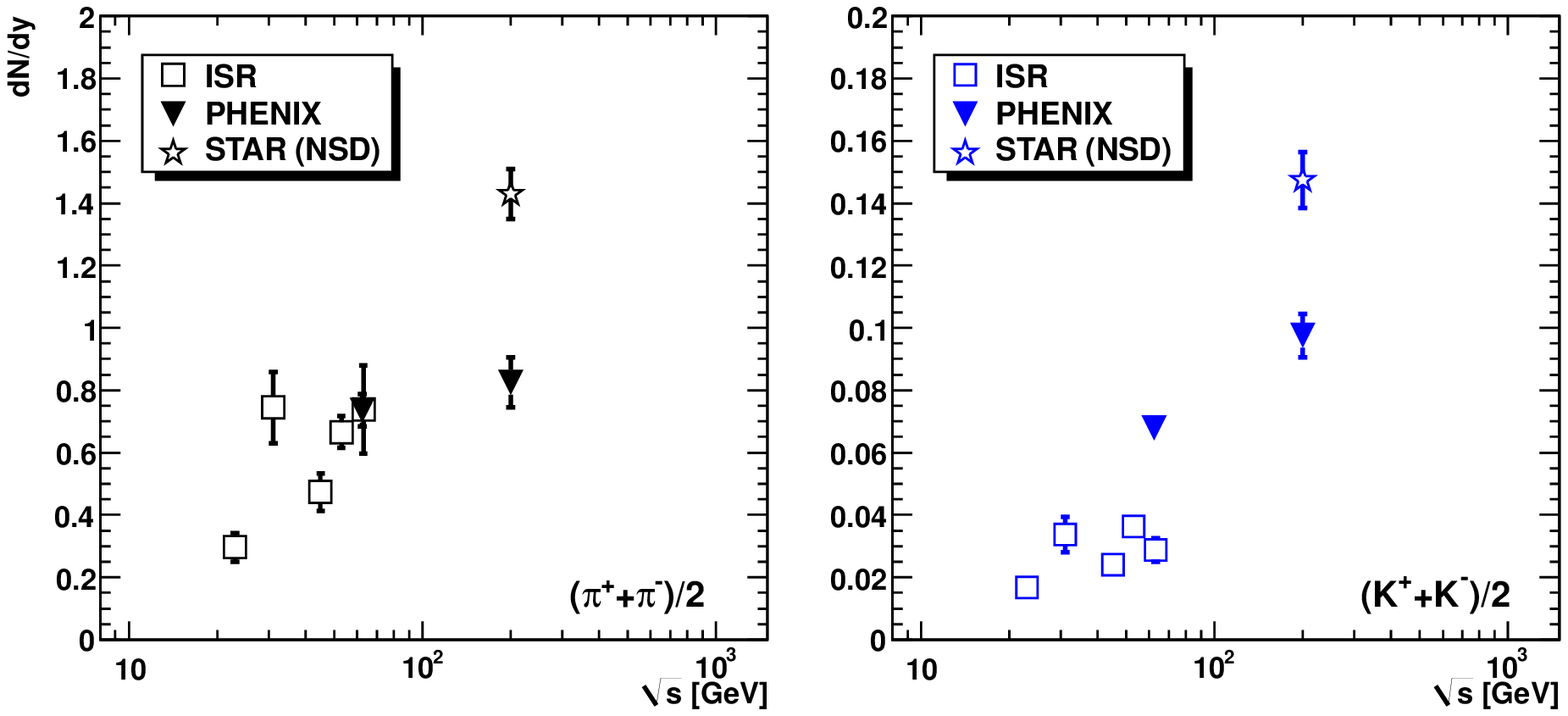}
\includegraphics[width=0.9\linewidth]{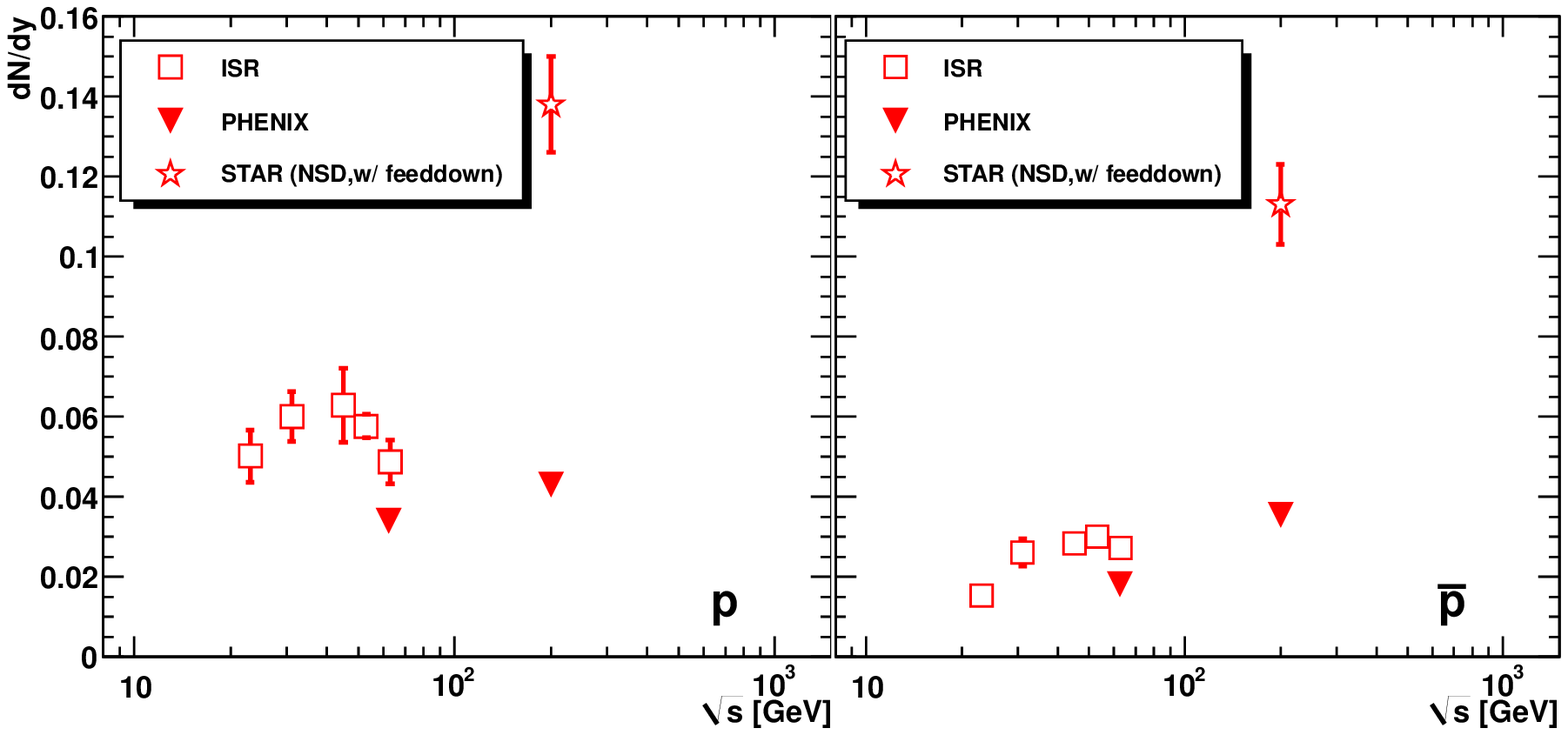}
\caption[]{(color online) 
(upper) Particle yield $dN/dy$ at midrapidity for ($\pi^{+}+\pi^{-}$)/2, ($K^{+}+K^{-}$)/2 
as a function of $\sqs$ in $p+p$ collisions~\cite{Alper_npb87,Abelev:2008ez}.
The errors are statistical and systematic combined, but the statistical errors are negligible.
The $dN/dy$ from STAR is determined for NSD $p+p$ events.
(lower) Similar plots for $p$ and $\pbar$ with feed-down correction applied to our data.
The $dN/dy$ from STAR is determined for NSD $p+p$ events, and is not corrected for weak-decay feed down.}
\label{fig:dndy_vs_roots}
\end{figure*}

Figure~\ref{fig:dndy_vs_roots} shows
$dN/dy$ as a function of collision energy for each particle 
species.  Our results on $dN/dy$ are consistent with the those at ISR 
energies~\cite{Alper_npb87}. It should be noted that STAR quotes the 
nonsingle diffractive (NSD) multiplicity while our measurement quotes 
the inelastic multiplicity, normalizing the integrated measured 
inclusive cross section by the total inelastic cross section~\cite{UA5}. 
At $\sqrt{s}$ = 200 GeV, the inelastic cross section ($\sigma^{INEL}$) 
is 42 mb~\cite{Miller:2007ri}, and the single diffractive (SD) cross 
section is almost equal to the double diffractive (DD) cross section, 
$\sigma_{NN}^{SD} \approx \sigma_{NN}^{DD} \approx$ 4 
mb~\cite{Kopeliovich:2003tz}. As the single diffractive refers only to 
the projectile proton in a $p+p$ fixed target measurement, one has to 
subtract the SD cross section for each proton from the inelastic cross 
section to determine the NSD cross section~\cite{B_Kopeliovich_private}. 
The resulting NSD cross section ($\sigma^{NSD}$) should be 
$42-2\times 4$mb = 34 mb. The ratio of the NSD multiplicity to the inelastic 
multiplicity is: $\sigma^{INEL}/ \sigma^{NSD}$ = 42/34 = 1.24, i.e. the 
NSD multiplicity is 24\% higher than the inelastic multiplicity, and 
this effect can be actually seen in the experimental 
data~\cite{Aamodt:2010ft}.

We would like to point out also that the NSD charged particle 
multiplicity at $\sqs = 200$ GeV by STAR is $\approx$ 20\% larger than 
other NSD results~\cite{Aamodt:2010ft}. By taking this fact and the 
difference between NSD and inelastic into account, one can naturally 
understand $\approx$ 50\% difference on yields between STAR and the 
present analysis, for pions and kaons, as shown in 
Fig.~\ref{fig:dndy_vs_roots}.
For protons and antiprotons the difference between STAR and the present 
analysis is bigger than those in pions and kaons. In addition to the 
effects we have mentioned above, the weak decay feed-down correction can 
contribute to it, since we remove $p$ and $\pbar$ from the weak decay 
(see Section~\ref{sec:analysis:feed}), while STAR does not.

\begin{figure}[htb]
\includegraphics[width=1.1\linewidth]{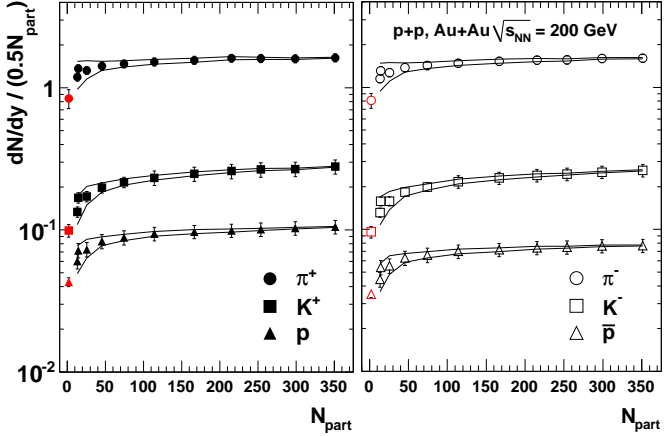}
\caption[]{(color online) 
Particle yield per unit rapidity ($dN/dy$) per participant
pair (0.5 $N_{\rm part}$) as a function of $N_{\rm part}$ for pions, kaons, protons and
antiprotons in $p+p$ (present analysis, red color) and Au$+$Au collisions~\cite{PPG026} 
at $\sqrt{s_{NN}}$~=~200~GeV. The left (right) panel shows the $dN/dy$ 
for positive 
(negative) particles. The error bars represent the quadratic sum of statistical 
errors and systematic errors from cut conditions. The lines represent the effect 
of the systematic error on $N_{\rm part}$ which affects all curves in the same way.}
\label{fig:dndy_Npart}
\end{figure}

Figure~\ref{fig:dndy_Npart} shows the collision centrality dependence of 
$dN/dy$ per participant pair (0.5 $N_{\rm part}$) in $p+p$ (present 
analysis) and Au$+$Au~\cite{PPG026} collisions at $\sqrt{s_{NN}}$ = 200 
GeV. The error bars on each point represent the quadratic sum of the 
statistical errors and systematic errors from cut conditions. The 
statistical errors are negligible. The lines represent the effect of the 
systematic error on $N_{\rm part}$ which affects all curves in the same 
way. The data indicate that $dN/dy$ per participant pair increases for 
all particle species with $N_{\rm part}$ up to $\approx$ 100, and 
saturates from the midcentral to the most central collisions. 
As seen in Fig.~\ref{fig:meanpt_Npart} for $\meanpt$, the $dN/dy$ values 
in $p+p$ are consistent with the expectation from the $N_{\rm part}$ 
dependence in Au$+$Au.

%%%%%%%%%%%%%%%%%%%%%%%%%%%%%%%%%%%%%%%%%%%%%%%%%%%%%%%%%%%%% Fig_16

%%%%%%%%%%%%%%%%%%%%%%%%%%%%%%%%%%%%%%%%%%%%%%%%%%%%%%%%%%%%% Fig_17

%%%%%%%%%%%%%%%%%%%%%%%%%%%%%%%%%%%%%%%%%%%%%%%%%%%%%%%%%%%%% Fig_18

%\clearpage

%%%%%%%%%%%%%%%%%%%%%%%%%%%%%%%%%%%%%%%%%%%%%%%%%%%%%%%%%%%%%%%%%%%% Fig_19
\begin{figure}[htb]
\includegraphics[width=1.1\linewidth]{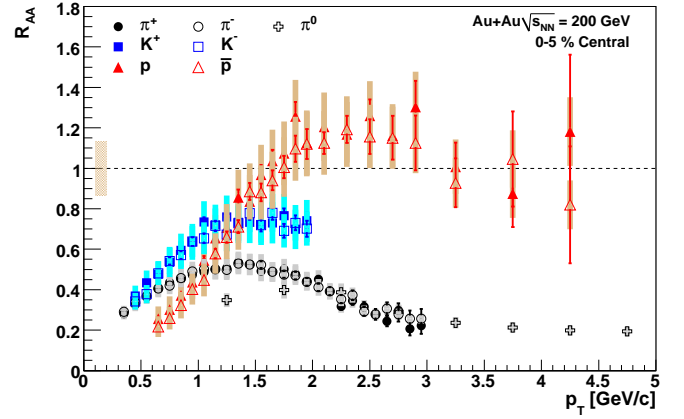}
\caption{(color online) 
$R_{\rm AA}$ of $\pi^{\pm}$, $\pi^{0}$, $K^{\pm}$, $p$, and $\pbar$ in Au$+$Au collisions at 
$\sqrt{s_{NN}}~=$~200 GeV at 0--5~\% collision centrality. The data for identified charged hadrons in
 Au$+$Au at $\sqrt{s_{NN}}$ = 200 GeV 
is taken from ~\cite{PPG026} and the data for $p+p$ is taken form the 
present analysis at $\sqs$ = 200 GeV.
The neutral pions (PHENIX) is taken from ~\cite{PPG080}. 
The statistical uncertainties are shown as bars, and the systematic uncertainties are shown as shaded boxes
on each data point. The overall normalization uncertainty on $R_{\rm AA}$ (13.8\%) is shown in the shaded box 
around unity (at $\pt = 0.1$ GeV/$c$)), which is the quadratic sum of (1)
uncertainty of $p+p$ inelastic cross section (9.7\%) and (2) uncertainty $\langle N_{\rm coll} \rangle$ (9.9\%).
}
\label{fig:raa_run2auau200_run5pp200}
\end{figure}

%%%%%%%%%%%%%%%%%%%%%%%%%%%%%%%%%%%%%%%%%%%%%%%%%%%%%%%%%%%%%%%%%%%% Fig_20
\begin{figure}[t]
\includegraphics[width=1.05\linewidth]{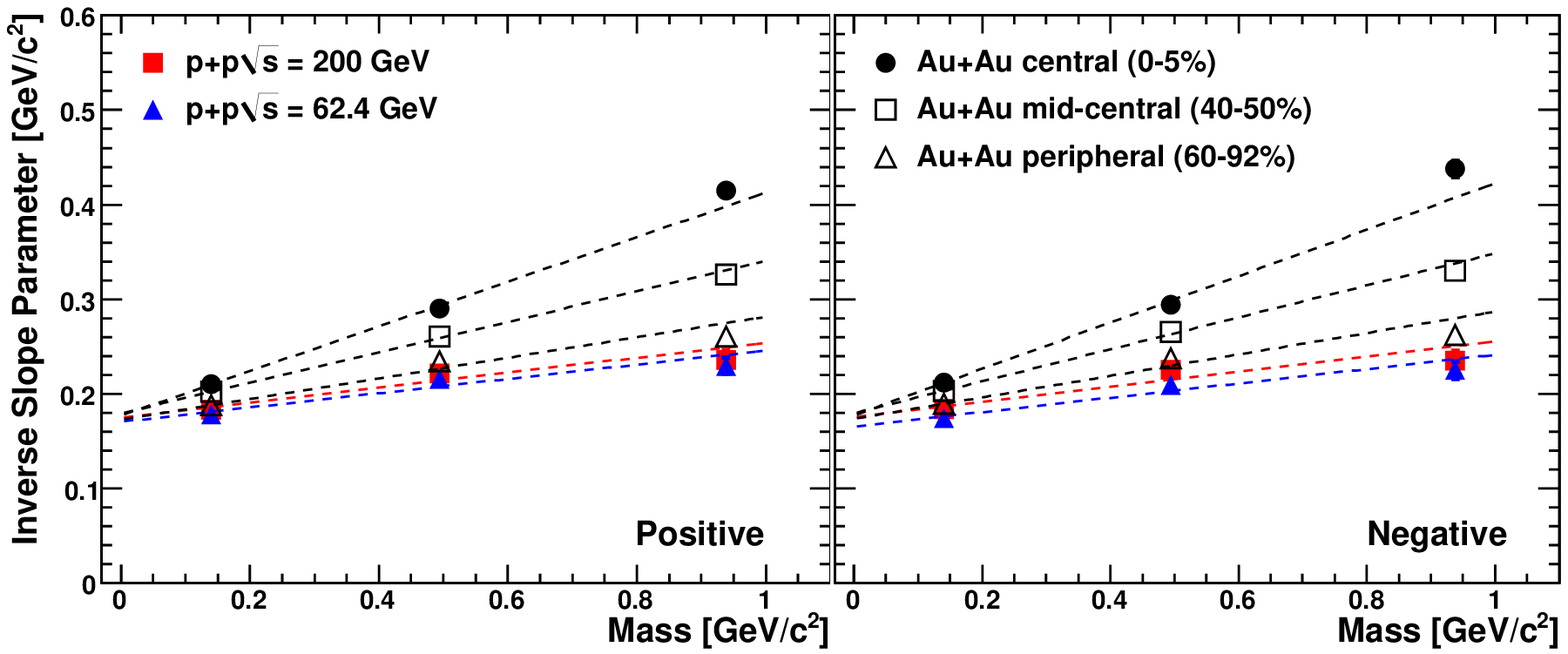}
\caption{(color online) 
Mass dependence of inverse slope parameter $T_{\rm inv}$ in 
$m_T -m$ spectra for (left) positive and (right) negative hadrons in 
$p+p$ collisions at $\sqs$~=~200 and 62.4~GeV, as well as for 
peripheral, midcentral and central in Au$+$Au collisions at 
$\sqrt{s_{NN}}$~=~200~GeV~\cite{PPG026}. The errors are statistical and 
systematic combined, smaller than symbols. The statistical errors are 
negligible.  The fit ranges are 0.2--1.0~GeV/$c^{2}$ for pions and 
0.1--1.0~GeV/$c^{2}$ for kaons, protons, and antiprotons in 
$m_T-m$~\cite{PPG026}. The dotted lines represent a linear fit of the results 
for each data set, as a function of mass using Eq.~(\ref{eq:flow}).}
\label{fig:slope_mass_pp_vs_AuAu}
\end{figure}

%%\clearpage

%\clearpage
%---------------
% R_AA
%---------------
\subsection{Nuclear Modification Factor $R_{\rm AA}$}
\label{sec:RAA}

In order to quantify the modification effect in nucleus-nucleus (A$+$A) 
collisions with respect to nucleon-nucleon collisions, the nuclear 
modification factor $R_{\rm AA}$ is used. $R_{\rm AA}$ is the ratio 
between the yield in A$+$A scaled by the average number of binary 
nucleon-nucleon collisions ($\langle N_{\rm coll} \rangle$) and the 
yield in $p+p$, as defined by the following equation:

\begin{equation}
  R_{\rm AA}(p_T)\,=\,
  \frac{(1/N^{\rm evt}_{\rm AA})\,{\rm d}^2N_{\rm AA}/{\rm d}p_T dy}
  {\langle T_{\rm AA} \rangle
    \,\times\, {\rm d}^2\sigma_{\rm pp}/{\rm d}p_T dy}
\label{eq:R_AA}
\end{equation}
where $\langle T_{\rm AA} \rangle$ is the nuclear thickness function, 
defined as follows:  $\langle T_{\rm AA} \rangle = \langle N_{\rm 
coll}\rangle/\sigma_{\rm pp}^{\rm inel}$. For the total A$+$A interaction 
cross section $\sigma^{\rm int}_{\rm AA}$ (minimum bias A$+$A collisions), 
$\langle T_{\rm AA} \rangle =A^2/\sigma^{\rm int}_{\rm AA}$.

In general, $R_{\rm AA}$ is expressed as a function of $\pt$ and 
collision centrality for A$+$A collisions. Due to the dominance of hard 
scatterings of partons at high $\pt$, $R_{\rm AA}$ is expected to be 
around unity above $\pt \approx$ 2 GeV/$c$, if there is no yield 
modification by the nucleus in A$+$A.  If there is a suppression 
(enhancement), $R_{\rm AA}$ is less than (greater than) unity. 
For the total A$+$A interaction cross section at a given $p_T$ integrated 
over centrality (minimum bias A$+$A collisions) $\sigma_{\rm AA}(p_T) = 
A^2 \sigma_{\rm pp}(p_T)$ and $R_{\rm AA}(p_T) \equiv 1.0$.

Figure~\ref{fig:raa_run2auau200_run5pp200} shows the $R_{\rm AA}$ of 
$\pi^{\pm}$, $\pi^{0}$, $K^{\pm}$, $p$, and $\pbar$ in Au$+$Au 
collisions at $\sqrt{s_{NN}}~=$~200 GeV at 0--5~\% collision centrality. 
The data for identified charged hadrons in Au$+$Au at $\sqrt{s_{NN}}$ = 
200 GeV are taken from ~\cite{PPG026} measured by the PHENIX experiment, 
and those for $p+p$ are taken from the present analysis at $\sqs$ = 200 
GeV. The $R_{\rm AA}$ for neutral pions is taken from ~\cite{PPG080}. 
The overall normalization uncertainty on $R_{\rm AA}$ (13.8\%) is shown 
as a shaded box around unity (at $\pt = 0.1$ GeV/$c$)), which is the 
quadratic sum of (1) the uncertainty of $p+p$ inelastic cross section 
(9.7\%) and (2) the uncertanty $\langle N_{\rm coll} \rangle$ (9.9\%).

For pions $R_{\rm AA}$ is largely suppressed by a factor of $\approx$ 5, 
compared to $p+p$. This suppression effect is understood to be due to jet 
quenching or energy loss of partons in hot and dense medium created in 
Au$+$Au central collisions at RHIC energies~\cite{PPG014,Adcox:2004mh}. 
For kaons there is a similar trend as pions over a more limited $\pt$ 
range.  For protons and antiprotons there is an enhancement in 
$\pt =$ 2--4 GeV/$c$.  As reported in~\cite{PPG026,PPG015,PPG030}, 
possible explanations of the observed enhancements include the quark 
recombination model~\cite{Fries:2003vb,Hwa:2002tu,Greco:2003xt} and/or 
strong partonic and hadronic radial flow~\cite{Hirano:2003pw}.

%%%%%%%%%%%%%%%%%
% 5. DISCUSSION %
%%%%%%%%%%%%%%%%%
\section{DISCUSSION}
\label{sec:discussion}

In this section, we discuss (1) soft particle production at low $\pt$, 
including the possibility of radial flow in $p+p$ collisions, and (2) the 
transition from soft to hard process, and hadron fragmentation at 
high $\pt$, where we show the $x_T$ scaling of measured spectra, and make 
a comparison with NLO and NLL pQCD calculations.

%%%%%%%%%%%%%%%%%%%%%%%%%%%%%%%%%%%%%%%%%%%%%%%%%%%%  Table_VIII
\begin{table}
\caption{Inverse slope parameter $T_{\rm inv}$ for $\pi^{\pm}$, $K^{\pm}$, 
$p$ and $\pbar$ for $p+p$ collisions at $\sqs = 200$ and 62.4 GeV. The 
fit ranges are 0.2--1.0~GeV/$c^{2}$ for pions and 0.1--1.0~GeV/$c^{2}$ 
for kaons, protons, and antiprotons in $m_T-m$. These fit ranges are 
chosen in order to perform a comparison with $T_{\rm inv}$ in Au$+$Au 
collisions at RHIC~\cite{PPG026}. The errors are statistical and 
systematic combined, but the statistical errors are negligible.}
 \begin{ruledtabular} \begin{tabular}{cccc}
$\sqrt{s}$ & hadron  & $T_{\rm inv}$  & $\chi^{2}$/NDF \\ 
(GeV) & & (MeV/$c^2$) &  \\ \hline 
200  & $\pi^{+}$ &  183  $\pm$  4  & 12.9/6   \\
        & $\pi^{-}$ &  184  $\pm$  4  & 7.5/6   \\
        & $K^{+}$   &  221  $\pm$  5  & 10.0/8   \\
        & $K^{-}$   &  225  $\pm$  6  & 12.4/8   \\
        & $p$       &  236  $\pm$ 10  & 2.3/10   \\
        & $\pbar$   &  235  $\pm$ 10  & 1.2/10   \\
62.4 & $\pi^{+}$ & 178   $\pm$ 4     & 5.7/6   \\
         & $\pi^{-}$ & 174   $\pm$ 3     & 9.8/6   \\
         & $K^{+}$   & 216   $\pm$ 5     & 3.0/8   \\
         & $K^{-}$   & 209   $\pm$ 5     & 5.3/8   \\
         & $p$       & 230   $\pm$ 8     & 1.4/9   \\
         & $\pbar$   & 225   $\pm$ 9    & 2.0/9   \\
\end{tabular} \end{ruledtabular}
\label{tab:t_inv_auau_range}
\end{table}

%%%%%%%%%%%%%%%%%%%%%%%%%%%%%%%%%%%%%%%%%%%%%%%%%%%%  Table_IX
\begin{table}
\caption{The extracted fit parameters of the freeze-out temperature 
($T_0$) and the measure of the strength of the average radial 
transverse flow ($\ut$) using Eq.~(\ref{eq:flow}).  The fit results 
shown here are for positive and negative particles, and for the two 
different energies.}
\begin{ruledtabular} \begin{tabular}{ccccc}
$\sqrt{s}$ & $\pm$ & $T_{0}$  &  $\ut$  &  $\chi^{2}$/NDF \\ 
(GeV) & & (MeV/$c^2$) & &   \\ \hline 
200 & Positive & 175 $\pm$ 5 & 0.28 $\pm$ 0.02 & 4.1/1 \\
    & Negative & 176 $\pm$ 5 & 0.28 $\pm$ 0.02 & 6.0/1 \\
62.4 & Positive & 170 $\pm$ 5 & 0.27 $\pm$ 0.02 & 5.4/1 \\
     & Negative & 165 $\pm$ 4 & 0.28 $\pm$ 0.02 & 3.8/1 \\
\end{tabular} \end{ruledtabular}
\label{tab:mt_slope_pp}
\end{table}

%%%%%%%%%%%%%%%%%%%
% 5.1 Radial Flow %
%%%%%%%%%%%%%%%%%%%
\subsection{Radial flow}
\label{sec:flow}

In heavy ion collisions at RHIC energies, it is found that the inverse 
slope parameter ($T_{\rm inv}$) of $\mt-m$ spectra has a clear dependence on 
the hadron mass, i.e. heavier particles have larger inverse 
slope parameters~\cite{PPG009,PPG026}.   $T_{\rm inv}$ increases almost 
linearly as a function of particle mass; is largest when the nucleus-nucleus 
collision has a small impact parameter (central collisions); and is 
smallest for the collisions with a large impact parameter (peripheral 
collisions), as shown in Fig.~\ref{fig:slope_mass_pp_vs_AuAu}.

This experimental observation can be interpreted as the existence of a 
radial flow generated by violent nucleon-nucleon collisions in two 
colliding nuclei and developed both in the quark-gluon plasma (QGP) 
phase and in hadronic rescatterings~\cite{Hirano:2003pw}. The radial 
flow velocity increases the transverse momentum of particles 
proportional to their mass, thus $T_{\rm inv}$ increases linearly as a 
function of particle mass.  It is interesting to determine whether or 
not such an expansion is observed in high energy $p+p$ 
collisions~\cite{Alexopoulos:1990hn}.

Figure~\ref{fig:slope_mass_pp_vs_AuAu} shows the mass dependence of 
inverse slope parameter $T_{\rm inv}$ in $m_T -m$ spectra for positive 
(left) and negative (right) particles in $p+p$ collisions at 
$\sqs$~=~200 and 62.4~GeV (also shown in Fig.~\ref{fig:tinv_vs_mass}), 
as well as for peripheral, midcentral and central in Au$+$Au collisions 
at $\sqrt{s_{NN}}$~=~200~GeV~\cite{PPG026}. The fit ranges are 
$m_T-m$~=~0.2--1.0~GeV/$c^{2}$ for pions, and 
$m_T-m$~=~0.1--1.0~GeV/$c^{2}$ for kaons, 
protons, and antiprotons, which are chosen in order to perform a fair 
comparison with $T_{\rm inv}$ in Au$+$Au collisions at RHIC~\cite{PPG026}. 
The values of $T_{\rm inv}$ in $p+p$ for these fit ranges (see 
Table~\ref{tab:t_inv_auau_range}) are all lower by roughly one standard 
deviation from the values in Table~\ref{tab:tinv_fit} for the common fit 
range of $m_T-m$~=~0.3--1.0~GeV/$c^{2}$.

In general, the inverse slope parameters increase with increasing 
particle mass in both Au$+$Au and $p+p$ collisions at 200 GeV.  
However, this increase is only modest in $p+p$ collisions and slightly 
weaker than in 60--92\% central Au$+$Au collisions at 200 GeV.
 %
% (deleted the following liens: 2010.0701, by TC)
% For $p+p$ cases at both beam energies, the values of $T_{\rm inv}$ increase by 
% $\sim$1\% if the fit range is changed from $0.1 < \mt-m < 1.0$ GeV/$c^2$ to 
% $0.3 < \mt-m < 1.0$ GeV/$c^2$.
% Such a fit range dependence is also reported by the other experiments, e.g. $T_{\rm inv}$ of charged kaons 
% at $\sqs = 540$ GeV~\cite{compilation_kaon} changes from 158.9 $\pm$ 8.1 MeV to 215.7 $\pm$ 18.4 MeV, 
% if the fit range is changed from $0.28 < \pt < 1.22$ GeV/$c$ to $0.45 < \pt < 1.05$ GeV/$c$. 
%
Also note that there is a mean multiplicity dependence of the transverse 
momentum spectra in $p+p$ collisions~\cite{e735_1993}, that is not 
discussed in the present paper.

We use a radial flow picture~\cite{Schnedermann:1993ws,Csorgo:2001xm}  
with the fitting function:
\begin{equation}
T = T_{0} + m \ut^{2},
\label{eq:flow}
\end{equation}
where $T_0$ is a hadron freeze-out temperature and $\ut$ is a 
``measure'' of the strength of the (average radial) transverse flow. 
The relationship between the averaged transverse velocity 
($\langle \beta_t \rangle$) and $\ut$ is given by 
\begin{equation}
\langle \beta_t \rangle = \ut / \sqrt{1 + \ut^2}.
\end{equation}

The dotted lines in Fig.~\ref{fig:slope_mass_pp_vs_AuAu} represent the 
linear fit to the $p+p$ collisions at $\sqs = 200$ and 62.4 GeV, that 
are compared to those in Au$+$Au collisions at $\sqsn = 200$ GeV in three 
different collision centrality classes. The fit results in $p+p$ are 
also given in Table~\ref{tab:mt_slope_pp}. For Au$+$Au most central data 
(0--5\%), $\ut \approx$ 0.49 $\pm 0.07$, while in $p+p$, $\ut \approx$ 
0.28 at both 62.4 GeV and 200 GeV. While this radial flow model is 
consistent with the data in central and midcentral Au$+$Au, i.e. 
$\pi/K/p$ points are on a straight line, it does not give a good 
description of either peripheral Au$+$Au or $p+p$ collisions 
(poor $\chi^2$ in Table~\ref{tab:mt_slope_pp}).  This can be interpreted 
that radial flow is absent in $p+p$, where the $\pi/K/p$ points are 
obviously not on a straight line (Fig.~\ref{fig:tinv_vs_mass}), and 
that the radial flow develops only for a larger system.

% For mT-m = 0.3-1.0 GeV 
%
% \begin{ruledtabular} \begin{tabular}{cccc}
% \bf {Particle} &  $\sqs = 200$ GeV &  $\sqs = 62.4$ GeV \\ \hline
% $\pi^{+}$  & 191   $\pm$  5   & 185   $\pm$ 5    \\
% $\pi^{-}$  & 189   $\pm$  5   & 182   $\pm$ 4    \\
% $K^{+}$    & 233   $\pm$  7   & 224   $\pm$ 7    \\
% $K^{-}$    & 239   $\pm$  8   & 217   $\pm$ 7    \\
% $p$        & 245   $\pm$ 14   & 230   $\pm$ 10   \\
% $\pbar$    & 242   $\pm$ 14   & 237   $\pm$ 11   \\ \hline \hline
% \bf {Fit parameter}  &  $\sqs = 200$ GeV &  $\sqs = 62.4$ GeV \\ \hline
% $T_{0}^{(+)}$         &  182  $\pm$ 7    &   179 $\pm$ 6   \\
% $T_{0}^{(-)}$         &  181  $\pm$ 7    &   173 $\pm$ 5   \\
% $\ut^{(+)}$           &  0.29 $\pm$ 0.03 &   0.26  $\pm$ 0.03  \\
% $\ut^{(-)}$           &  0.29 $\pm$ 0.03 &   0.28  $\pm$ 0.02  \\
% \end{tabular} \end{ruledtabular}
% \label{tab:mt_slope_pp}
% \end{table}

%\clearpage

%-----------------
% 5.2 xT scaling 
%-----------------
\subsection{$\xt$ scaling}
\label{sec:xt}

From the measurements of $\pt$ spectra of hadrons in $p+p$ collisions, 
it is known that fragmentation of hard scattered partons is the dominant 
production mechanism of high $\pt$ hadrons. It has been predicted 
theoretically from general principles, that such a production mechanism 
leads to a data scaling behavior called ``$\xt$ 
scaling''~\cite{xt_scaling_1}, where the scaling variable is defined as 
$\xt = 2\pt/\sqrt{s}$. Such a data scaling behavior was seen first on 
preliminary ISR data at CERN as reported in~\cite{xt_scaling_1}.

In the kinematic range corresponding to the $\xt$ scaling limit, the 
invariant cross section near midrapidity can be written as:
 \begin{equation}
{E \frac{d^{3}\sigma}{dp^{3}}} = 
\frac{1}{\pt^{\neff}} F(\xt) = \frac{1}{\sqrt{s}^{\neff}} G(\xt),
\label{eq:xt_scaling}
 \end{equation}
where $F(\xt)$ and $G(\xt)$ are universal scaling functions. The 
parameter $\neff$ is characteristic for the type of interaction between 
constituent partons. For example for single photon or vector gluon 
exchange, $\neff$ = 4~\cite{BBK1971}. Because of higher order effects, 
the running of the strong coupling constant $\alpha_s = \alpha_s(Q^2)$, 
the evolution of the parton distribution functions and fragmentation 
functions, and nonvanishing transverse momentum $k_T$ of the 
initial-state, $\neff$ in general is not a constant but a function of 
$\xt$ and $\sqrt{s}$, i.e. $\neff = \neff(\xt, \sqrt{s})$. This $\neff$ 
corresponds to the logarithmic variation of yield ratios at the same 
$\xt$ for different $\sqs$~\cite{Arleo:2009ch}. Note that the $\xt$ 
scaling power $\neff$ is different from the exponent $n$ that 
characterizes the power-law behavior of the single particle invariant 
spectrum at high $\pt$ .

The value of $n_{\rm eff}$ depends on both the value of $\sqs$ and the 
range of $x_T$ and depending on the reaction tends to settle at an 
asymptotic value between 6 and 4.5 where hard-scattering dominates and 
higher twist effects are small. This fact can also be used to determine 
the transition between soft and hard particle production mechanisms.

Earlier measurements of $\neff(\xt,\sqrt{s})$ in $p+p$ collisions found 
values in the range of 5--8~\cite{Darriulat:1980nk, xt_scaling_1, 
xt_scaling_2, xt_scaling_3, xt_scaling_4, xt_scaling_5}. Here we present 
the PHENIX results for the $\xt$ scaling of pions, protons, and 
antiprotons and compare them with earlier data measured at various 
different values of $\sqs$. Due to the limited $\pt$ range of our kaon 
measurements, kaons are not included in these comparisons.

%%%%%%%%%%%%%%%%%%%%%%%%%%%%%%%%%%%%%%%%%%%%%%%%%%%%%%%%%%%%%%%% Fig_21
\begin{figure*}[htb]
\includegraphics[width=0.48\linewidth]{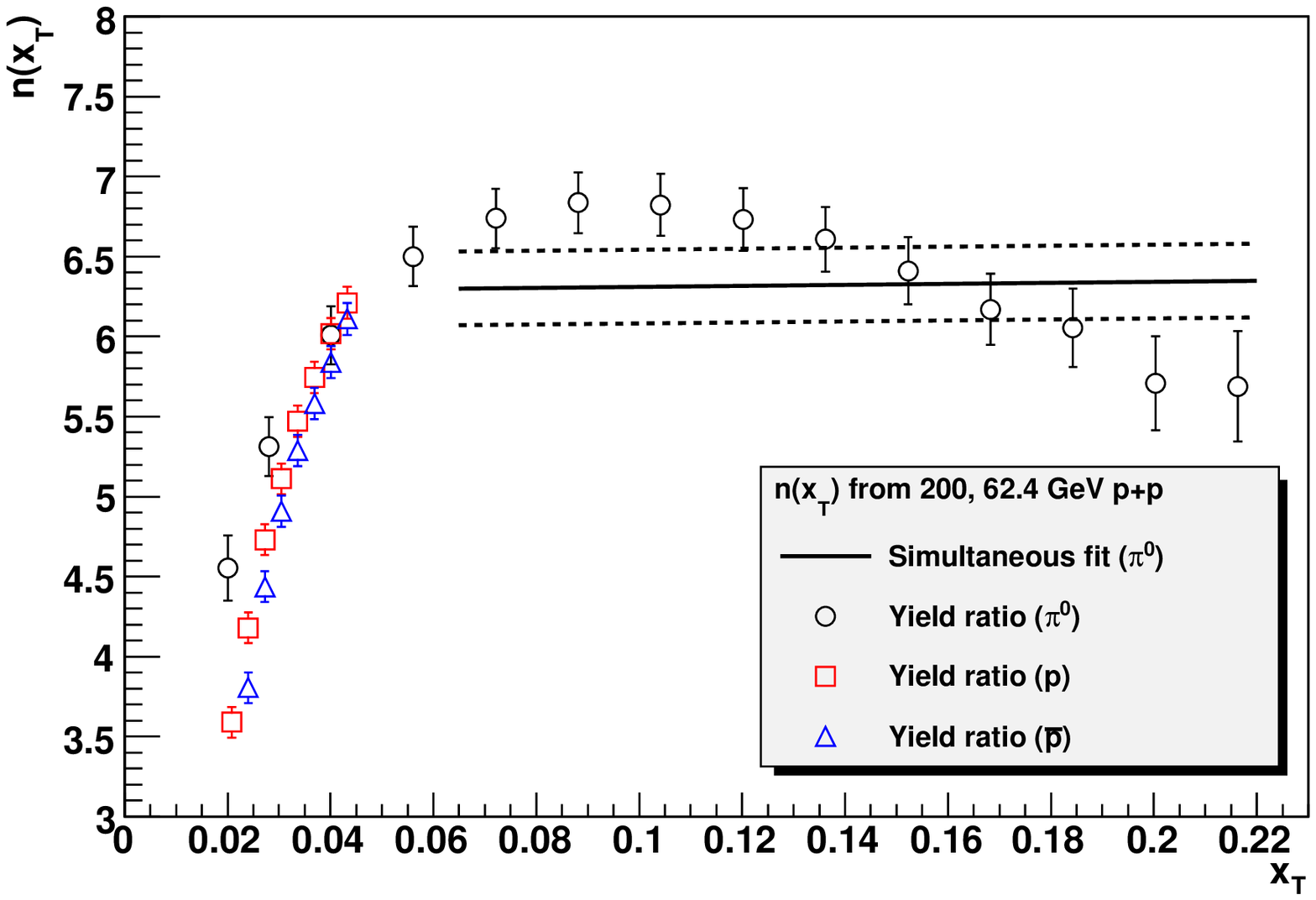}
\includegraphics[width=0.48\linewidth]{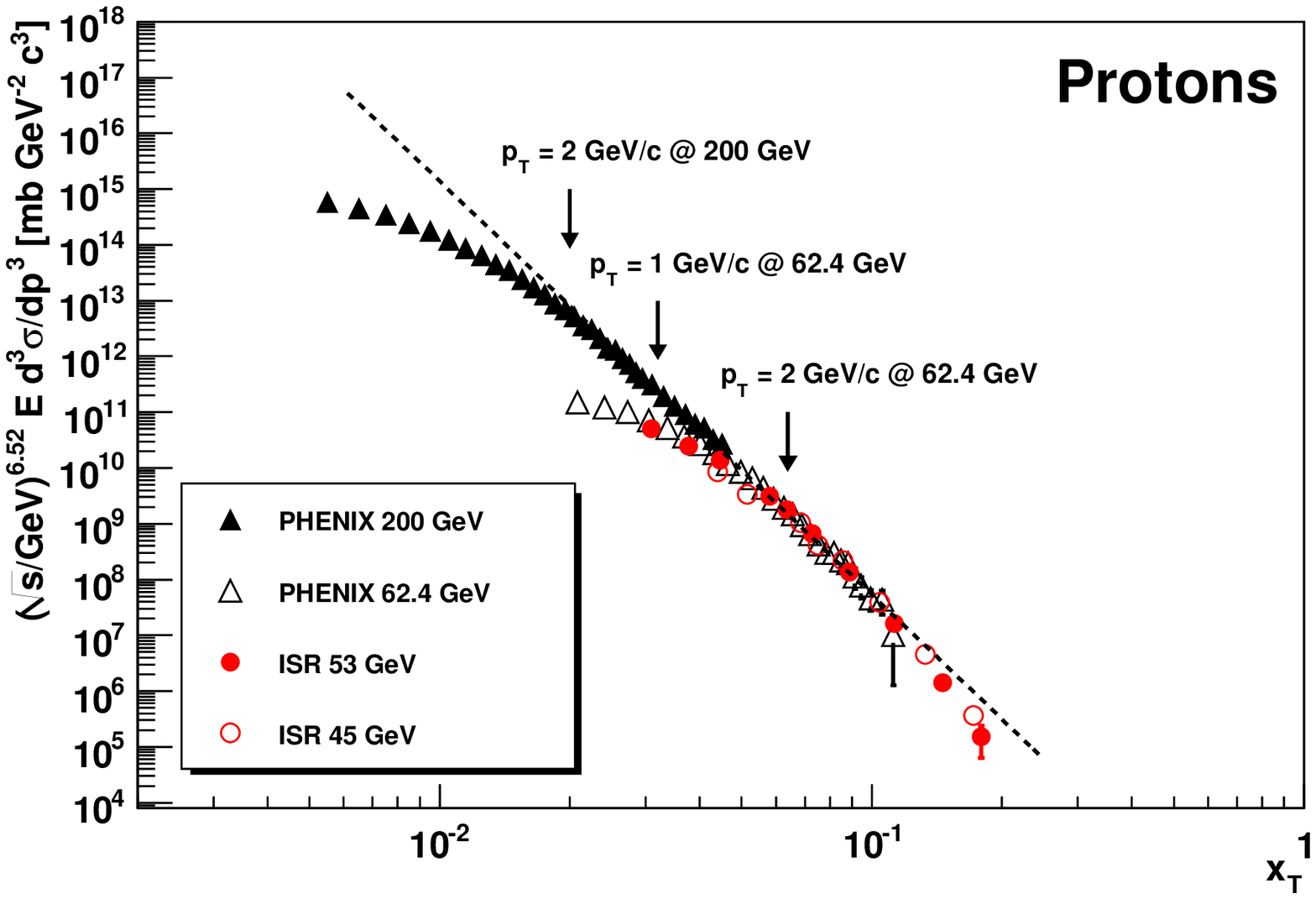}
\includegraphics[width=0.48\linewidth]{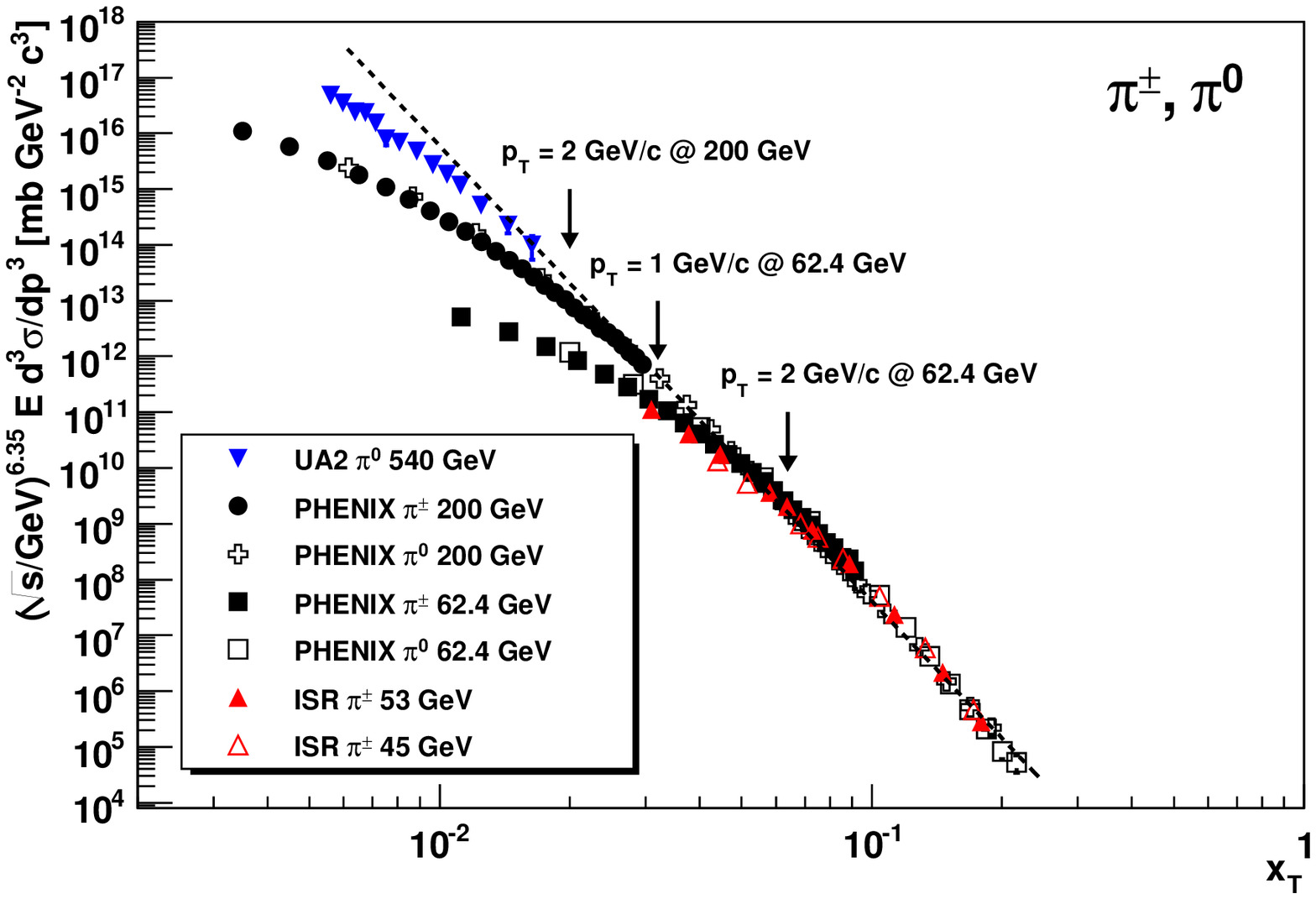}
\includegraphics[width=0.48\linewidth]{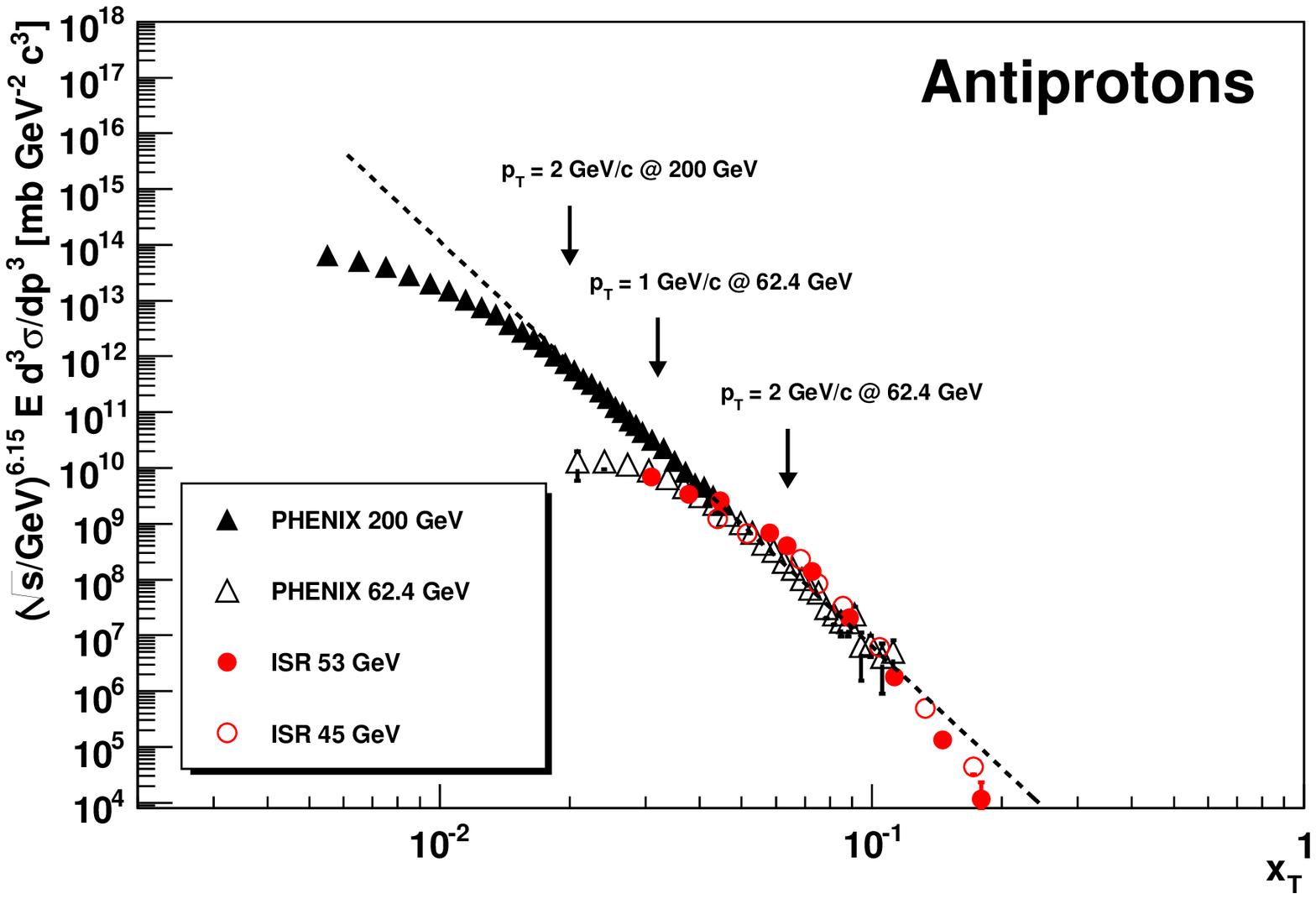}
\caption{(color online) 
(upper left) $\xt$-scaling power $\neff$ as determined from the 
ratios of yields as a function of $\xt$, for (open circle) neutral 
pions, (open square) protons, and (open triangle) antiprotons using 
$p+p$ data at $\sqs =$ 200 and 62.4 GeV energies. The error of each data 
point is from the systematic and statistical errors of $\pt$ spectra. 
The other plots show $\xt$ spectra for (lower left) pions ($\pi^{\pm}$, 
$\piz$), (upper right) protons, and (lower right) antiprotons in $p+p$ 
collisions at different $\sqs$ at midrapidity. Only statistical 
uncertainties are shown.  The dashed curves are the fitting results. }
 \label{fig:n_vs_xt_all}
\end{figure*}
%\label{fig:xt_pion_all}
%\label{fig:xt_proton_all}
%\label{fig:xt_antiproton_all}

%%%%%%%%%%%%%%%%%%%%%%%%%%%%%%%%%%%%%%%%%%%%%%%%%%%%%%%%%%%%%%%% Fig_22
\begin{figure}[htb]
\includegraphics[width=1.01\linewidth]{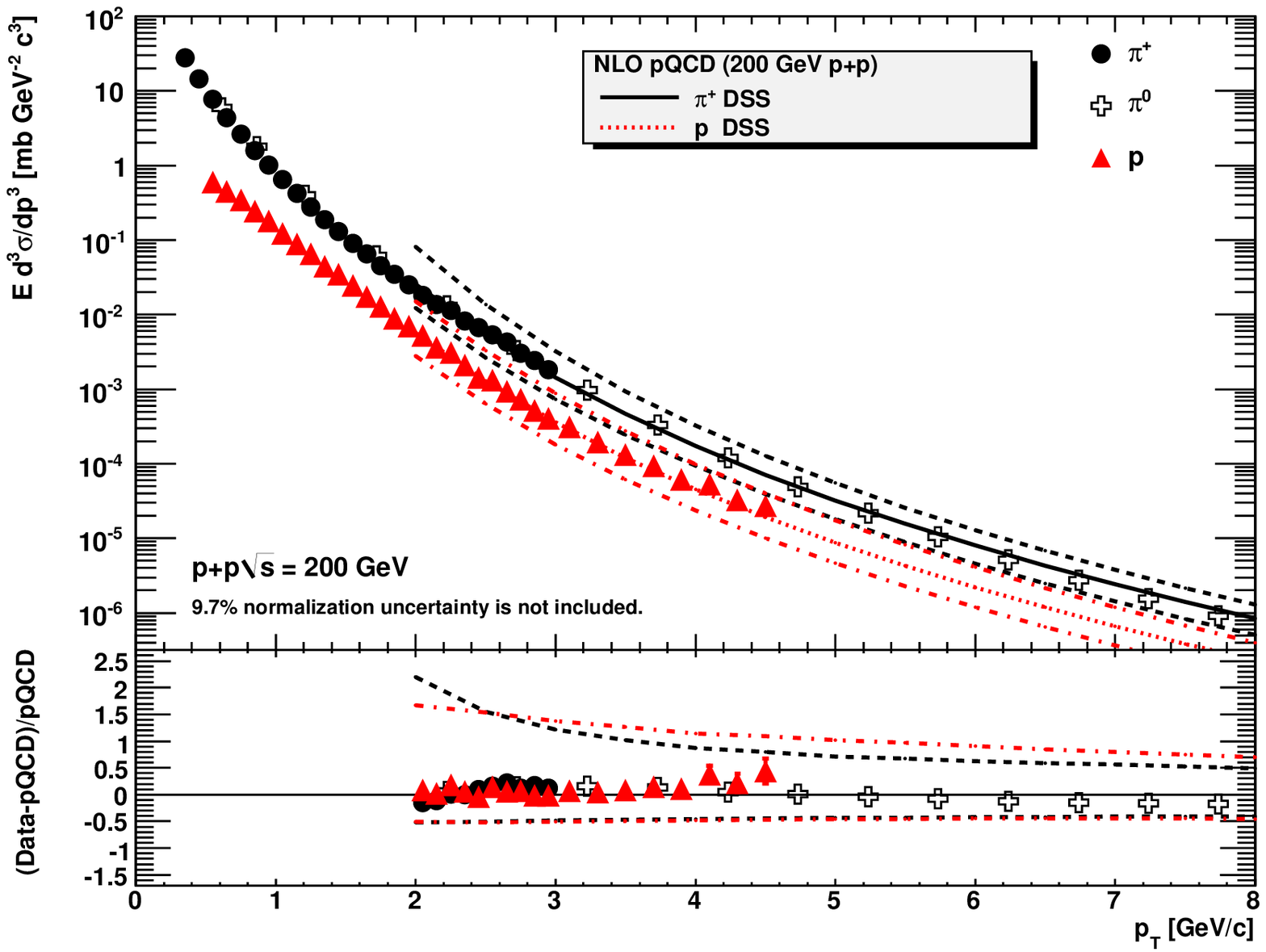}
\includegraphics[width=1.01\linewidth]{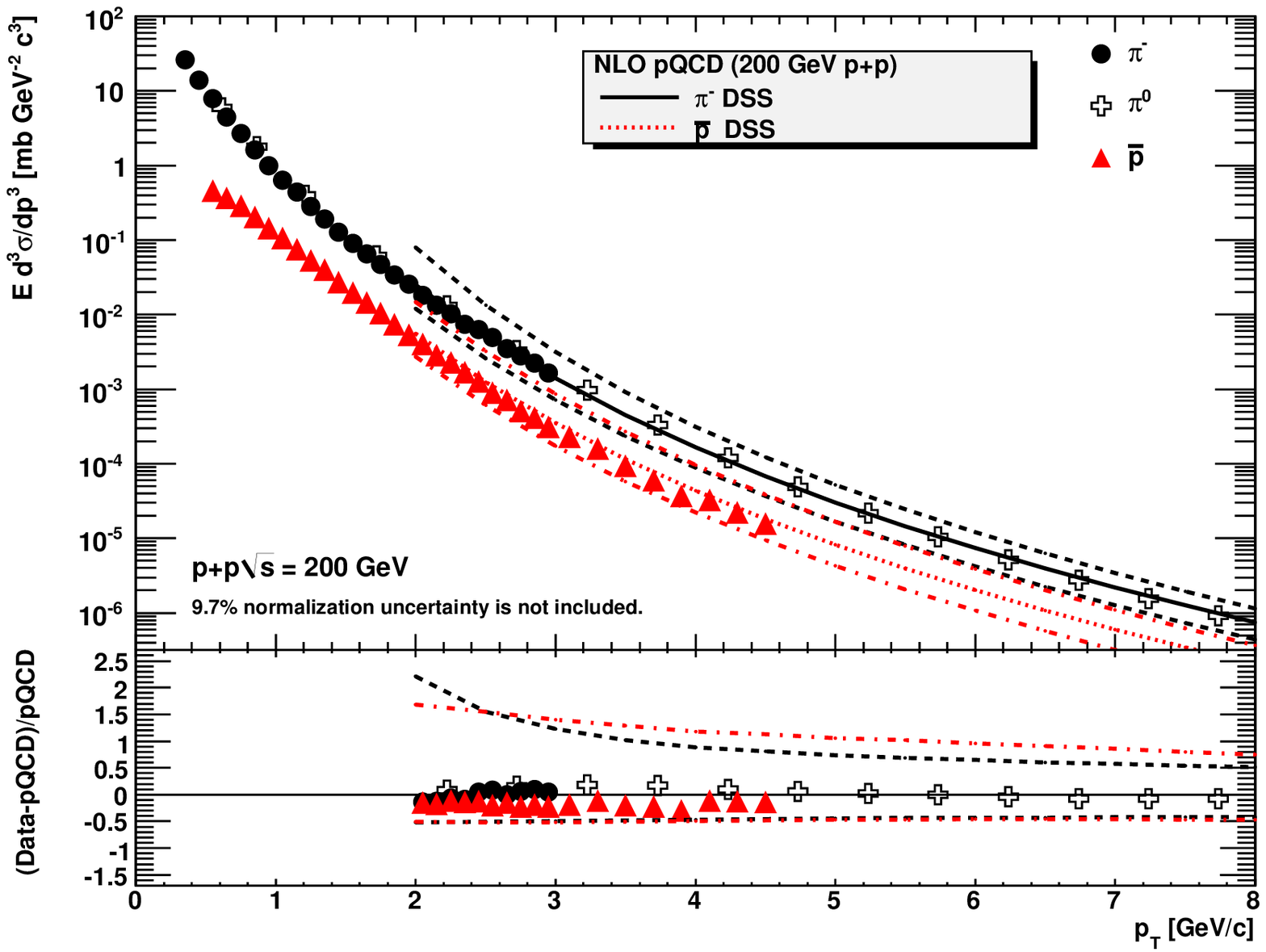}
\caption{(color online) 
Transverse momentum distributions for 
(upper) positive and (lower) negative particles at $\sqs =$ 200 GeV in $p+p$ collisions.
Only statistical uncertainties are shown.
The normalization uncertainty (9.7\%) is not included.
NLO pQCD calculations (FF: DSS) are also shown.
Solid lines are for $\mu$ = $\pt$, and dashed lines are for $\mu$ = $\pt$/2, 2$\pt$.
The lower panel in each plot shows the ratio of (data - pQCD 
result)/pQCD result for each particle species.
}
\label{fig:pt_DSS_run5pp200}
\end{figure}

%%%%%%%%%%%%%%%%%%%%%%%%%%%%%%%%%%%%%%%%%%%%%%%%%%%%%%%%%%%%%% Fig_23
\begin{figure*}[tbh]
\includegraphics[width=0.48\linewidth]{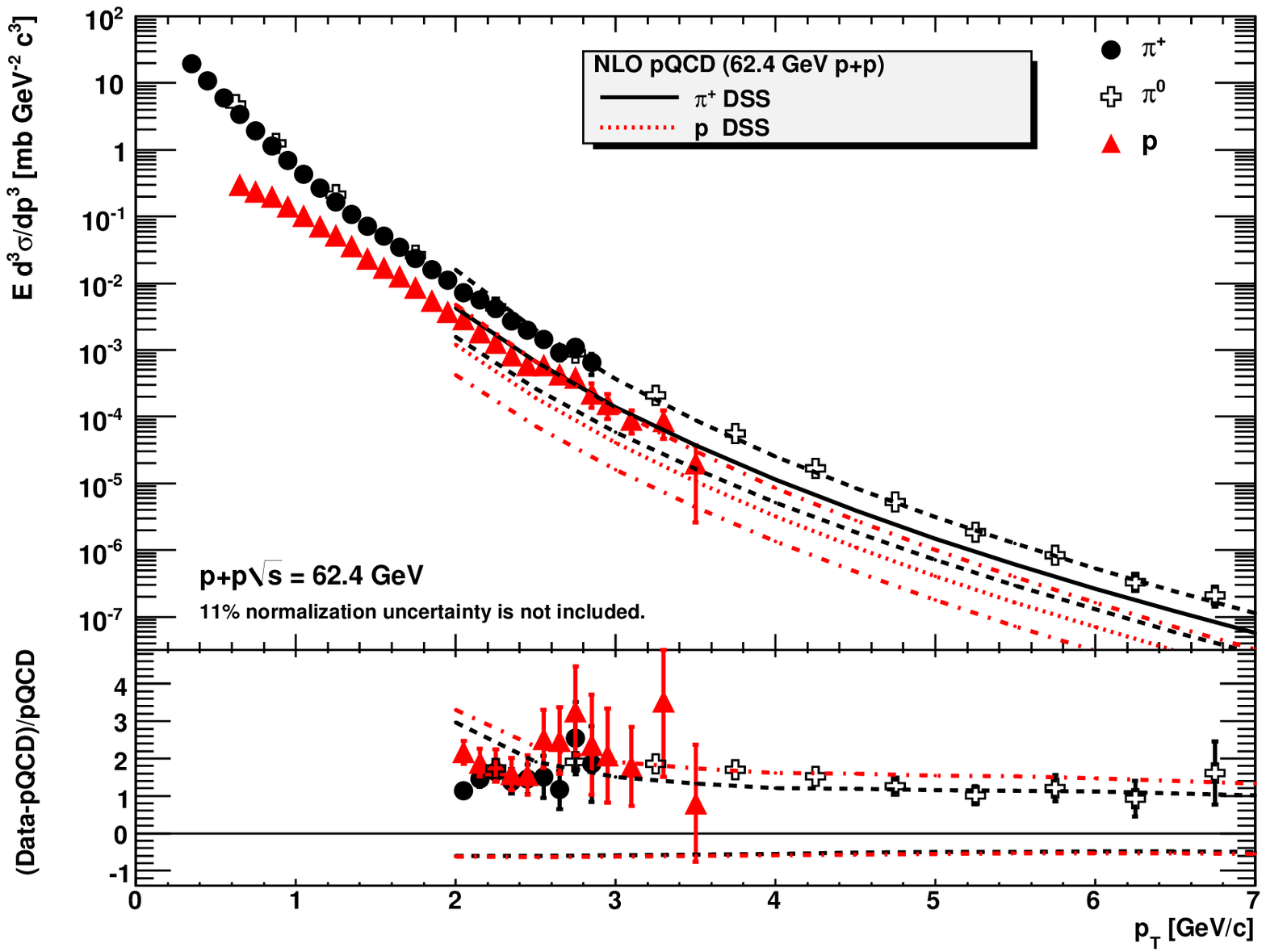}
\includegraphics[width=0.48\linewidth]{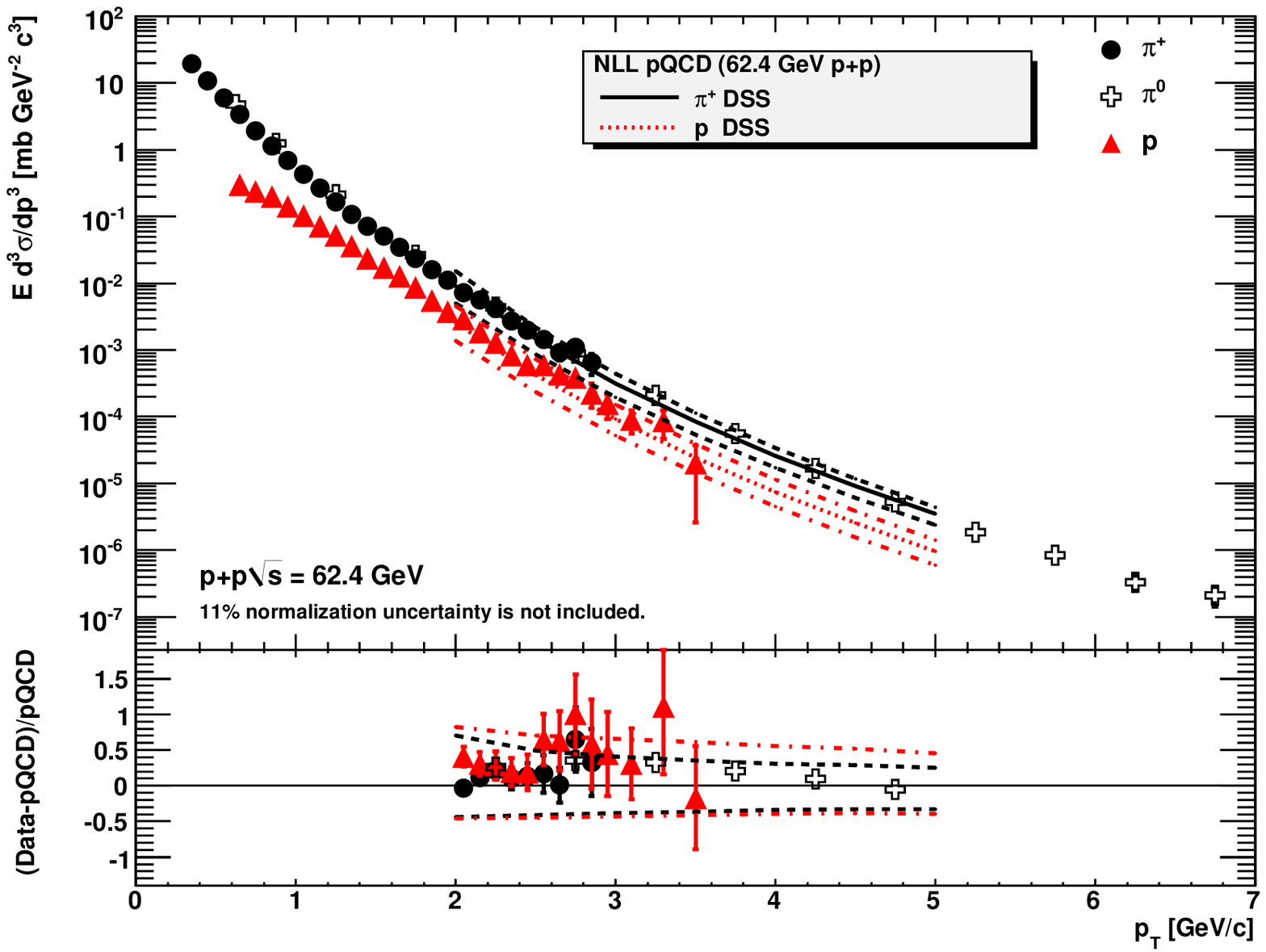}
\includegraphics[width=0.48\linewidth]{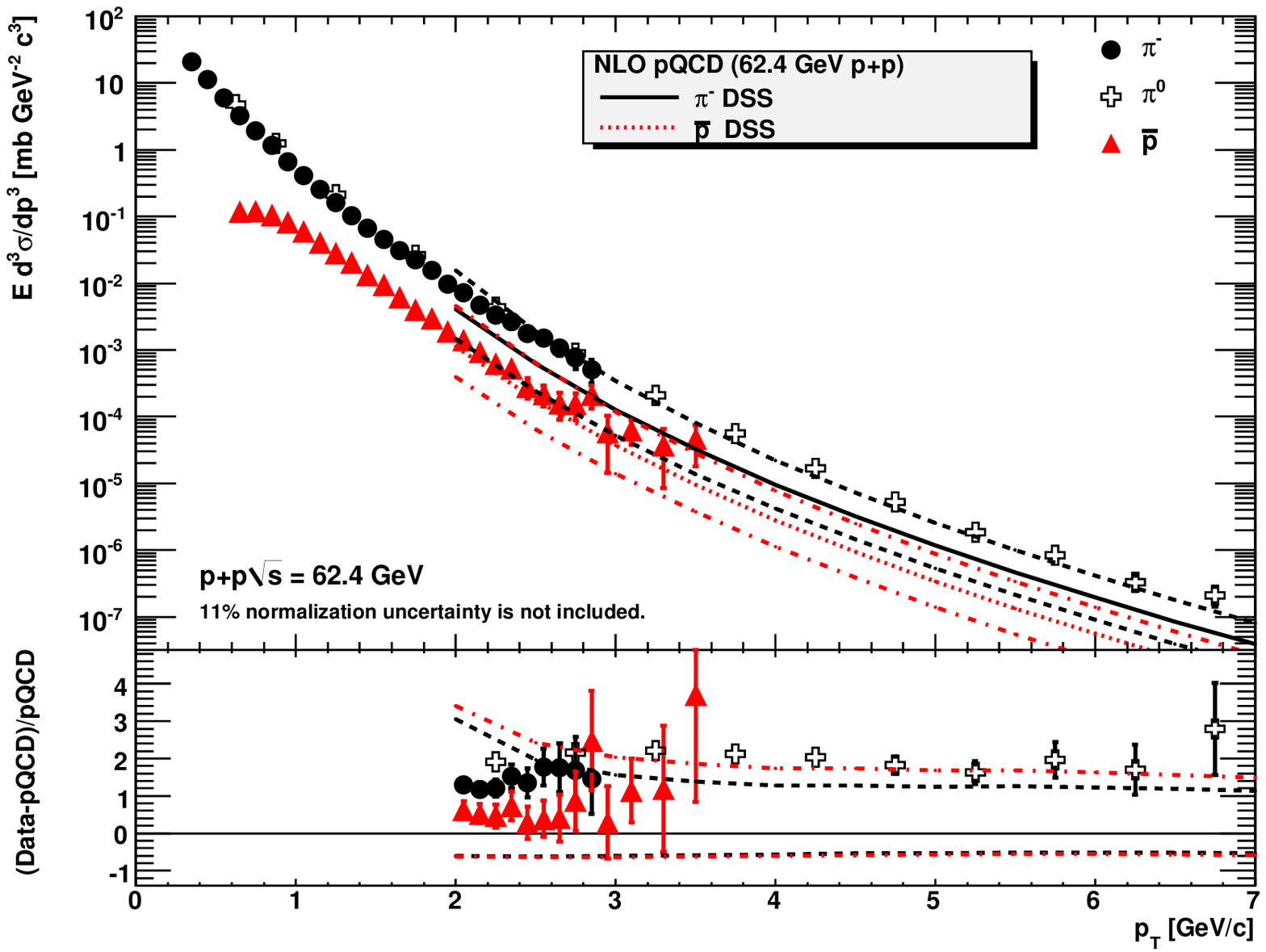}
\includegraphics[width=0.48\linewidth]{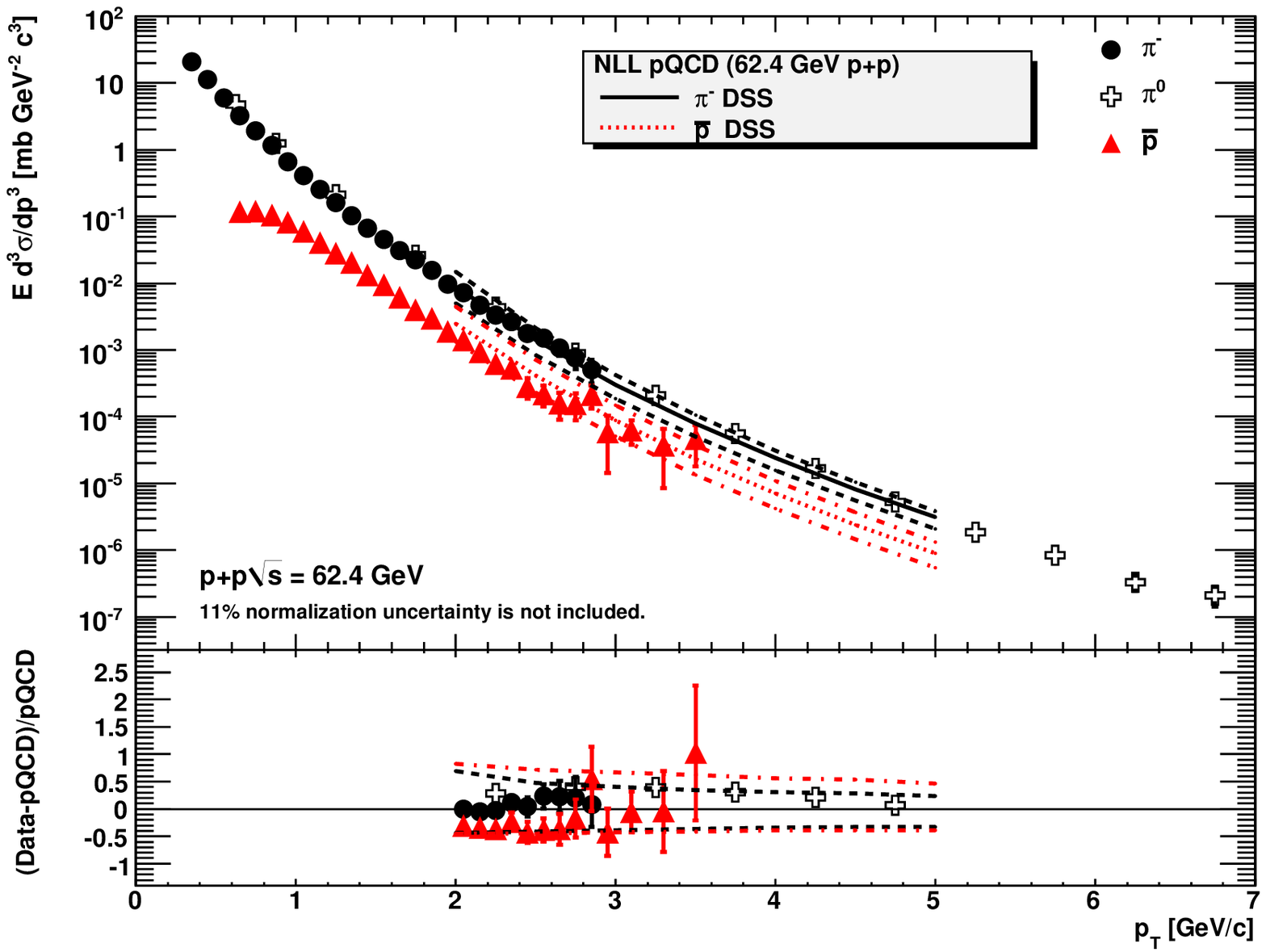}
\caption{(color online) 
Transverse momentum distributions for (upper) positive 
and (lower) negative particles at $\sqs =$ 62.4 GeV in $p+p$ collisions.
Only statistical uncertainties are shown.
The normalization uncertainty (11\%) is not included.
(left) NLO and (right) NLL pQCD calculations (FF: DSS) are also shown.
Solid lines are for $\mu$ = $\pt$, and dashed lines are for $\mu$ = $\pt$/2, 2$\pt$.
The lower panel in each plot shows the ratio of (data - pQCD result)/pQCD result 
for each particle species.
}
\label{fig:pt_DSS_run6pp62}
\end{figure*}

We have evaluated the $\xt$-scaling power $\neff$ using two different 
methods that are both based on Eq.~(\ref{eq:xt_scaling}):

\begin{description}

\item {Method 1} is based on the linear variation of the logarithm of
the ratio of the yields at different $\sqs$:
  \begin{equation}
  {n_{eff}(\xt)} = \frac{\log(Yield(\xt, 62.4)/Yield(\xt, 200))}{\log(200/62.4)} .
  \label{eq:neff_yield_ratio}
  \end{equation}
The $\neff(\xt)$ is shown in Fig.~\ref{fig:n_vs_xt_all} as a function of $\xt$
for neutral pions, protons, and antiprotons for $p+p$ collisions at RHIC.

\item {Method 2} is based on fitting the $\xt$ distributions for a given 
type of particle
measured at different energies.
A common fitting function is defined as follows:
  \begin{equation}
  {E \frac{d^{3}\sigma}{dp^{3}}} = 
\Bigl( \frac{A}{\sqrt{s}} \Bigr)^{\neff} \xt^{-m},
  \label{eq:neff_simul_fit}
  \end{equation}
limiting the fitting region to the high transverse momentum region ($\pt > 2$ GeV/$c$).

\end{description}

The $\xt$ distributions for pions, protons, and antiprotons are shown 
in Fig.~\ref{fig:n_vs_xt_all}.  PHENIX data are presented together with 
earlier data of~\cite{pi0_ua2,PPG087,Adare:2007dg,Alper_npb87}.
Dashed curves show the fitting results.
The obtained $\neff$ values are summarized in Table~\ref{tab:xt_neff}.

The exponent $\neff$ of the $\xt$ scaling is found to have similar 
values for different particles, in the range of 6.3--6.5 for pions, 
protons, and antiprotons. The data points deviate from the $\xt$ scaling 
in the transverse momentum region of $\pt <$ 2 GeV/$c$. This scaling 
violation may be interpreted as a transition from hard to soft 
multiparticle production. For the highest $\xt$ points for protons and 
antiprotons (but not for pions) the asymptotic $\xt$ curve gets steeper. 
Further measurements at larger $\xt$, possibly at lower center-of-mass  
energies are needed to clarify this point.

%%%%%%%%%%%%%%%%%%%%%%%%%%%%%%%%%%%%%%%%%%%%%%%%%%%%  Table_X
\begin{table}[htb]
\caption{Summary of $\xt$-scaling power $\neff$ in $p+p$ collisions.
The errors are systematic error from the fitting.}
\begin{ruledtabular} \begin{tabular}{ccccc}
hadron   & $A$  &  $\neff$ &  $m$  &  $\chi^{2}$/NDF \\ \hline 
$\pi$    & 0.82    $\pm$ 0.08      & 6.35   $\pm$ 0.23  & 8.16   $\pm$ 0.22   & 156/31   \\
$p$      & 1.12    $\pm$ 0.17      & 6.52   $\pm$ 0.59  & 7.41   $\pm$ 0.29   & 40/38   \\
$\pbar$  & 0.84    $\pm$ 0.04      & 6.15   $\pm$ 0.05  & 7.26   $\pm$ 0.07   & 30/38   \\
\end{tabular} \end{ruledtabular}
\label{tab:xt_neff}
\end{table}
%

%\clearpage

%---------------
% 5.3 NLO and NLL pQCD 
%---------------
\subsection{Comparison to NLO and NLL pQCD calculations}
\label{sec:pQCD}
%% --- introduction of pQCD ---

In Figs.~\ref{fig:pt_DSS_run5pp200} to \ref{fig:pt_DSS_run6pp62}, our 
results for pion, proton and antiproton $\pt$ spectra at $\sqs=$ 200 and 
62.4 GeV in $p+p$ collisions are compared to the NLO pQCD calculations. 
Because of the limited $\pt$ reach in the measurements, the results for 
charged kaons are not compared to the NLO pQCD calculations. In these NLO 
pQCD calculations for $\eta<1$ from W. Vogelsang~\cite{W_Vogelsang_private} 
the cross section is factorized into initial parton distribution functions 
(PDFs) in the colliding protons, short-distance partonic hard-scattering 
cross sections which can be evaluated using perturbative QCD, and 
parton-to-hadron fragmentation functions (FFs).

%% --- PDF and FF functions ---

For the description of the initial parton distributions, the Coordinated 
Theoretical-Experimental Project on QCD (CTEQ6M5)~\cite{CTEC_PDF} PDFs 
are used.  Different scales $\mu$ = $\pt$/2, $\pt$, 2$\pt$ are utilized, 
where $\mu$ represents equal factorization, renormalization, and 
fragmentation scales.  These provide initial conditions for the pQCD 
cross section calculations.  Partons are then fragmented to hadrons with 
the help of the de Florian-Sassot-Stratmann (DSS) set of fragmentation 
functions which have the charge separation~\cite{DSS_FF}.  There are 
several other FFs, such as the ``Albino-Kniehl-Kramer'' (AKK) set of 
FFs~\cite{AKK_FF}, and the ``Kniehl-Kramer-Potter'' (KKP) set of 
FF~\cite{KKP_FF}.  Only the results for DSS FFs are shown in this paper, 
because they give better agreement with our measurements than other FFs.  
For example, in $p+p$ collisions at \sqs = 200 GeV the yields for 
($p+\pbar$)/2 in AKK (KKP) FFs are a factor of two smaller (larger) than 
the present measurement.

%% --- comparison to data (general) ---

It is known that pion production in $\sqrt{s}$ = 200 GeV $p+p$ 
collisions is reasonably well described by pQCD down to $\pt \sim$ 2 
GeV/$c$ and up to $\pt \sim$ 20 
GeV/$c$~\cite{Adare:2007dg,Adams:2006nd}. On the other hand, there are 
large variations in the $p$ and $\pbar$ yields among various 
fragmentation functions~\cite{Adams:2006nd} as we mentioned above. From 
the comparisons between baryon data and pQCD calculations at both 
\sqs = 200 and 62.4 GeV, it is potentially interesting to obtain a 
constraint on the fragmentation function, particularly the gluon 
fragmentation function for $p$ and $\pbar$.

%% --- comparison to our data ---

For the DSS fragmentation function, there is good agreement between the 
data and NLO pQCD calculations for pions and protons at 200 GeV,
but not so good agreement with $\pbar$.  It is more clearly shown in 
Fig.~\ref{fig:ratio_negpos_run5pp200_run6pp62} that the $\pbar/p$ ratio 
at 200 GeV is not correctly described with NLO + DSS framework,
which indicates that there is still room to improve the DSS 
fragmentation functions.
The left-side plots of Fig.~\ref{fig:pt_DSS_run6pp62} show that 
for 62.4 GeV NLO + DSS pQCD calculations underestimate yields by a 
factor of two or three for all species.  However, as it is still on the 
edge of the scale uncertainty of the NLO calculation, NLO pQCD agrees 
with the data within the large uncertainties.

%% --- NLL calculation ---

As shown in \cite{PPG087}, the NLL calculations 
give much better agreement with the data for $\pi^0$ in $p+p$ 
collisions at $\sqs =$~62.4 GeV.  This means the resummed calculation is 
necessary to describe the cross section at 62.4 GeV. On the other hand, 
the resummation for \sqs~=~200 GeV is not reliable, since the 
resummation can be done for a larger $\xt = 2\pt/\sqrt{s}$, which is not 
accessible for \sqs~=~200 GeV data due to the $\pt$ limitation. 
The right-side plots of Fig.~\ref{fig:pt_DSS_run6pp62} show the $\pt$ 
distributions for 
$\pi^{\pm}$, $p$, and $\pbar$ in $p+p$ collisions at 62.4 GeV, together 
with the results of NLL pQCD calculations. The DSS FFs are used. It is 
found that the agreement between NLL pQCD and data is better than those 
for NLO pQCD.

%% --- need for high pt ----
The presented $\pt$ spectra extend to the semi-hard 3--4 GeV/$c$ region 
for pions and (anti)protons, which make them useful as a baseline to 
study in further detail the nuclear modification factor in A$+$A 
collisions. More detailed measurements at larger $\pt$ are necessary for 
the further understanding of FFs and its particle species dependence at 
each beam energy.

%\clearpage

%%%%%%%%%%%%%%
% 6. SUMMARY %
%%%%%%%%%%%%%%
\section{SUMMARY AND CONCLUSION}
\label{sec:summary}
%% --- data presentation ---

We presented transverse momentum distributions and yields for 
$\pi^{\pm}$, $K^{\pm}$, $p$ and $\pbar$ in $p+p$ collisions at 
\sqs~=~200 and 62.4~GeV at midrapidity, which provide an important 
baseline for heavy-ion-collision measurements at RHIC.  The inverse 
slope parameter $T_{\rm inv}$, mean transverse momentum $\meanpt$, and yield 
per unit rapidity $dN/dy$ are compared to the measurements at different 
\sqs in $p+p$ and $p+\pbar$ collisions.  While $T_{\rm inv}$ and $\meanpt$ 
show a similar value for all particle species between 200 and 62.4 GeV, 
$dN/dy$ shows a relatively large difference, especially for kaons and 
antiprotons, between 200 and 62.4 GeV.  The $\pbar/p$ ratio is $\sim$0.8 
at 200 GeV, and $\sim$0.5 at 62.4 GeV and the $\pt$ dependence of the 
$p/\pi^+$ ($p/\pi^0$) ratios varies between 62.4 and 200 GeV. Together 
with the measured $dN/dy$ this gives insight on baryon transport 
and production at midrapidity.

We also analyzed the scaling properties of identified particle spectra, 
such as the $\mt$ scaling and $\xt$ scaling.   Baryons and mesons are 
split in the $\mt$ spectral shape at both 200 and 62.4 GeV. This 
splitting can be understood as the difference of hard production yields 
between baryons and mesons. The $\xt$ scaling power $\neff$ shows 
similar values for pions, protons, and antiprotons.

We also compared the results in $p+p$ collisions at 200 GeV with the 
those in Au$+$Au at 200 GeV in the same experiment.  It is found that 
$T_{\rm inv}$, $\meanpt$, and $dN/dy$ change smoothly from $p+p$ 
to Au$+$Au, and all the values in $p+p$ are consistent with expectations 
from the $N_{\rm part}$ dependence in Au$+$Au.  On the nuclear 
modification factor $R_{\rm AA}$, there is a large suppression for 
pions, while there is an enhancement for protons and antiprotons at 
$\pt=$2--4 GeV/$c$.  The observed suppression can be understood by the 
energy loss of partons in hot and dense medium created in Au$+$Au central 
collisions at RHIC energies~\cite{PPG014,Adcox:2004mh}.  Possible 
explanations of the observed enhancements for protons and antiprotons 
include quark recombination~\cite{Fries:2003vb,Hwa:2002tu,Greco:2003xt} 
and/or strong partonic and hadronic radial flow~\cite{Hirano:2003pw}.

%% --- hard part (pQCD) ---

Identified particle spectra are extended to the semi-hard 3--4 GeV/$c$ 
region for pions and (anti)protons, which make it possible to study in 
further detail the nuclear modification factor of identified particles 
in A$+$A collisions. NLO pQCD calculations with DSS fragmentation 
function show good agreement for pions and protons at 200 GeV while 
there is less good agreement with $\pbar$.  This indicates that 
fragmentation functions can be improved further.

For 62.4 GeV, NLO pQCD calculations underestimate by a factor of two or 
three the yields for all particle species. On the other hand, NLL pQCD 
calculation gives a better agreement with the data. This suggests that 
resummed calculations are necessary to describe the cross section at 
62.4 GeV.

%% --- comparison to models (Hydro, pQCD), soft/hard transition ---

From comparisons to some calculations such as NLO/NLL pQCD framework, 
one can discuss the mechanism of soft and hard particle production in 
$p+p$ collisions. There is a transition between these two regions 
(``soft-hard transition'') at $\pt$ $\sim$2 GeV/$c$ for pions, and at 
$\pt$ $\sim$3 GeV/$c$ for (anti)protons, or equivalently, $\mt-m$ = 
1--2 GeV/$c^2$ for all particle species at both energies. The fractions 
of soft and hard components gradually change in the transition region. 
The new measurements presented in this work indicate that the behavior 
of $T_{\rm inv}$ and $\meanpt$ of identified particles in $p+p$ collisions 
needs improved systematics of measurement at both higher and lower 
\sqs to clarify the precise \sqs dependences.

% See NOTES below for information about reference citations, figures 
% and tables, using wide text for equations, landscape figures, etc.

~\\
%%%%%%%%%%%%%%%%%%%%%%%%%  Acknowledgements 
\begin{acknowledgments}

%\section{Acknowledgements}   % Run-6 long from for PRC, PLB, etc.

We thank the staff of the Collider-Accelerator and Physics
Departments at Brookhaven National Laboratory and the staff of
the other PHENIX participating institutions for their vital
contributions.  We acknowledge support from the 
Office of Nuclear Physics in the
Office of Science of the Department of Energy,
the National Science Foundation, 
a sponsored research grant from Renaissance Technologies LLC, 
Abilene Christian University Research Council, 
Research Foundation of SUNY, 
and Dean of the College of Arts and Sciences, Vanderbilt University 
(U.S.A),
Ministry of Education, Culture, Sports, Science, and Technology
and the Japan Society for the Promotion of Science (Japan),
Conselho Nacional de Desenvolvimento Cient\'{\i}fico e
Tecnol{\'o}gico and Funda\c c{\~a}o de Amparo {\`a} Pesquisa do
Estado de S{\~a}o Paulo (Brazil),
Natural Science Foundation of China (People's Republic of China),
Ministry of Education, Youth and Sports (Czech Republic),
Centre National de la Recherche Scientifique, Commissariat
{\`a} l'{\'E}nergie Atomique, and Institut National de Physique
Nucl{\'e}aire et de Physique des Particules (France),
Ministry of Industry, Science and Technologies,
Bundesministerium f\"ur Bildung und Forschung, Deutscher
Akademischer Austausch Dienst, and Alexander von Humboldt Stiftung (Germany),
Hungarian National Science Fund, OTKA (Hungary), 
Department of Atomic Energy and Department of Science and Technology (India),
Israel Science Foundation (Israel), 
National Research Foundation and WCU program of the 
Ministry Education Science and Technology (Korea),
Ministry of Education and Science, Russia Academy of Sciences,
Federal Agency of Atomic Energy (Russia),
VR and the Wallenberg Foundation (Sweden), 
the U.S. Civilian Research and Development Foundation for the
Independent States of the Former Soviet Union, 
the U.S.-Hungarian Fulbright Foundation for Educational Exchange,
and the U.S.-Israel Binational Science Foundation.

\end{acknowledgments}

%%%%%%%%%%%%%%%%%%%%%%%%%%%  References 

%\bibliography{ppg101x0}  

%==========
% Appendix 
%==========

\appendix
\section{Table of cross sections}
\label{appendix}
The cross sections for $\pi^{\pm}$, $K^{\pm}$, $p$ and $\pbar$ in 
$p+p$ collisions at $\sqs$~=~200 and 62.4 ~GeV at midrapidity are tabulated
in Tables~\ref{tab:pion-200} -- ~\ref{tab:ppbar-feeddown-62}. 
Statistical and systematic uncertainties are also shown. 
The normalization uncertainty (9.7\% for 200 GeV, 11\% for 62.4 GeV) is not included.
For protons and antiprotons, there are two kinds of tables, i.e. with and without
the feed-down weak decay corrections.

\clearpage

%%%%%%%%%%%%%

%%%%%%%%%%%%%%%%%%%%%%%%%%%%%%%%%%%%%%%%%%%%%%%%%%%%  Table_XI
\begingroup \squeezetable
\begin{table*}[htb]
\caption{$\pi^{+}$ and $\pi^{-}$ cross sections ($E \frac{d^3\sigma}{dp^{3}}$ [mb GeV$^{-2}c^3$]) in $p+p$ collisions at $\sqrt{s} =$ 200 GeV.
Statistical (2nd column) and systematic (3rd column) uncertainties are shown for each particle species.
The normalization uncertainty (9.7\%) is not included.}
\begin{ruledtabular} \begin{tabular}{ccc}
$\pt$ [GeV/$c$] & $\pi^{+}$ &  $\pi^{-}$ \\ \hline
0.35 & 2.77e+01 $\pm$ 3.0e-01 $\pm$ 1.9e+00 & 2.63e+01 $\pm$ 3.7e-01 $\pm$ 1.8e+00 \\ 
0.45 & 1.45e+01 $\pm$ 1.5e-01 $\pm$ 1.0e+00 & 1.40e+01 $\pm$ 2.0e-01 $\pm$ 9.8e-01 \\ 
0.55 & 7.76e+00 $\pm$ 8.6e-02 $\pm$ 5.4e-01 & 7.91e+00 $\pm$ 1.2e-01 $\pm$ 5.5e-01 \\ 
0.65 & 4.39e+00 $\pm$ 5.3e-02 $\pm$ 3.1e-01 & 4.44e+00 $\pm$ 7.0e-02 $\pm$ 3.1e-01 \\ 
0.75 & 2.65e+00 $\pm$ 3.5e-02 $\pm$ 1.9e-01 & 2.69e+00 $\pm$ 4.6e-02 $\pm$ 1.9e-01 \\ 
0.85 & 1.59e+00 $\pm$ 2.2e-02 $\pm$ 1.1e-01 & 1.60e+00 $\pm$ 2.9e-02 $\pm$ 1.1e-01 \\ 
0.95 & 1.01e+00 $\pm$ 1.5e-02 $\pm$ 7.1e-02 & 9.83e-01 $\pm$ 1.9e-02 $\pm$ 6.9e-02 \\ 
1.05 & 6.45e-01 $\pm$ 1.1e-02 $\pm$ 4.5e-02 & 6.30e-01 $\pm$ 1.3e-02 $\pm$ 4.4e-02 \\ 
1.15 & 4.18e-01 $\pm$ 7.2e-03 $\pm$ 2.9e-02 & 4.36e-01 $\pm$ 9.5e-03 $\pm$ 3.1e-02 \\ 
1.25 & 2.76e-01 $\pm$ 5.0e-03 $\pm$ 1.9e-02 & 2.79e-01 $\pm$ 6.3e-03 $\pm$ 2.0e-02 \\ 
1.35 & 1.88e-01 $\pm$ 3.6e-03 $\pm$ 1.3e-02 & 1.90e-01 $\pm$ 4.4e-03 $\pm$ 1.3e-02 \\ 
1.45 & 1.29e-01 $\pm$ 2.6e-03 $\pm$ 9.0e-03 & 1.29e-01 $\pm$ 3.1e-03 $\pm$ 9.0e-03 \\ 
1.55 & 9.07e-02 $\pm$ 1.9e-03 $\pm$ 6.4e-03 & 9.05e-02 $\pm$ 2.3e-03 $\pm$ 6.3e-03 \\ 
1.65 & 6.52e-02 $\pm$ 1.4e-03 $\pm$ 4.6e-03 & 6.47e-02 $\pm$ 1.7e-03 $\pm$ 4.5e-03 \\ 
1.75 & 4.48e-02 $\pm$ 9.9e-04 $\pm$ 3.1e-03 & 4.69e-02 $\pm$ 1.2e-03 $\pm$ 3.3e-03 \\ 
1.85 & 3.45e-02 $\pm$ 8.1e-04 $\pm$ 2.4e-03 & 3.40e-02 $\pm$ 9.3e-04 $\pm$ 2.4e-03 \\ 
1.95 & 2.49e-02 $\pm$ 6.1e-04 $\pm$ 1.7e-03 & 2.56e-02 $\pm$ 7.4e-04 $\pm$ 1.8e-03 \\ 
2.05 & 1.83e-02 $\pm$ 4.7e-04 $\pm$ 1.3e-03 & 1.81e-02 $\pm$ 5.5e-04 $\pm$ 1.3e-03 \\ 
2.15 & 1.37e-02 $\pm$ 3.8e-04 $\pm$ 9.6e-04 & 1.33e-02 $\pm$ 4.3e-04 $\pm$ 9.3e-04 \\ 
2.25 & 1.13e-02 $\pm$ 3.5e-04 $\pm$ 7.9e-04 & 1.03e-02 $\pm$ 3.6e-04 $\pm$ 7.2e-04 \\ 
2.35 & 8.21e-03 $\pm$ 2.8e-04 $\pm$ 5.7e-04 & 7.48e-03 $\pm$ 2.8e-04 $\pm$ 5.2e-04 \\ 
2.45 & 6.73e-03 $\pm$ 2.5e-04 $\pm$ 4.7e-04 & 6.34e-03 $\pm$ 2.7e-04 $\pm$ 4.4e-04 \\ 
2.55 & 5.39e-03 $\pm$ 2.3e-04 $\pm$ 3.8e-04 & 4.96e-03 $\pm$ 2.3e-04 $\pm$ 3.5e-04 \\ 
2.65 & 4.27e-03 $\pm$ 2.0e-04 $\pm$ 3.0e-04 & 3.47e-03 $\pm$ 1.8e-04 $\pm$ 2.4e-04 \\ 
2.75 & 3.02e-03 $\pm$ 1.6e-04 $\pm$ 2.1e-04 & 2.82e-03 $\pm$ 1.6e-04 $\pm$ 2.0e-04 \\ 
2.85 & 2.45e-03 $\pm$ 1.4e-04 $\pm$ 1.7e-04 & 2.23e-03 $\pm$ 1.5e-04 $\pm$ 1.6e-04 \\ 
2.95 & 1.82e-03 $\pm$ 1.2e-04 $\pm$ 1.3e-04 & 1.66e-03 $\pm$ 1.2e-04 $\pm$ 1.2e-04 \\ 
\end{tabular} \end{ruledtabular}
\label{tab:pion-200}
\end{table*}\endgroup

%%\clearpage
\begingroup \squeezetable
%%%%%%%%%%%%%%%%%%%%%%%%%%%%%%%%%%%%%%%%%%%%%%%%%%%%  Table_XII
\begin{table*}[htb]
\caption{$K^{+}$ and $K^{-}$ cross sections ($E \frac{d^3\sigma}{dp^{3}}$ [mb GeV$^{-2}c^3$]) in $p+p$ collisions at $\sqrt{s} =$ 200 GeV. 
Statistical (2nd column) and systematic (3rd column) uncertainties are shown for each particle species.
The normalization uncertainty (9.7\%) is not included.}
\begin{ruledtabular} \begin{tabular}{ccc}
$\pt$ [GeV/$c$] & $K^{+}$ &  $K^{-}$ \\ \hline
0.45 & 1.96e+00 $\pm$ 5.0e-02 $\pm$ 1.4e-01 & 1.89e+00 $\pm$ 7.0e-02 $\pm$ 1.3e-01 \\ 
0.55 & 1.35e+00 $\pm$ 3.0e-02 $\pm$ 9.4e-02 & 1.37e+00 $\pm$ 4.3e-02 $\pm$ 9.6e-02 \\ 
0.65 & 8.71e-01 $\pm$ 1.9e-02 $\pm$ 6.1e-02 & 8.28e-01 $\pm$ 2.3e-02 $\pm$ 5.8e-02 \\ 
0.75 & 5.86e-01 $\pm$ 1.3e-02 $\pm$ 4.1e-02 & 5.60e-01 $\pm$ 1.6e-02 $\pm$ 3.9e-02 \\ 
0.85 & 3.95e-01 $\pm$ 8.7e-03 $\pm$ 2.8e-02 & 3.87e-01 $\pm$ 1.1e-02 $\pm$ 2.7e-02 \\ 
0.95 & 2.60e-01 $\pm$ 5.8e-03 $\pm$ 1.8e-02 & 2.54e-01 $\pm$ 7.3e-03 $\pm$ 1.8e-02 \\ 
1.05 & 1.72e-01 $\pm$ 3.9e-03 $\pm$ 1.2e-02 & 1.83e-01 $\pm$ 5.5e-03 $\pm$ 1.3e-02 \\ 
1.15 & 1.26e-01 $\pm$ 3.0e-03 $\pm$ 8.9e-03 & 1.16e-01 $\pm$ 3.5e-03 $\pm$ 8.1e-03 \\ 
1.25 & 8.52e-02 $\pm$ 2.1e-03 $\pm$ 6.0e-03 & 8.97e-02 $\pm$ 2.8e-03 $\pm$ 6.3e-03 \\ 
1.35 & 6.08e-02 $\pm$ 1.5e-03 $\pm$ 4.3e-03 & 6.23e-02 $\pm$ 2.0e-03 $\pm$ 4.4e-03 \\ 
1.45 & 4.59e-02 $\pm$ 1.2e-03 $\pm$ 3.2e-03 & 4.27e-02 $\pm$ 1.4e-03 $\pm$ 3.0e-03 \\ 
1.55 & 3.29e-02 $\pm$ 9.0e-04 $\pm$ 2.3e-03 & 3.21e-02 $\pm$ 1.1e-03 $\pm$ 2.2e-03 \\ 
1.65 & 2.39e-02 $\pm$ 6.6e-04 $\pm$ 1.7e-03 & 2.23e-02 $\pm$ 7.4e-04 $\pm$ 1.6e-03 \\ 
1.75 & 1.86e-02 $\pm$ 5.3e-04 $\pm$ 1.3e-03 & 1.81e-02 $\pm$ 6.2e-04 $\pm$ 1.3e-03 \\ 
1.85 & 1.49e-02 $\pm$ 4.4e-04 $\pm$ 1.0e-03 & 1.36e-02 $\pm$ 4.7e-04 $\pm$ 9.5e-04 \\ 
1.95 & 1.13e-02 $\pm$ 3.5e-04 $\pm$ 7.9e-04 & 1.03e-02 $\pm$ 3.7e-04 $\pm$ 7.2e-04 \\ 
\end{tabular} \end{ruledtabular}
\label{tab:kaon-200}
\end{table*}\endgroup

%%\clearpage

\begingroup \squeezetable
%%%%%%%%%%%%%%%%%%%%%%%%%%%%%%%%%%%%%%%%%%%%%%%%%%%%  Table_XIII
\begin{table*}[htb]
\caption{$p$ and $\pbar$ cross sections ($E \frac{d^3\sigma}{dp^{3}}$ [mb GeV$^{-2}c^3$]) in $p+p$ collisions at $\sqrt{s} =$ 200 GeV. 
Statistical (2nd column) and systematic (3rd column) uncertainties are shown for each particle species.
The normalization uncertainty (9.7\%) is not included. Feed-down weak decay corrections are not applied.}
\begin{ruledtabular} \begin{tabular}{ccc}
$\pt$ [GeV/$c$] & $p$ &  $\overline{p}$ \\ \hline
0.55 & 1.02e+00 $\pm$ 2.0e-02 $\pm$ 6.2e-02 & 7.88e-01 $\pm$ 1.6e-02 $\pm$ 5.5e-02 \\ 
0.65 & 7.40e-01 $\pm$ 1.4e-02 $\pm$ 4.5e-02 & 6.04e-01 $\pm$ 1.2e-02 $\pm$ 4.2e-02 \\ 
0.75 & 5.58e-01 $\pm$ 1.1e-02 $\pm$ 3.4e-02 & 4.62e-01 $\pm$ 9.1e-03 $\pm$ 3.2e-02 \\ 
0.85 & 3.77e-01 $\pm$ 7.7e-03 $\pm$ 2.3e-02 & 3.18e-01 $\pm$ 6.3e-03 $\pm$ 2.2e-02 \\ 
0.95 & 2.73e-01 $\pm$ 5.9e-03 $\pm$ 1.6e-02 & 2.18e-01 $\pm$ 4.4e-03 $\pm$ 1.5e-02 \\ 
1.05 & 1.80e-01 $\pm$ 4.0e-03 $\pm$ 1.1e-02 & 1.58e-01 $\pm$ 3.3e-03 $\pm$ 1.1e-02 \\ 
1.15 & 1.27e-01 $\pm$ 2.9e-03 $\pm$ 7.6e-03 & 1.08e-01 $\pm$ 2.4e-03 $\pm$ 7.6e-03 \\ 
1.25 & 9.18e-02 $\pm$ 2.2e-03 $\pm$ 5.5e-03 & 7.54e-02 $\pm$ 1.7e-03 $\pm$ 5.3e-03 \\ 
1.35 & 6.24e-02 $\pm$ 1.6e-03 $\pm$ 3.7e-03 & 5.58e-02 $\pm$ 1.3e-03 $\pm$ 3.9e-03 \\ 
1.45 & 4.80e-02 $\pm$ 1.3e-03 $\pm$ 2.9e-03 & 3.73e-02 $\pm$ 8.9e-04 $\pm$ 2.6e-03 \\ 
1.55 & 3.32e-02 $\pm$ 9.1e-04 $\pm$ 2.0e-03 & 2.68e-02 $\pm$ 6.6e-04 $\pm$ 1.9e-03 \\ 
1.65 & 2.31e-02 $\pm$ 6.5e-04 $\pm$ 1.4e-03 & 1.93e-02 $\pm$ 4.9e-04 $\pm$ 1.4e-03 \\ 
1.75 & 1.70e-02 $\pm$ 5.0e-04 $\pm$ 1.0e-03 & 1.39e-02 $\pm$ 3.7e-04 $\pm$ 9.8e-04 \\ 
1.85 & 1.17e-02 $\pm$ 3.6e-04 $\pm$ 7.0e-04 & 9.69e-03 $\pm$ 2.6e-04 $\pm$ 6.8e-04 \\ 
1.95 & 8.98e-03 $\pm$ 2.9e-04 $\pm$ 5.4e-04 & 6.94e-03 $\pm$ 1.9e-04 $\pm$ 4.9e-04 \\ 
2.05 & 6.68e-03 $\pm$ 2.3e-04 $\pm$ 4.0e-04 & 5.12e-03 $\pm$ 1.5e-04 $\pm$ 3.6e-04 \\ 
2.15 & 4.62e-03 $\pm$ 1.6e-04 $\pm$ 2.8e-04 & 3.61e-03 $\pm$ 1.1e-04 $\pm$ 2.5e-04 \\ 
2.25 & 3.91e-03 $\pm$ 1.5e-04 $\pm$ 2.4e-04 & 2.90e-03 $\pm$ 9.2e-05 $\pm$ 2.0e-04 \\ 
2.35 & 2.63e-03 $\pm$ 1.0e-04 $\pm$ 1.6e-04 & 2.09e-03 $\pm$ 7.0e-05 $\pm$ 1.5e-04 \\ 
2.45 & 1.79e-03 $\pm$ 7.4e-05 $\pm$ 1.1e-04 & 1.58e-03 $\pm$ 5.5e-05 $\pm$ 1.1e-04 \\ 
2.55 & 1.62e-03 $\pm$ 7.0e-05 $\pm$ 1.0e-04 & 1.10e-03 $\pm$ 4.2e-05 $\pm$ 7.8e-05 \\ 
2.65 & 1.15e-03 $\pm$ 5.4e-05 $\pm$ 7.2e-05 & 8.85e-04 $\pm$ 3.7e-05 $\pm$ 6.3e-05 \\ 
2.75 & 8.89e-04 $\pm$ 4.4e-05 $\pm$ 5.6e-05 & 6.22e-04 $\pm$ 2.8e-05 $\pm$ 4.4e-05 \\ 
2.85 & 6.38e-04 $\pm$ 3.5e-05 $\pm$ 4.1e-05 & 5.07e-04 $\pm$ 2.4e-05 $\pm$ 3.6e-05 \\ 
2.95 & 4.97e-04 $\pm$ 3.0e-05 $\pm$ 3.2e-05 & 3.80e-04 $\pm$ 2.0e-05 $\pm$ 2.7e-05 \\ 
3.05 & 4.13e-04 $\pm$ 2.6e-05 $\pm$ 2.7e-05 & 3.13e-04 $\pm$ 1.7e-05 $\pm$ 2.3e-05 \\ 
3.10 & 3.80e-04 $\pm$ 1.8e-05 $\pm$ 2.5e-05 & 2.75e-04 $\pm$ 1.1e-05 $\pm$ 2.0e-05 \\ 
3.30 & 2.33e-04 $\pm$ 1.3e-05 $\pm$ 1.6e-05 & 1.92e-04 $\pm$ 9.0e-06 $\pm$ 1.4e-05 \\ 
3.50 & 1.57e-04 $\pm$ 1.0e-05 $\pm$ 1.1e-05 & 1.12e-04 $\pm$ 6.5e-06 $\pm$ 8.6e-06 \\ 
3.70 & 1.11e-04 $\pm$ 8.9e-06 $\pm$ 8.3e-06 & 7.16e-05 $\pm$ 5.2e-06 $\pm$ 5.8e-06 \\ 
3.90 & 7.25e-05 $\pm$ 7.2e-06 $\pm$ 5.8e-06 & 4.40e-05 $\pm$ 4.0e-06 $\pm$ 3.8e-06 \\ 
4.10 & 6.23e-05 $\pm$ 6.7e-06 $\pm$ 5.3e-06 & 3.81e-05 $\pm$ 3.9e-06 $\pm$ 3.6e-06 \\ 
4.30 & 3.83e-05 $\pm$ 5.5e-06 $\pm$ 3.6e-06 & 2.63e-05 $\pm$ 3.3e-06 $\pm$ 2.8e-06 \\ 
4.50 & 3.22e-05 $\pm$ 5.2e-06 $\pm$ 3.3e-06 & 1.82e-05 $\pm$ 2.8e-06 $\pm$ 2.2e-06 \\ 
\end{tabular} \end{ruledtabular}
\label{tab:ppbar-200}
\end{table*} \endgroup

%%\clearpage

\begingroup \squeezetable
%%%%%%%%%%%%%%%%%%%%%%%%%%%%%%%%%%%%%%%%%%%%%%%%%%%%  Table_XIV
\begin{table*}[htb]
\caption{$p$ and $\pbar$ cross sections ($E \frac{d^3\sigma}{dp^{3}}$ [mb GeV$^{-2}c^3$]) in $p+p$ collisions at $\sqrt{s} =$ 200 GeV. 
Statistical (2nd column) and systematic (3rd column) uncertainties are shown for each particle species.
The normalization uncertainty (9.7\%) is not included. Feed-down weak decay corrections are applied.}
\begin{ruledtabular} \begin{tabular}{ccc}
$\pt$ [GeV/$c$] & $p$ &  $\overline{p}$ \\ \hline
0.55 & 5.93e-01 $\pm$ 1.1e-02 $\pm$ 1.4e-01 & 4.56e-01 $\pm$ 9.2e-03 $\pm$ 1.1e-01 \\ 
0.65 & 4.45e-01 $\pm$ 8.4e-03 $\pm$ 9.4e-02 & 3.63e-01 $\pm$ 7.0e-03 $\pm$ 7.8e-02 \\ 
0.75 & 3.47e-01 $\pm$ 6.9e-03 $\pm$ 6.6e-02 & 2.87e-01 $\pm$ 5.6e-03 $\pm$ 5.6e-02 \\ 
0.85 & 2.42e-01 $\pm$ 4.9e-03 $\pm$ 4.2e-02 & 2.04e-01 $\pm$ 4.0e-03 $\pm$ 3.6e-02 \\ 
0.95 & 1.80e-01 $\pm$ 3.9e-03 $\pm$ 2.9e-02 & 1.44e-01 $\pm$ 2.9e-03 $\pm$ 2.4e-02 \\ 
1.05 & 1.22e-01 $\pm$ 2.7e-03 $\pm$ 1.8e-02 & 1.06e-01 $\pm$ 2.2e-03 $\pm$ 1.6e-02 \\ 
1.15 & 8.77e-02 $\pm$ 2.0e-03 $\pm$ 1.2e-02 & 7.48e-02 $\pm$ 1.6e-03 $\pm$ 1.1e-02 \\ 
1.25 & 6.46e-02 $\pm$ 1.6e-03 $\pm$ 8.4e-03 & 5.31e-02 $\pm$ 1.2e-03 $\pm$ 7.1e-03 \\ 
1.35 & 4.47e-02 $\pm$ 1.1e-03 $\pm$ 5.5e-03 & 4.00e-02 $\pm$ 9.4e-04 $\pm$ 5.1e-03 \\ 
1.45 & 3.49e-02 $\pm$ 9.4e-04 $\pm$ 4.1e-03 & 2.72e-02 $\pm$ 6.5e-04 $\pm$ 3.3e-03 \\ 
1.55 & 2.45e-02 $\pm$ 6.8e-04 $\pm$ 2.7e-03 & 1.98e-02 $\pm$ 4.9e-04 $\pm$ 2.3e-03 \\ 
1.65 & 1.73e-02 $\pm$ 4.9e-04 $\pm$ 1.9e-03 & 1.45e-02 $\pm$ 3.7e-04 $\pm$ 1.6e-03 \\ 
1.75 & 1.28e-02 $\pm$ 3.8e-04 $\pm$ 1.3e-03 & 1.06e-02 $\pm$ 2.8e-04 $\pm$ 1.2e-03 \\ 
1.85 & 8.92e-03 $\pm$ 2.7e-04 $\pm$ 8.9e-04 & 7.42e-03 $\pm$ 2.0e-04 $\pm$ 7.9e-04 \\ 
1.95 & 6.95e-03 $\pm$ 2.2e-04 $\pm$ 6.8e-04 & 5.37e-03 $\pm$ 1.5e-04 $\pm$ 5.6e-04 \\ 
2.05 & 5.21e-03 $\pm$ 1.8e-04 $\pm$ 4.9e-04 & 4.00e-03 $\pm$ 1.2e-04 $\pm$ 4.0e-04 \\ 
2.15 & 3.63e-03 $\pm$ 1.3e-04 $\pm$ 3.4e-04 & 2.84e-03 $\pm$ 8.7e-05 $\pm$ 2.8e-04 \\ 
2.25 & 3.10e-03 $\pm$ 1.2e-04 $\pm$ 2.8e-04 & 2.30e-03 $\pm$ 7.3e-05 $\pm$ 2.2e-04 \\ 
2.35 & 2.10e-03 $\pm$ 8.2e-05 $\pm$ 1.9e-04 & 1.67e-03 $\pm$ 5.6e-05 $\pm$ 1.6e-04 \\ 
2.45 & 1.44e-03 $\pm$ 6.0e-05 $\pm$ 1.3e-04 & 1.27e-03 $\pm$ 4.4e-05 $\pm$ 1.2e-04 \\ 
2.55 & 1.31e-03 $\pm$ 5.7e-05 $\pm$ 1.1e-04 & 8.89e-04 $\pm$ 3.4e-05 $\pm$ 8.3e-05 \\ 
2.65 & 9.31e-04 $\pm$ 4.4e-05 $\pm$ 8.0e-05 & 7.19e-04 $\pm$ 3.0e-05 $\pm$ 6.6e-05 \\ 
2.75 & 7.26e-04 $\pm$ 3.6e-05 $\pm$ 6.2e-05 & 5.08e-04 $\pm$ 2.3e-05 $\pm$ 4.6e-05 \\ 
2.85 & 5.23e-04 $\pm$ 2.9e-05 $\pm$ 4.4e-05 & 4.16e-04 $\pm$ 2.0e-05 $\pm$ 3.8e-05 \\ 
2.95 & 4.09e-04 $\pm$ 2.4e-05 $\pm$ 3.4e-05 & 3.13e-04 $\pm$ 1.6e-05 $\pm$ 2.8e-05 \\ 
3.05 & 3.41e-04 $\pm$ 2.2e-05 $\pm$ 2.9e-05 & 2.58e-04 $\pm$ 1.4e-05 $\pm$ 2.3e-05 \\ 
3.10 & 3.14e-04 $\pm$ 1.5e-05 $\pm$ 2.6e-05 & 2.28e-04 $\pm$ 9.3e-06 $\pm$ 2.0e-05 \\ 
3.30 & 1.94e-04 $\pm$ 1.1e-05 $\pm$ 1.6e-05 & 1.60e-04 $\pm$ 7.5e-06 $\pm$ 1.4e-05 \\ 
3.50 & 1.32e-04 $\pm$ 8.6e-06 $\pm$ 1.1e-05 & 9.42e-05 $\pm$ 5.4e-06 $\pm$ 8.6e-06 \\ 
3.70 & 9.35e-05 $\pm$ 7.5e-06 $\pm$ 8.2e-06 & 6.03e-05 $\pm$ 4.4e-06 $\pm$ 5.6e-06 \\ 
3.90 & 6.13e-05 $\pm$ 6.1e-06 $\pm$ 5.6e-06 & 3.72e-05 $\pm$ 3.4e-06 $\pm$ 3.6e-06 \\ 
4.10 & 5.28e-05 $\pm$ 5.7e-06 $\pm$ 5.1e-06 & 3.24e-05 $\pm$ 3.3e-06 $\pm$ 3.4e-06 \\ 
4.30 & 3.26e-05 $\pm$ 4.7e-06 $\pm$ 3.3e-06 & 2.23e-05 $\pm$ 2.8e-06 $\pm$ 2.6e-06 \\ 
4.50 & 2.75e-05 $\pm$ 4.4e-06 $\pm$ 3.0e-06 & 1.56e-05 $\pm$ 2.4e-06 $\pm$ 2.0e-06 \\ 
\end{tabular} \end{ruledtabular}
\label{tab:ppbar-feeddown-200}
\end{table*} \endgroup

%%\clearpage
%===== 62.4 GeV data tables
%
\begingroup \squeezetable
%%%%%%%%%%%%%%%%%%%%%%%%%%%%%%%%%%%%%%%%%%%%%%%%%%%%  Table_XV
\begin{table*}[htb]
\caption{$\pi^{+}$ and $\pi^{-}$ cross sections ($E \frac{d^3\sigma}{dp^{3}}$ [mb GeV$^{-2}c^3$]) in $p+p$ collisions at $\sqrt{s} =$ 62.4 GeV. 
Statistical (2nd column) and systematic (3rd column) uncertainties are shown for each particle species.
The normalization uncertainty (11\%) is not included.}
\begin{ruledtabular} \begin{tabular}{ccc}
$\pt$ [GeV/$c$] & $\pi^{+}$ &  $\pi^{-}$ \\ \hline
0.35 & 1.96e+01 $\pm$ 1.8e-01 $\pm$ 1.4e+00 & 2.08e+01 $\pm$ 1.5e-01 $\pm$ 1.3e+00 \\ 
0.45 & 1.07e+01 $\pm$ 1.1e-01 $\pm$ 7.5e-01 & 1.12e+01 $\pm$ 8.6e-02 $\pm$ 6.7e-01 \\ 
0.55 & 5.95e+00 $\pm$ 6.3e-02 $\pm$ 4.2e-01 & 5.94e+00 $\pm$ 4.9e-02 $\pm$ 3.6e-01 \\ 
0.65 & 3.38e+00 $\pm$ 3.9e-02 $\pm$ 2.4e-01 & 3.25e+00 $\pm$ 3.0e-02 $\pm$ 1.9e-01 \\ 
0.75 & 1.91e+00 $\pm$ 2.4e-02 $\pm$ 1.3e-01 & 1.92e+00 $\pm$ 2.0e-02 $\pm$ 1.2e-01 \\ 
0.85 & 1.13e+00 $\pm$ 1.6e-02 $\pm$ 7.9e-02 & 1.15e+00 $\pm$ 1.3e-02 $\pm$ 6.9e-02 \\ 
0.95 & 6.86e-01 $\pm$ 1.0e-02 $\pm$ 4.8e-02 & 6.68e-01 $\pm$ 8.4e-03 $\pm$ 4.0e-02 \\ 
1.05 & 4.30e-01 $\pm$ 7.2e-03 $\pm$ 3.0e-02 & 4.06e-01 $\pm$ 5.7e-03 $\pm$ 2.4e-02 \\ 
1.15 & 2.65e-01 $\pm$ 4.9e-03 $\pm$ 1.9e-02 & 2.53e-01 $\pm$ 4.0e-03 $\pm$ 1.5e-02 \\ 
1.25 & 1.66e-01 $\pm$ 3.5e-03 $\pm$ 1.2e-02 & 1.60e-01 $\pm$ 2.9e-03 $\pm$ 9.6e-03 \\ 
1.35 & 1.08e-01 $\pm$ 2.6e-03 $\pm$ 7.5e-03 & 1.03e-01 $\pm$ 2.1e-03 $\pm$ 6.2e-03 \\ 
1.45 & 7.20e-02 $\pm$ 1.9e-03 $\pm$ 5.0e-03 & 6.74e-02 $\pm$ 1.6e-03 $\pm$ 4.0e-03 \\ 
1.55 & 5.04e-02 $\pm$ 1.5e-03 $\pm$ 3.5e-03 & 4.54e-02 $\pm$ 1.2e-03 $\pm$ 2.7e-03 \\ 
1.65 & 3.48e-02 $\pm$ 1.2e-03 $\pm$ 2.4e-03 & 3.07e-02 $\pm$ 9.7e-04 $\pm$ 1.8e-03 \\ 
1.75 & 2.33e-02 $\pm$ 9.5e-04 $\pm$ 1.6e-03 & 2.25e-02 $\pm$ 8.3e-04 $\pm$ 1.4e-03 \\ 
1.85 & 1.58e-02 $\pm$ 7.8e-04 $\pm$ 1.1e-03 & 1.55e-02 $\pm$ 6.8e-04 $\pm$ 9.3e-04 \\ 
1.95 & 1.11e-02 $\pm$ 6.7e-04 $\pm$ 7.8e-04 & 9.63e-03 $\pm$ 5.2e-04 $\pm$ 5.8e-04 \\ 
2.05 & 7.13e-03 $\pm$ 5.2e-04 $\pm$ 5.0e-04 & 7.23e-03 $\pm$ 4.7e-04 $\pm$ 4.3e-04 \\ 
2.15 & 5.63e-03 $\pm$ 5.0e-04 $\pm$ 4.0e-04 & 4.72e-03 $\pm$ 3.9e-04 $\pm$ 2.9e-04 \\ 
2.25 & 4.22e-03 $\pm$ 4.3e-04 $\pm$ 3.0e-04 & 3.32e-03 $\pm$ 3.4e-04 $\pm$ 2.0e-04 \\ 
2.35 & 2.69e-03 $\pm$ 3.7e-04 $\pm$ 1.9e-04 & 2.67e-03 $\pm$ 3.3e-04 $\pm$ 1.7e-04 \\ 
2.45 & 1.96e-03 $\pm$ 3.1e-04 $\pm$ 1.4e-04 & 1.75e-03 $\pm$ 2.9e-04 $\pm$ 1.1e-04 \\ 
2.55 & 1.45e-03 $\pm$ 3.3e-04 $\pm$ 1.1e-04 & 1.49e-03 $\pm$ 2.7e-04 $\pm$ 9.8e-05 \\ 
2.65 & 9.07e-04 $\pm$ 2.2e-04 $\pm$ 7.0e-05 & 1.07e-03 $\pm$ 2.5e-04 $\pm$ 7.3e-05 \\ 
2.75 & 1.09e-03 $\pm$ 3.0e-04 $\pm$ 8.6e-05 & 7.62e-04 $\pm$ 2.5e-04 $\pm$ 5.4e-05 \\ 
2.85 & 6.48e-04 $\pm$ 2.3e-04 $\pm$ 5.3e-05 & 5.10e-04 $\pm$ 2.0e-04 $\pm$ 3.7e-05 \\ 
\end{tabular} \end{ruledtabular}
\label{tab:pion-62}
\end{table*} \endgroup

%%\clearpage

\begingroup \squeezetable
%%%%%%%%%%%%%%%%%%%%%%%%%%%%%%%%%%%%%%%%%%%%%%%%%%%%  Table_XVI
\begin{table*}[htb]
\caption{$K^{+}$ and $K^{-}$ cross sections ($E \frac{d^3\sigma}{dp^{3}}$ [mb GeV$^{-2}c^3$]) in $p+p$ collisions at $\sqrt{s} =$ 62.4 GeV. 
Statistical (2nd column) and systematic (3rd column) uncertainties are shown for each particle species.
The normalization uncertainty (11\%) is not included.}
\begin{ruledtabular} \begin{tabular}{ccc}
$\pt$ [GeV/$c$] & $K^{+}$ &  $K^{-}$ \\ \hline
0.45 & 1.18e+00 $\pm$ 2.7e-02 $\pm$ 8.2e-02 & 1.06e+00 $\pm$ 1.9e-02 $\pm$ 7.4e-02 \\ 
0.55 & 8.18e-01 $\pm$ 1.8e-02 $\pm$ 5.7e-02 & 7.48e-01 $\pm$ 1.3e-02 $\pm$ 5.2e-02 \\ 
0.65 & 6.07e-01 $\pm$ 1.3e-02 $\pm$ 4.3e-02 & 5.30e-01 $\pm$ 9.6e-03 $\pm$ 3.7e-02 \\ 
0.75 & 3.72e-01 $\pm$ 8.4e-03 $\pm$ 2.6e-02 & 3.43e-01 $\pm$ 6.7e-03 $\pm$ 2.4e-02 \\ 
0.85 & 2.50e-01 $\pm$ 6.1e-03 $\pm$ 1.8e-02 & 2.14e-01 $\pm$ 4.6e-03 $\pm$ 1.5e-02 \\ 
0.95 & 1.73e-01 $\pm$ 4.7e-03 $\pm$ 1.2e-02 & 1.40e-01 $\pm$ 3.4e-03 $\pm$ 9.8e-03 \\ 
1.05 & 1.12e-01 $\pm$ 3.3e-03 $\pm$ 7.8e-03 & 9.05e-02 $\pm$ 2.5e-03 $\pm$ 6.3e-03 \\ 
1.15 & 7.94e-02 $\pm$ 2.7e-03 $\pm$ 5.6e-03 & 6.17e-02 $\pm$ 1.9e-03 $\pm$ 4.3e-03 \\ 
1.25 & 4.88e-02 $\pm$ 1.9e-03 $\pm$ 3.4e-03 & 4.35e-02 $\pm$ 1.5e-03 $\pm$ 3.0e-03 \\ 
1.35 & 3.41e-02 $\pm$ 1.5e-03 $\pm$ 2.4e-03 & 2.84e-02 $\pm$ 1.2e-03 $\pm$ 2.0e-03 \\ 
1.45 & 2.45e-02 $\pm$ 1.2e-03 $\pm$ 1.7e-03 & 1.96e-02 $\pm$ 9.2e-04 $\pm$ 1.4e-03 \\ 
1.55 & 1.63e-02 $\pm$ 9.5e-04 $\pm$ 1.1e-03 & 1.34e-02 $\pm$ 7.6e-04 $\pm$ 9.4e-04 \\ 
1.65 & 1.28e-02 $\pm$ 8.0e-04 $\pm$ 9.1e-04 & 9.77e-03 $\pm$ 6.2e-04 $\pm$ 7.0e-04 \\ 
1.75 & 9.56e-03 $\pm$ 6.8e-04 $\pm$ 7.1e-04 & 6.65e-03 $\pm$ 4.8e-04 $\pm$ 4.9e-04 \\ 
1.85 & 6.34e-03 $\pm$ 5.4e-04 $\pm$ 5.0e-04 & 4.87e-03 $\pm$ 4.2e-04 $\pm$ 3.8e-04 \\ 
1.95 & 5.28e-03 $\pm$ 5.1e-04 $\pm$ 4.4e-04 & 3.45e-03 $\pm$ 3.7e-04 $\pm$ 2.9e-04 \\ 
\end{tabular} \end{ruledtabular}
\label{tab:kaon-62}
%~\\
%~\\
\caption{$p$ and $\pbar$ cross sections ($E \frac{d^3\sigma}{dp^{3}}$ [mb GeV$^{-2}c^3$]) in $p+p$ collisions at $\sqrt{s} =$ 62.4 GeV.
Statistical (2nd column) and systematic (3rd column) uncertainties are shown for each particle species.
The normalization uncertainty (11\%) is not included. Feed-down weak decay corrections are not applied.}
\begin{ruledtabular} \begin{tabular}{ccc}
$\pt$ [GeV/$c$] & $p$ &  $\overline{p}$ \\ \hline
0.65 & 4.63e-01 $\pm$ 7.1e-03 $\pm$ 4.2e-02 & 3.09e-01 $\pm$ 4.6e-03 $\pm$ 2.2e-02 \\ 
0.75 & 3.28e-01 $\pm$ 5.4e-03 $\pm$ 3.0e-02 & 2.19e-01 $\pm$ 3.6e-03 $\pm$ 1.5e-02 \\ 
0.85 & 2.49e-01 $\pm$ 4.5e-03 $\pm$ 2.2e-02 & 1.59e-01 $\pm$ 2.9e-03 $\pm$ 1.1e-02 \\ 
0.95 & 1.69e-01 $\pm$ 3.4e-03 $\pm$ 1.5e-02 & 1.10e-01 $\pm$ 2.3e-03 $\pm$ 7.7e-03 \\ 
1.05 & 1.20e-01 $\pm$ 2.7e-03 $\pm$ 1.1e-02 & 7.50e-02 $\pm$ 1.8e-03 $\pm$ 5.3e-03 \\ 
1.15 & 8.12e-02 $\pm$ 2.1e-03 $\pm$ 7.3e-03 & 4.95e-02 $\pm$ 1.4e-03 $\pm$ 3.5e-03 \\ 
1.25 & 5.81e-02 $\pm$ 1.7e-03 $\pm$ 5.2e-03 & 3.32e-02 $\pm$ 1.1e-03 $\pm$ 2.3e-03 \\ 
1.35 & 3.95e-02 $\pm$ 1.4e-03 $\pm$ 3.6e-03 & 2.37e-02 $\pm$ 9.4e-04 $\pm$ 1.7e-03 \\ 
1.45 & 2.55e-02 $\pm$ 9.9e-04 $\pm$ 2.3e-03 & 1.53e-02 $\pm$ 7.1e-04 $\pm$ 1.1e-03 \\ 
1.55 & 1.84e-02 $\pm$ 8.4e-04 $\pm$ 1.7e-03 & 1.07e-02 $\pm$ 6.0e-04 $\pm$ 7.5e-04 \\ 
1.65 & 1.37e-02 $\pm$ 7.2e-04 $\pm$ 1.2e-03 & 7.03e-03 $\pm$ 4.7e-04 $\pm$ 4.9e-04 \\ 
1.75 & 9.31e-03 $\pm$ 5.8e-04 $\pm$ 8.4e-04 & 4.49e-03 $\pm$ 3.7e-04 $\pm$ 3.1e-04 \\ 
1.85 & 5.90e-03 $\pm$ 4.4e-04 $\pm$ 5.3e-04 & 3.39e-03 $\pm$ 3.4e-04 $\pm$ 2.4e-04 \\ 
1.95 & 4.02e-03 $\pm$ 3.6e-04 $\pm$ 3.6e-04 & 2.12e-03 $\pm$ 2.4e-04 $\pm$ 1.5e-04 \\ 
2.05 & 3.11e-03 $\pm$ 3.1e-04 $\pm$ 2.8e-04 & 1.58e-03 $\pm$ 2.2e-04 $\pm$ 1.1e-04 \\ 
2.15 & 1.99e-03 $\pm$ 2.5e-04 $\pm$ 1.8e-04 & 1.04e-03 $\pm$ 1.7e-04 $\pm$ 7.3e-05 \\ 
2.25 & 1.37e-03 $\pm$ 2.1e-04 $\pm$ 1.2e-04 & 6.99e-04 $\pm$ 1.5e-04 $\pm$ 4.9e-05 \\ 
2.35 & 8.94e-04 $\pm$ 1.5e-04 $\pm$ 8.0e-05 & 5.90e-04 $\pm$ 1.3e-04 $\pm$ 4.1e-05 \\ 
2.45 & 6.34e-04 $\pm$ 1.3e-04 $\pm$ 5.7e-05 & 3.13e-04 $\pm$ 1.1e-04 $\pm$ 2.2e-05 \\ 
2.55 & 6.33e-04 $\pm$ 1.4e-04 $\pm$ 5.7e-05 & 2.43e-04 $\pm$ 8.3e-05 $\pm$ 1.7e-05 \\ 
2.65 & 4.56e-04 $\pm$ 1.2e-04 $\pm$ 4.1e-05 & 1.80e-04 $\pm$ 7.9e-05 $\pm$ 1.3e-05 \\ 
2.75 & 4.11e-04 $\pm$ 1.1e-04 $\pm$ 3.7e-05 & 1.74e-04 $\pm$ 7.5e-05 $\pm$ 1.2e-05 \\ 
2.85 & 2.40e-04 $\pm$ 9.5e-05 $\pm$ 2.2e-05 & 2.39e-04 $\pm$ 9.1e-05 $\pm$ 1.7e-05 \\ 
2.95 & 1.63e-04 $\pm$ 6.6e-05 $\pm$ 1.5e-05 & 6.57e-05 $\pm$ 5.0e-05 $\pm$ 4.7e-06 \\ 
3.10 & 9.65e-05 $\pm$ 3.7e-05 $\pm$ 8.9e-06 & 7.07e-05 $\pm$ 2.8e-05 $\pm$ 5.1e-06 \\ 
3.30 & 9.05e-05 $\pm$ 4.1e-05 $\pm$ 8.5e-06 & 4.14e-05 $\pm$ 3.2e-05 $\pm$ 3.1e-06 \\ 
3.50 & 2.13e-05 $\pm$ 1.9e-05 $\pm$ 2.0e-06 & 5.21e-05 $\pm$ 3.2e-05 $\pm$ 4.0e-06 \\
\end{tabular} \end{ruledtabular}
\label{tab:ppbar-62}
\end{table*} \endgroup

%%\clearpage

\begingroup \squeezetable
%%%%%%%%%%%%%%%%%%%%%%%%%%%%%%%%%%%%%%%%%%%%%%%%%%%%  Table_XVII
\begin{table*}[htb]
\caption{$p$ and $\pbar$ cross sections ($E \frac{d^3\sigma}{dp^{3}}$ [mb GeV$^{-2}c^3$]) in $p+p$ collisions at $\sqrt{s} =$ 62.4 GeV.
Statistical (2nd column) and systematic (3rd column) uncertainties are shown for each particle species.
The normalization uncertainty (11\%) is not included. Feed-down weak decay corrections are applied.}
\begin{ruledtabular} \begin{tabular}{ccc}
$\pt$ [GeV/$c$] & $p$ &  $\overline{p}$ \\ \hline
0.65 & 2.95e-01 $\pm$ 4.5e-03 $\pm$ 6.6e-02 & 1.18e-01 $\pm$ 1.8e-03 $\pm$ 6.5e-02 \\ 
0.75 & 2.38e-01 $\pm$ 3.9e-03 $\pm$ 3.8e-02 & 1.20e-01 $\pm$ 2.0e-03 $\pm$ 3.4e-02 \\ 
0.85 & 1.96e-01 $\pm$ 3.5e-03 $\pm$ 2.5e-02 & 1.05e-01 $\pm$ 1.9e-03 $\pm$ 1.9e-02 \\ 
0.95 & 1.40e-01 $\pm$ 2.8e-03 $\pm$ 1.6e-02 & 8.12e-02 $\pm$ 1.7e-03 $\pm$ 1.1e-02 \\ 
1.05 & 1.03e-01 $\pm$ 2.3e-03 $\pm$ 1.1e-02 & 5.91e-02 $\pm$ 1.4e-03 $\pm$ 6.6e-03 \\ 
1.15 & 7.18e-02 $\pm$ 1.9e-03 $\pm$ 7.1e-03 & 4.07e-02 $\pm$ 1.1e-03 $\pm$ 4.0e-03 \\ 
1.25 & 5.23e-02 $\pm$ 1.5e-03 $\pm$ 5.1e-03 & 2.81e-02 $\pm$ 9.4e-04 $\pm$ 2.6e-03 \\ 
1.35 & 3.60e-02 $\pm$ 1.2e-03 $\pm$ 3.4e-03 & 2.04e-02 $\pm$ 8.1e-04 $\pm$ 1.8e-03 \\ 
1.45 & 2.34e-02 $\pm$ 9.1e-04 $\pm$ 2.2e-03 & 1.33e-02 $\pm$ 6.2e-04 $\pm$ 1.1e-03 \\ 
1.55 & 1.70e-02 $\pm$ 7.7e-04 $\pm$ 1.6e-03 & 9.42e-03 $\pm$ 5.2e-04 $\pm$ 7.8e-04 \\ 
1.65 & 1.27e-02 $\pm$ 6.7e-04 $\pm$ 1.2e-03 & 6.19e-03 $\pm$ 4.1e-04 $\pm$ 5.1e-04 \\ 
1.75 & 8.67e-03 $\pm$ 5.4e-04 $\pm$ 8.1e-04 & 3.97e-03 $\pm$ 3.3e-04 $\pm$ 3.2e-04 \\ 
1.85 & 5.51e-03 $\pm$ 4.1e-04 $\pm$ 5.1e-04 & 3.00e-03 $\pm$ 3.0e-04 $\pm$ 2.4e-04 \\ 
1.95 & 3.76e-03 $\pm$ 3.3e-04 $\pm$ 3.5e-04 & 1.88e-03 $\pm$ 2.2e-04 $\pm$ 1.5e-04 \\ 
2.05 & 2.91e-03 $\pm$ 2.9e-04 $\pm$ 2.7e-04 & 1.41e-03 $\pm$ 2.0e-04 $\pm$ 1.1e-04 \\ 
2.15 & 1.86e-03 $\pm$ 2.4e-04 $\pm$ 1.7e-04 & 9.24e-04 $\pm$ 1.5e-04 $\pm$ 7.4e-05 \\ 
2.25 & 1.28e-03 $\pm$ 2.0e-04 $\pm$ 1.2e-04 & 6.21e-04 $\pm$ 1.3e-04 $\pm$ 5.0e-05 \\ 
2.35 & 8.39e-04 $\pm$ 1.4e-04 $\pm$ 7.7e-05 & 5.25e-04 $\pm$ 1.2e-04 $\pm$ 4.2e-05 \\ 
2.45 & 5.95e-04 $\pm$ 1.2e-04 $\pm$ 5.5e-05 & 2.78e-04 $\pm$ 9.4e-05 $\pm$ 2.2e-05 \\ 
2.55 & 5.94e-04 $\pm$ 1.3e-04 $\pm$ 5.5e-05 & 2.16e-04 $\pm$ 7.4e-05 $\pm$ 1.7e-05 \\ 
2.65 & 4.28e-04 $\pm$ 1.1e-04 $\pm$ 3.9e-05 & 1.60e-04 $\pm$ 7.0e-05 $\pm$ 1.3e-05 \\ 
2.75 & 3.86e-04 $\pm$ 1.1e-04 $\pm$ 3.6e-05 & 1.55e-04 $\pm$ 6.6e-05 $\pm$ 1.2e-05 \\ 
2.85 & 2.25e-04 $\pm$ 8.9e-05 $\pm$ 2.1e-05 & 2.13e-04 $\pm$ 8.1e-05 $\pm$ 1.7e-05 \\ 
2.95 & 1.54e-04 $\pm$ 6.2e-05 $\pm$ 1.4e-05 & 5.85e-05 $\pm$ 4.4e-05 $\pm$ 4.7e-06 \\ 
3.10 & 9.06e-05 $\pm$ 3.4e-05 $\pm$ 8.5e-06 & 6.30e-05 $\pm$ 2.5e-05 $\pm$ 5.2e-06 \\ 
3.30 & 8.50e-05 $\pm$ 3.8e-05 $\pm$ 8.1e-06 & 3.69e-05 $\pm$ 2.8e-05 $\pm$ 3.1e-06 \\ 
3.50 & 2.00e-05 $\pm$ 1.7e-05 $\pm$ 2.0e-06 & 4.64e-05 $\pm$ 2.8e-05 $\pm$ 4.0e-06 \\ 
\end{tabular} \end{ruledtabular}
\label{tab:ppbar-feeddown-62}
\end{table*} \endgroup

\end{document}